\documentclass[acmsmall]{acmart}

\usepackage{booktabs} % For formal tables
\usepackage{mathrsfs} % script font
\usepackage{bm}
\usepackage{breqn}
\usepackage{graphicx}
\usepackage{tikz}
\usetikzlibrary{patterns,shapes,backgrounds,shapes,positioning,petri,topaths}
\usepackage{pgfplots}
\usepackage{subcaption}
\usepackage[ruled,linesnumbered]{algorithm2e}
\usepackage{float}
\usepackage{adjustbox}
\usepackage{tabularx}
\usepackage{enumitem}
\usepackage[percent]{overpic}

\setcopyright{acmlicensed}
\acmJournal{PACMCGIT}
\acmYear{2021} \acmVolume{4} \acmNumber{3} \acmArticle{} \acmMonth{9} \acmPrice{15.00}\acmDOI{10.1145/3480143}

\setcopyright{rightsretained}
\newcommand{\R}{\mathbb{R}}
\makeatletter
\newcommand{\removelatexerror}{\let\@latex@error\@gobble}
\makeatother

\DeclareMathOperator*{\argmin}{arg\,min}

\DeclareMathOperator*{\tr}{tr}

\def\R{\mathbb{R}}

\def\vec#1{\mathbf{#1}} % Vector quantity

\def\egor#1{#1}

\def\new#1{#1}

\usepackage[utf8]{inputenc}

\begin{document}
\title{Volume Preserving Simulation of Soft Tissue with Skin}
%\titlenote{Produces the permission block, and copyright information}
%\subtitle{Extended Abstract}
%\subtitlenote{The full version of the author's guide is available as \texttt{acmart.pdf} document}

\author{Seung Heon Sheen}
\email{shsheen@cs.ubc.ca}
\affiliation{%
  \institution{University of British Columbia}
  \city{Vancouver}
  \state{British Columbia}
  \country{Canada}%
  %\postcode{V6T 1Z4}
}

\author{Egor Larionov}
\email{egor@cs.ubc.ca}
\affiliation{%
  \institution{University of British Columbia}
  \city{Vancouver}
  \state{British Columbia}
  \country{Canada}%
  %\postcode{V6T 1Z4}
}

\author{Dinesh K. Pai}
\email{pai@cs.ubc.ca}
\affiliation{%
\institution{University of British Columbia}%
\department{Computer Science}%
\city{Vancouver}%
\state{BC}%
\country{Canada}%
}%
\affiliation{%
\institution{Vital Mechanics Research}%
\city{Vancouver}%  
\state{BC}%
\country{Canada}%
}

\begin{abstract}
  Simulation of human soft tissues in contact with their environment
  is essential in many fields, including visual effects and apparel
  design.  Biological tissues are nearly incompressible. However,
  standard methods employ compressible elasticity models and achieve
  incompressibility indirectly by setting Poisson's ratio to be close
  to 0.5.  This approach can produce results that are plausible
  qualitatively but inaccurate quantatively. This approach also causes
  numerical instabilities and locking in coarse discretizations or
  otherwise poses a prohibitive restriction on the size of the time
  step.
  We propose a novel approach to alleviate these issues by
  replacing indirect volume preservation using Poisson's ratios with
  direct enforcement of zonal volume constraints, while controlling
  fine-scale volumetric deformation through a cell-wise compression penalty.
  To increase realism, we propose an epidermis model to mimic the
  dramatically higher surface stiffness on real skinned bodies.
  We demonstrate that our method produces stable realistic
  deformations with precise volume preservation but without locking
  artifacts. 
  Due to the volume preservation not being tied to mesh discretization, 
  our method also allows a resolution consistent simulation of incompressible materials.
  Our method improves the stability of the standard
  neo-Hookean model and the general compression recovery in the Stable
  neo-Hookean model.

  % Furthermore, standard models for soft tissue simulation lack
  % material properties of the skin, which is substantially more stiff
  % than the underlying soft tissue.

  %    \todo{We generalize the notion of the Poisson's ratio from the traditional definition associated with incompressibility of the material, where $\nu = 0.5$ signifies a completely incompressible material. In our method, volume can be lost per cell but not everywhere.  }

    %Our method also significantly
    %improves the predictive qualities of volume preserving simulations with coarse meshes.
\end{abstract}

%
% The code below should be generated by the tool at
% http://dl.acm.org/ccs.cfm
% Please copy and paste the code instead of the example below.
%



\begin{teaserfigure}
  \centering
  \includegraphics[width=1.0\textwidth]{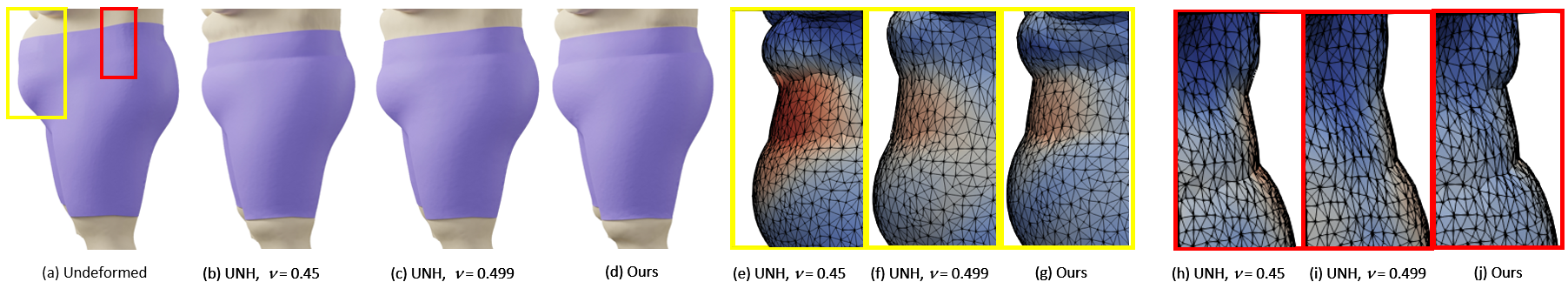}
  \caption{Simulating the fit of a tight garment requires realistic
    soft tissue deformation and illustrates the challenges of Finite
    Element Methods for human body simulations. Human tissues are soft
    but nearly incompressible. The neo-Hookean models popular in
    computer graphics enforce incompressibility indirectly, using a
    Poisson's ratio $\nu$ approaching 0.5; we call
    this standard approach ``Unconstrained neo-Hookean''
    (UNH). Naively using $\nu = 0.4545$ produces qualitatively
    reasonable deformation, but the body loses a significant amount
    ($\sim$19\%) of its volume, mostly in regions of high compression
    (see (b), and close-ups (e,h)). Using $\nu = 0.499$ preserves body
    volume up to an error of 0.5\%, but produces limited deformations
    due to locking (c,f,i).  By contrast, in our ``Constrained
    neo-Hookean'' (CNH) method, we can designate volume preserving
    zones to match anatomical compartments, and exactly conserve
    volume within each zone while avoiding problems with volumetric
    locking. This results in more realistic displacement of soft
    tissues (d,g,j).
    \\
    In addition to visual differences, volume preservation can lead to
    significant differences in the predictions of how well the garment
    fits; UNH with $\nu = 0.4545$ predicts a waistband circumference 7
    cm smaller than that with $\nu = 0.499$, and 4.5cm smaller than
    with our method.  Our results can improve predictions of human
    soft tissue mechanics in applications ranging from virtual try-ons
    to visual effects.  }
  \label{fig:teaser}
\end{teaserfigure}

% Upper waistband 134.13 140.77 138.60
% Lower 143.57 151.09 148.09
% (/ (+ (- 134.13 138.60)  (- 143.57 148.09)) 2) -4.4950000000000045
% (/ (+ (- 134.13 140.77)  (- 143.57 151.09)) 2)-7.0800000000000125

\maketitle

\section{Introduction}

Elastic materials are ubiquitous in everyday life. Many objects we
interact with are organic in nature such as plants, animals, food, and
most importantly our own bodies. Interestingly, most organic solids
are nearly incompressible (due to their high water content), which
makes them particularly difficult to simulate. Human soft tissue, for
instance, is essentially incompressible, with a Poisson's ratio close
to 0.5 \cite{fung2013biomechanics}. As a result, much of contemporary
research in computer graphics focuses on robust simulation of
incompressible hyperelastic solids (see Section~\ref{sec:related}). We
focus on the popular neo-Hookean models, which are relatively simple
while including non-linearity and the temptation to control
incompressibility by setting Poisson's ratio $\nu \simeq 0.5$.

% Unfortunately, this approach for soft tissue simulation enforces
% incompressibility indirectly with a per-element energy term, which
% means that the material parameters like Poisson's ratio are
% dependent on the resolution of the simulation mesh. In addition,
% high Poisson's ratios cause numerical instability and locking when
% coupled with a coarse discretization \cite{Irving:2007}.

However, it is impossible to emulate true incompressibility and
extremely difficult to simulate even near-incompressibility with this
approach.  This is because as the material approaches
incompressibility $\nu \rightarrow 0.5$, the first Lam\'e parameter
% (which penalizes volume change)
$\lambda \rightarrow +\infty$ (see Section~\ref{sec:back} for
background). The numerical and visual artifacts arising from the failure
to correctly enforce incompressibility is known as {\em \new{volumetric} locking}
\new{(for ease of discussion, we will simply refer to it as `locking'.)}

There are multiple approaches to tackle this issue: the simplest being
just using higher-order elements \cite{longva:2020} or hexahedral elements. However, the
increased computational cost and difficulty of implementation might
not be desirable. Another class of popular methods is non-conforming
finite elements (such as the Discontinous Galerkin class of methods),
where the additional or non-conforming degrees of freedom allow
significant deformation and therefore reduce the stiffness of the
system.  The last approach includes methods that seek to remove these
element-wise constraints through Mixed Finite Elements or coarsened
constraints, both of which are related to our method.

Our core idea is to tease apart the concept of {\em
		incompressibility}, a constraint on a derivative (the deformation
gradient) from the related concept of {\em volume preservation}, a
constraint on an integral (the volume of a finite region of material
that we call a ``zone''). Incompressibility is enforced per element in
the standard neo-Hookean models, usually implicitly, using an energy
term. By contrast, we enforce volume preservation as an explicit
constraint on the volume of a zone. Volume preservation gives us
considerable flexibility to choose larger zones that span multiple
elements, zones that are independent of discretization, and zones that
are aligned with meaningful anatomical tissue compartments (muscles,
abdomen, breast, etc.). Zones may also overlap (e.g., we can preserve
both the total volume of a body, and volumes of important tissue
compartments).

A second key idea is that since volume preservation is already
enforced using constraints, we can use much smaller values of
$\lambda$ or $\nu$, thereby avoiding locking and related numerical
instabilities. This can, of course, lead to volume loss per element
but that will be compensated by volume gain in other elements in a
zone to preserve volume. In other words, our simulation mesh may be
viewed as a type of Arbitrary Lagrangian-Eulerian (ALE) mesh, in which
volume is never lost but allowed to flow from one cell to another.
Locking is always aggravated when using a coarse simulation mesh,
but our method allows a stable simulation of volume preserving materials
with a coarse mesh.
To our knowledge, this technique has not yet been closely studied with
FEM simulations in computer graphics.

Note that $\lambda$ now controls only element volume, rather than the
incompressibility of the material. In the rest of the paper we will
repurpose the first Lam\'e parameter $\lambda$ to control volume
change \emph{per element}, instead of \emph{zonal} volume change.
This allows us to penalize extreme volume loss per element.
We address this in Section~\ref{sec:penalty} with an amendment to the
volume penalty term found in compressible elastic energy
models. Additionally, the proposed correction improves the compression
response when using invertible energies.
%Compared to the Stable neo-Hookean
%model \cite{Smith:2018}, our penalty method stays faithful to the
%classical neo-Hookean compression penalty, while providing reliable
%inversion recovery during extreme
%compression.% \todo{Add note about comparison with Stomakhin}

Human bodies are covered by a layer of skin, a complex multi-layered structure. The outer layer comprising the epidermis is much stiffer than the underlying tissues, and significantly affects the quality of deformation. We propose a simple model of the epidermis and show that this
extension contributes heavily to the appearance of realistic tissue deformation.

\begin{figure}[tb]
	\centering
	\begin{subfigure}{.2\linewidth}
		\centering
		\includegraphics[width=1.0\textwidth]{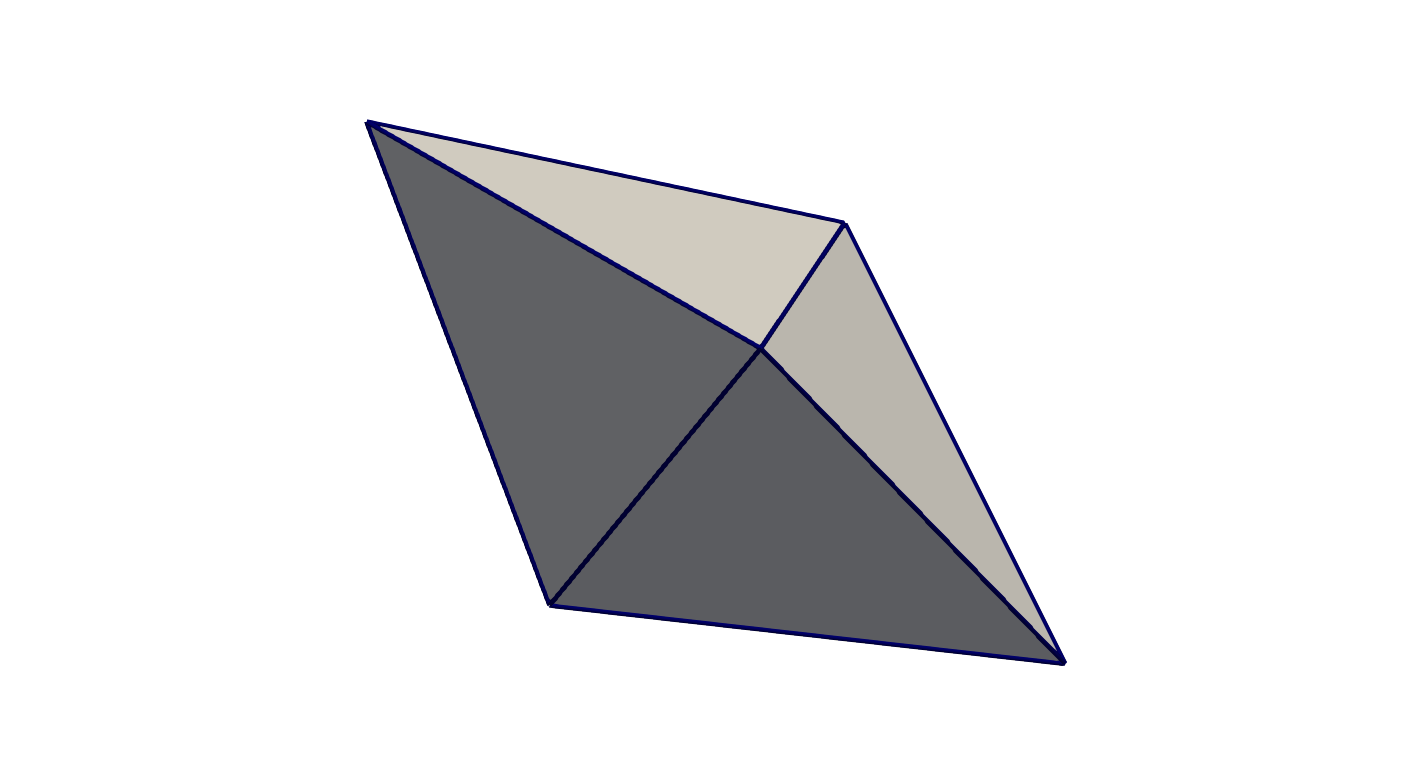}
		\caption*{(a)}
		\label{sfig:simple_ref}
	\end{subfigure}%
	\begin{subfigure}{.2\linewidth}
		\centering
		\includegraphics[width=1.0\textwidth]{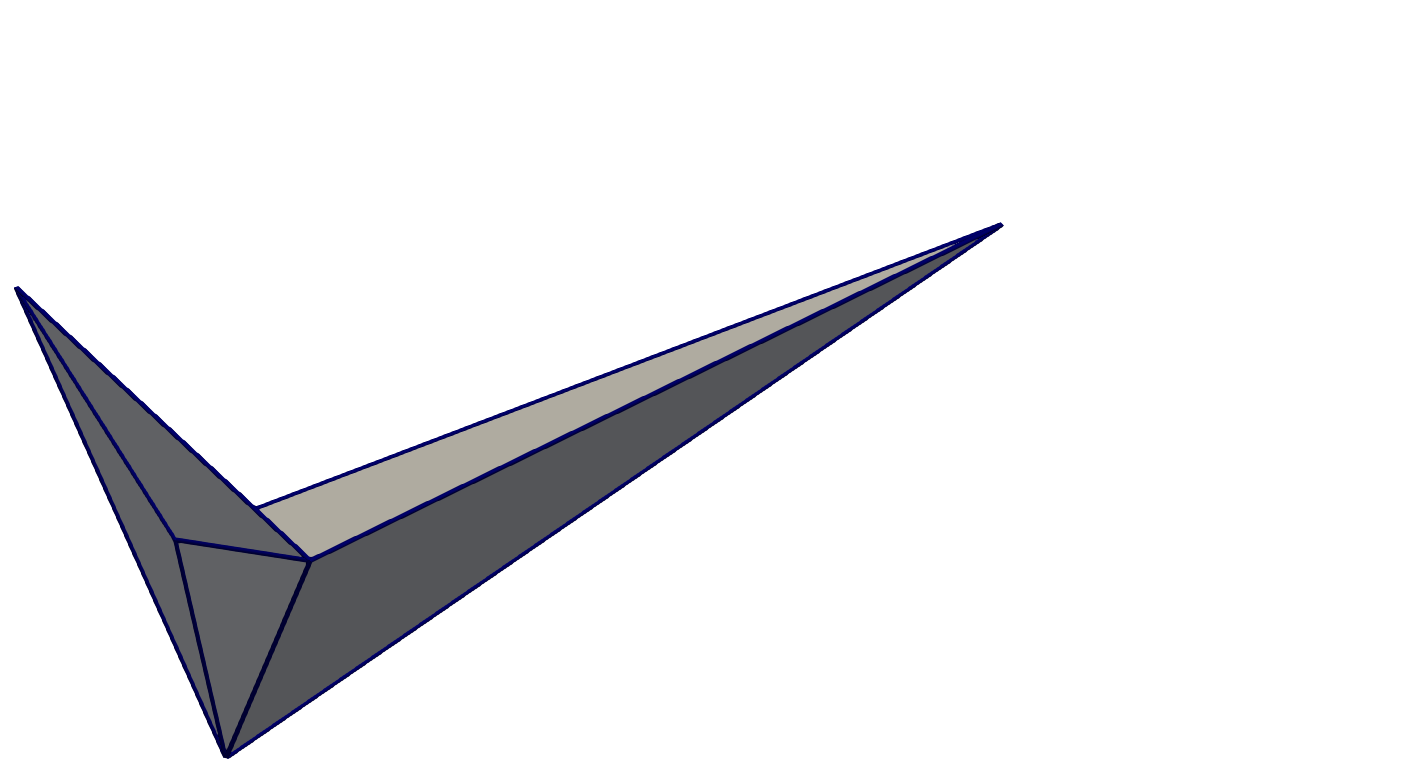}
		\caption*{(b)}
		\label{sfig:simple}
	\end{subfigure}%
	\begin{subfigure}{.2\linewidth}
		\centering
		\includegraphics[width=1.0\textwidth]{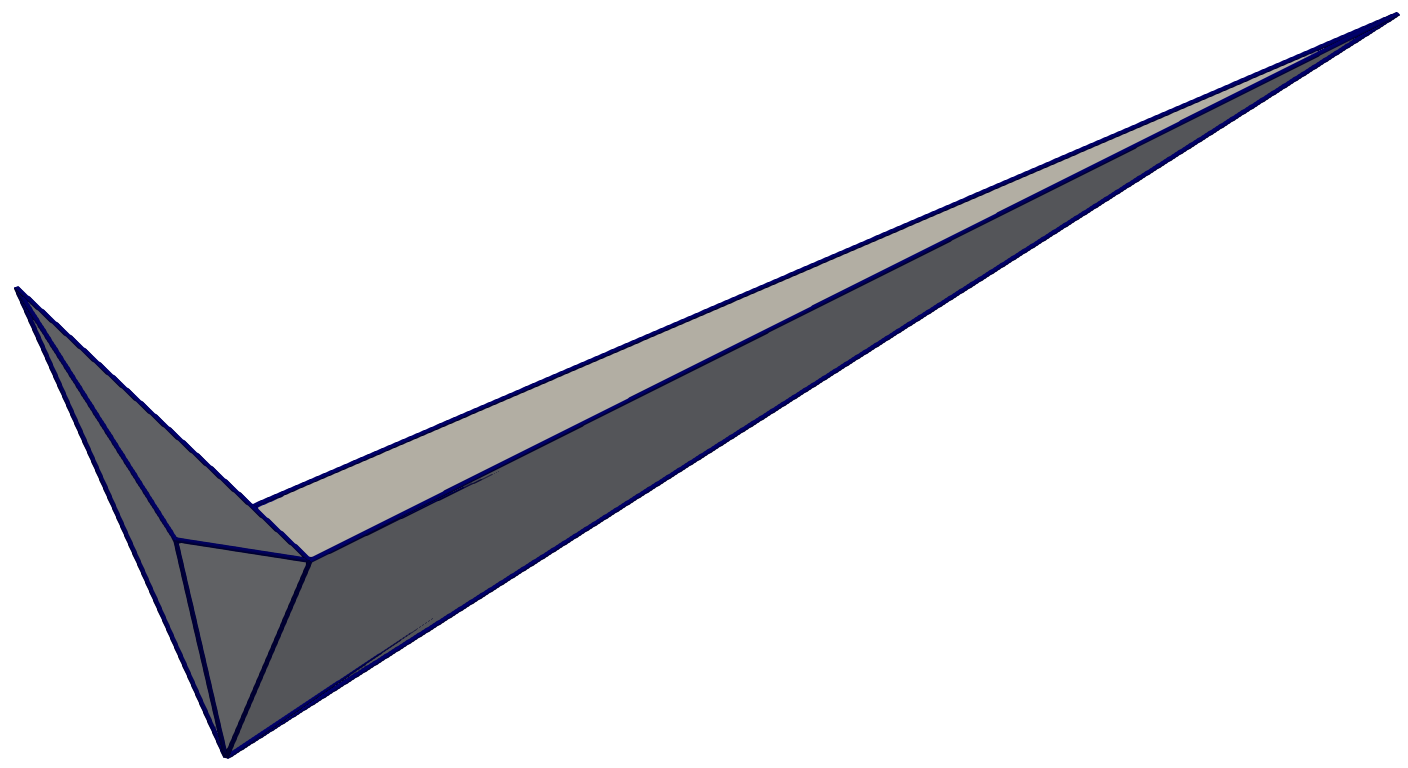}
		\caption*{(c)}
		\label{sfig:simple_vc}
	\end{subfigure}%
	\caption{\textbf{Two Tet Simulation}: (a) The reference state of the mesh, where two
		tetrahedra of equal volume are joined together by a face. The left tetrahedron is constrained
		by Dirichlet boundary conditions to be compressed into a plane. (b) With an unconstrained neo-Hookean (UNH) model Without the
		volume constraint, the total volume of the final mesh is $46.628\%$ of the original. (c) With a volume constraint (CNH model), the tetrahedron on the right inflates to twice
		the volume to keep the total volume constant.
	}
	\label{fig:twotets}
\end{figure}

A simple illustration of these ideas is given in
Figure~\ref{fig:twotets}. It illustrates the more general scenario in
which locking artifacts increase at lower mesh resolutions, whereas our
volume preservation is independent of the discretization of the zones.

% there
% are two tetrahedral elements connected at one triangle face; one
% element is flattened, while the other is allowed to deform. When
% compression is penalized locally on each element, the total volume of
% the system will never be preserved, and so the two element system will
% necessarily lose volume. 

% In the context of computer graphics, we are most interested in
% conserving the aesthetic properties visible on the surface of the
% solid.  With this in mind, we propose to constrain the volume of the
% solid zonally, and relax the Poisson's ratio of the material. This
% allows volume in the same zone to flow from one element to another,
% which removes locking artifacts.  This technique turns the purely
% Lagrangian approach to FEM simulations into an arbitrary
% Lagrangian-Eulerian formulation (ALE) because the FEM mesh becomes
% loosely coupled to the solid material.  

% DKP This could be moved to Results
Figure~\ref{fig:teaser} shows the practical relevance of good volume
preservation.  Closeups of the belly (yellow boxes) and side waist in
front view (red boxes) depict tissue displacement in false color, and
yield more insights. We see that the traditional Unconstrained
neo-Hookean (UNH) model compresses under the waistband by losing volume,
without significantly extruding tissue away (e,h), whereas our method
extrudes tissue more realistically, producing a sharp bulge (g,j) due to
volume preservation.  Increasing $\nu$ doesn't help the UNH models since
locking reduces the deformation (f,i).

	{\em Contributions:} we propose a new approach for simulating human
tissues and other soft objects that preserve volume, while avoiding
the common pitfalls of standard incompressible elasticity models. In
addition to avoiding locking artifacts, our zonal volume constraint
formulation makes deformation independent of discretization, and
allows zones to be aligned with meaningful anatomical tissue
compartments.  In addition, we repurpose the first Lam\'e parameter to
support inversion robustness and introduce a new form of the
local compression penalty.  We also extend the elastic energy potential to
model the stiff epidermis, and demonstrate its importance.  Finally,
we propose a simple but complete pipeline for assigning volumetric
zones using weights on the surface of the volumetric mesh, and
demonstrate the application of these methods to predicting the fit of
tight fitting garments.

%%% Local Variables:
%%% mode: latex
%%% TeX-master: "sigconf"
%%% End:

\section{Related Work}
\label{sec:related}

A number of recent contributions have significantly improved the
performance and behavior of hyperelastic solid simulations.  The
standard approach is using the Finite Element Method (FEM) on a
Lagrangian tetrahedral (or hexahedral) mesh \cite{Sifakis:2012}.  The
methods used for soft tissue simulations are typically split between
linear and non-linear hyperelasticity models.  A number of popular
elasticity models are used for soft tissue simulation including
neo-Hookean, St. Venant-Kirchhoff, and co-rotated elasticity.

Co-rotated elasticity \cite{Muller:2002,McAdams:2011}, has been tremendously successful in real-time
and interactive applications largely due to its simplicity.  However, it suffers from element
degeneration in large deformations \cite{Civit:2014} and has poor volume preservation properties
\cite{Smith:2018}.  Non-linear energy models like the neo-Hookean models \cite{BonetWood:2008}, have
been used to circumvent these issues at a larger computational cost; although in recent years,
neo-Hookean elasticity has also appeared in interactive simulations \cite{Liu:2017}.

Our work targets invariant-based non-linear hyperelastic models, such as neo-Hookean elasticity, for
their generality, superb handling of large deformations and inherent reflection stability.

Among non-linear elastic models are compressible and incompressible hyperelastic models. While
incompressible models \cite{Mooney:1940,Rivlin:1948} pose an explicit volume constraint on each
element, compressible models prevent severe compression using a penalty term \cite{BonetWood:2008}.
The most popular method for solving elasticity problems in computer graphics is the standard linear
FEM on a Lagrangian mesh because of its performance profile and versatility.
Unfortunately, imposing a severe penalty --- let alone a hard constraint --- for volume change on
each element can cause severe numerical difficulties and locking, especially in linear tetrahedral FEM.
% \todo{Need a reference to support this claim}.

In traditional FEM, volumetric locking is addressed by decoupling incompressibility from displacements with Mixed Finite Element Methods.
In computer graphics, Irving et al. \shortcite{Irving:2007} have addressed locking in tetrahedral meshes by
softly constraining the volume of the one ring around each vertex in a tetrahedral mesh using position
and velocity correction steps. This approach is an application of nodal strain elements \cite{bonet:1998},
where stress and strain, in this case their volumetric component, is nodally interpolated in a mixed FEM.
However, without additional stabilization of some sort, these types of mixed elements are known to be unstable
and the number of additional pressure variables are proportional to the number of nodes.
A further discussion of Mixed Finite Element Methods is presented in the next section.
Kaufmann et al. \shortcite{Kaufmann:2012} looked to solve locking by introducing additional degrees of freedom
to the system by using a Discontinuous Galerkin discretization. Smith et al. \shortcite{Smith:2018} proposed
the Stable neo-Hookean energy model to handle invertible elements as well as improve stability for high Poisson's ratios.

Irving et al. \shortcite{Irving:2007} proposed a method of (weakly) constraining volume in
each 1-ring neighborhood of every mesh vertex. This method is an implementation of the
Average Nodal Pressure element \cite{bonet:1998}, where the cell-wise constant pressure samples of the
one-ring neighbors are averaged on the nodes. By contrast, our method uses pressure samples
(i.e. the volume preserving zones) that are coarser and decoupled from the mesh topology.
Fine-scale cell-wise pressures are instead controlled through a local penalty, which avoids
additional pressure variables.  This way we can keep $\lambda$ low enough to
avoid locking, while simultaneously enforcing volume preservation.
Our method has significantly fewer constraints, compared to the total number of vertices, which
permits enforcing constraints exactly using constrained minimization to
solve the variational problem rather than using constraint projection.

A recent work in computer graphics by Fra\^{n}cu et al \shortcite{francu:2021} targets the problem
of locking by using a linear-linear mixed formulation. However, as noted by the authors, this
formulation does not satisfy the Bab\u{u}ska-Brezzi inf-sup condition and thus will result in
inaccurate pressure solutions. Specifically, their results show the same problems of
checkerboard patterns in the pressure modes. Although the spurious modes do not affect the
displacement results as the authors mention, it causes problems when a reasonable computation of
pressures is necessary, for example when one needs them to simulate frictional contact. Our method
does not suffer from this issue, since we avoid the spurious modes in pressure by sampling pressures
from a coarser scale compared to the displacements. Compare their Figure~15 to our Figure~\ref{fig:suspended-cubes}.
Our method also requires fewer constraints compared to theirs, since we only require one constraint
per ``zone", while they require a constraint per node.
% compare their fig 15 to our fig ?

Some works have targeted global volume preservation
\cite{Hong:2006,Hirota:2000,Promayon:1996,Diziol:2011}, however not in the context of volumetric
FEM.  Global volume constraints have also been applied in studies in skinning methods
\cite{Rohmer:2009}. Some also proposed using a sweep-based approach to conserve the volume
\cite{Yoon:2006,Angelidi:2004} or a vector field approach \cite{Funck:2007}. By contrast, we propose
zonal volume constraints for neo-Hookean type energy models for Lagrangian FEM simulations.

Finite element simulations also suffer from element inversions during severe deformation.
Inversion stability allows simulations to handle large deformations and permits taking large time
steps, which can improve simulation performance significantly.
A line of recent work has proposed methods for resolving element inversions by extending the energy
density function to the negative volume region.  Force filtering methods
\cite{Irving:2004,Teran:2005} have improved inversion stability but suffer from subtle problems
including invalid inversion recovery directions or derivative drift as thoroughly explored in
\cite{Smith:2018}. Stomakhin et al. \shortcite{Stomakhin:2012} propose a $C^1$ or $C^2$ extension of the
entire energy density function for low volume fractions, which resolves many of these problems.
However, filtering methods can be quite sensitive to appropriate specification of filtering thresholds
and reflection conventions \cite{wang:2016}. We instead follow a simpler approach similar to \cite{Smith:2018},
where we design a volumetric penalty term to satisfy necessary conditions for stability and inversion
robustness. Our penalty function improves upon the Stable neo-Hookean volumetric term by also introducing nonlinearity
to the stress, resulting in better inversion recovery and improved performance.

\section{Background}
\label{sec:back}

\subsection{Variational Elasticity}
In this section, we establish the context for our contributions by introducing FEM simulation of
hyperelastic materials as a variational problem.

Let $\Omega \subset \R^{3}$ be a union of mesh tetrahedra representing an elastic solid in its
undeformed configuration. Then let $\vec{x} \in \R^{3n}$ correspond to a stacked vector of mesh vertex
positions that prescribe the deformation of the solid, where $n$ is the total number of vertices
in the mesh. In an elasticity problem, we are
interested in finding the configuration $\vec{x}$ that results in the lowest potential energy
for the elastic solid $\Omega$ given a set of boundary conditions
and external forces. Mathematically, we may write the problem statement as
\begin{align}
 	\vec{x}^* := \argmin_{\vec{x}} W(\vec{x}),
  	\label{eq:energy_min}
\end{align}
where $W(\vec{x})$ represents the elastic work function for configuration $\vec{x}$. 
This formulation allows conservative external forces to be added as additional potentials in the
objective, however for the sake of simplicity we ignore external forces in the following sections.

With linear (constant strain) elements, $W$, which is the integral of energy density function $\Psi(\vec{x})$, can be written as the sum of volume-scaled per-element energies:
\begin{align}
  W(\vec{x}) = \int_{\Omega} \Psi(\vec{x}) := \sum_{e} V_{e} \Psi(\vec{F}_e(\vec{x})),
  \label{eq:total_energy}
\end{align}
where $V_e$ is the volume of element $e$ in the reference configuration, which depends on the element deformation gradient $\vec{F}_e$.  The choice of the energy density function $\Psi$ determines the hyperelastic energy model. 

For the time discretization, we may use any integration method. In the
dynamics examples below we adopt the implicit Euler time integrator
and add an inertial energy term to this minimization. However for
simplicity of exposition we focus on static FEM to describe our
approach to volume preservation.

\subsection{Incompressibility and Locking}
\label{subsec:IncompLocking}
There are two ways in which incompressibility could be enforced: either directly, as a constraint that the volume is preserved, or indirectly with a penalty term that powerfully resists compression. Since both ways are frequently referred to as ``incompressible,'' to avoid confusion we will refer to incompressible neo-Hookean models using the first method as {\em ``Constrained neo-Hookean''} (CNH), and those using the second method as
{\em ``Unconstrained neo-Hookean''} (UNH).

Most incompressible hyperelastic energy models used in graphics are of
the Unconstrained neo-Hookean type, and penalize element-wise volume change with a term scaled by
the first Lam\'e parameter $\lambda$, which depends on the Young's Modulus $E$ and Poisson's Ratio
$\nu$ as:
\begin{align}
  \lambda = \frac{E \nu}{(1+\nu)(1-2\nu)}.
  \label{eq:poisson2lame}
\end{align}

For instance, the most common version of such an energy density function
\cite{BonetWood:2008} is written as
\begin{align}
	\Psi_{\text{UNH}}(\vec{F}; \lambda, \mu) = \frac{\mu}{2}(I_C - 3) - \mu \log J + \frac{\lambda}{2} (\log J)^2
	\label{eq:neohookean}
\end{align}
where $I_C = \tr(\vec{F}^{\top}\vec{F})$, and $J = \det(\vec{F})$ represents the fraction of
volume after deformation. This means that when $J$ is close to zero (extreme compression),
$\Psi_{\text{UNH}}$ will generate large penalty forces to restore the element to reference
configuration. Another commonly used material model, co-rotated elasticity \cite{McAdams:2011} is written as
\begin{align*}
	\Psi_{\text{CR}}(\vec{F}; \lambda, \mu) = \mu\|\vec{F} - \vec{R}\|^2_F + \frac{\lambda}{2} \tr(\vec{S} - \vec{I})^2,
\end{align*}
where $\vec{R}$ and $\vec{S}$ form the polar decomposition: $\vec{F} = \vec{R}\vec{S}$ and $\vec{I}$
is the $3\times3$ identity matrix. Here, in a similar fashion, local compression is once again
penalized by $\lambda$. 
% However, as discussed in the introduction, high Poisson's ratios make the system stiff, which results in stiffness related issues such as instability, and artificial damping.

\new{
	There are multiple aspects of locking which are problematic for simulating volume preserving elastic solids.
	First, high Poisson's ratios make the system stiff, which results in stiffness related issues such as instability and artificial damping.  
	Somewhat related to this, when using linear tetrahedral elements and element-wise volume constraints, the resulting system becomes highly overconstrained.
	However, the main problem arises from the choice of the constitutive equation, when volumetric stress depends on $\lambda$.
	In classical FEM theory, C\'ea's Lemma dictates that the quasi-best approximation error depends not only on mesh discretization error, but also on $\lambda$. 
	Hence, when $\lambda \rightarrow +\infty$, the finite element solution can no longer be a reliable predictor of the solution of the PDE. 
	A more detailed explanation can be found in Braess \shortcite{Braess:2007}.
}

\subsection{Mixed Finite Element Methods}
It is often necessary to compute reliable solutions not only for displacements but also for
pressures (e.g., for frictional contact or fractures). For displacement-based one-field FEM,
pressure must be computed from the displacement variables $\vec{x}$. Specifically, cell-wise
hydrostatic pressure is usually computed as the negative of the divergence of the Cauchy stress
tensor. Since the Cauchy stress tensor is related to the derivative of the energy density function
$\Psi(\vec{x})$, the pressures computed from a one-field FEM mainly depend on the volume
term of $\Psi$. However, due to similar issues as discussed above, when the material is
incompressible and $\lambda \rightarrow +\infty$ the volume term stops being a reliable model for
volumetric stress.

One traditional way of decoupling incompressibility from $\lambda$ is by introducing an
additional pressure variable $\vec{p}$ that models the volumetric stress component of elements,
interpolated separately from displacement on the finite element mesh \cite{Bathe:2006}. This allows us to reformulate the variational problem as
\begin{align}
	\vec{x}^* & := \arg \max_{\vec{p}} \min_{\vec{x}} \int_{\Omega} \hat{\Psi}(\vec{x}) + \vec{p}^T \vec{c}(\vec{x}),
	\label{eq:mixed_variational}
\end{align}
where $\hat{\Psi}(\vec{x})$ is the deviatoric component of the displacement-based elastic potential,
and $\vec{c}(\vec{x})$ is a term that relates $\vec{p}$ to $\vec{x}$. This additional term can be
interpreted as a constraint on $\vec{p}$ to be proportional to the hydrostatic pressure computed
from the displacements $\vec{x}$. Then, $\vec{p}$ becomes the Lagrange multiplier for the constraint
$\vec{c}(\vec{x})$.  One implementation of this type of formulation is shown in Sussman et al.
\shortcite{Sussman:1987}.  These methods are known as the displacement-pressure Mixed Finite Element
Methods and are one of the most accurate ways to solve the problem.

With the additional degree of freedom, the Bab\u{u}ska-Brezzi inf-sup condition restricts the
choice of the space of finite element basis for the additional variable for the method to be stable \cite{bathe:2001}.
This condition dictates that the order of basis for the displacement variables must be higher than that
of the pressure
variables. Specifically, for conforming tetrahedral elements the lowest order finite element space choices are either
the Hood-Taylor elements ($P_2$ for $\vec{x}$ / $P_1$ for $\vec{p}$, where $P_k$ denotes the space of $k$-th order polynomials), or MINI ($P_1^+ / P_1$, where superscript $+$ denotes an enrichment of cubic bubble \cite{arnold:1984}).
Hence, Mixed FEM with a simple linear tetrahedral Finite Element basis for displacement is usually not valid for stable simulations.
This includes the Average Nodal Pressure elements proposed in Irving et al. \shortcite{Irving:2007}, where the Lagrange multipliers of 1-ring volume constraints can be interpreted as cell-wise constant pressure variables ($P_0$) being averaged on the nodes. Although this alleviates some of the problems arising from each element being constrained, it still fails to meet the inf-sup condition and spurious modes may occur without additional stabilization \cite{puso:2006}.

The intuition for our approach from mixed FEM is that, to achieve an efficient and stable computation of the additional pressure variables, one must sample the pressure variables in a coarser scale compared to the displacement variables. Then, we are able to split the pressure computation into a coarser and finer scale, to control the coarse-scale pressures as separate pressure variables as Lagrange multipliers for volume constraints, as in Eq.\ref{eq:mixed_variational}, and compute the fine-scale pressures from displacements. Therefore, we look to a much more efficient and simpler approach by enforcing a volume constraint for a few larger zones of elements, and modeling the element-wise local pressure as an additional local penalty term.

\section{Zonal Volume Constraint}

%\begin{figure}[t]
%	\centering
%	\begin{tikzpicture}
%		\useasboundingbox (-1,-1) rectangle (1,2); 
%		\draw (-1,0) node[fill, draw, circle ,minimum size=1.5mm,inner sep=0pt,outer sep=0pt, label=left:$\vec{x}_0$](p0){};
%		\draw (0.5,-0.8) node[fill, draw, circle ,minimum size=1.5mm,inner sep=0pt,outer sep=0pt, label=south:$\vec{x}_2$](p2){};
%		\draw (0.7,0.3) node[fill, draw, circle ,minimum size=1.5mm,inner sep=0pt,outer sep=0pt, label=north east:$\vec{x}_1$](p1){};
%		\draw (0.2,1.3) node[fill, draw, circle ,minimum size=1.5mm,inner sep=0pt,outer sep=0pt, label=north east:$\vec{x}_3$](p3){};
%		\draw[line width=1.2pt, -] (-1,0) -- (0.5,-0.8);
%		\draw[line width=1.2pt, -, opacity=0.5] (-1,0) -- (0.7,0.3);
%		\draw[line width=1.2pt, opacity=0.5] (0.5,-0.8) -- (0.7,0.3);
%		\draw[line width=1.2pt, -] (-1,0) -- (0.2,1.3);
%		\draw[line width=1.2pt] (0.2,1.3) -- (0.5,-0.8);
%		\draw[line width=1.2pt, opacity=0.5] (0.2,1.3) -- (0.7,0.3);
%	\end{tikzpicture}	
%  \caption{A sample tetrahedral element, with 4 vertices at positions $\vec{x}_0$, $\vec{x}_1$, $\vec{x}_2$, and
%  $\vec{x}_3$.}
%	\label{fig:f2}
%\end{figure}

To solve the problem of volumetric locking, instead of enforcing a per-tetrahedron
near-incompressibility with high Poisson's ratio,
%we propose to enforce incompressibility through constraining volumes of zones defined as local sets of finite elements. Specifically, we add zonal volume constraints to the general energy minimization problem defined in equation~\eqref{eq:energy_min}. 
we adopt the approach of the Mixed Finite Element Method. Essentially,
we solve the saddle-point system as a constrained minimization with constraint function
$\vec{c}(\vec{x}) = \vec{0}$. Specifically, our constraint enforces the total volumes of zones defined as
local sets of finite elements to be preserved.
Compared to other mixed elements, our approach is much more efficient and easier to implement while showing
comparable results. Moreover, our approach provides the modeling flexibility of choosing zones that
are aligned with anatomical compartments (see \egor{Section~\ref{sec:zoning}}).

Each constraint is simply formulated as the requirement that the total volume of all elements in a
specified zone of the deformed mesh is equal to the initial volume. That is, for $j$-th zone $G_j$, the zonal
volume constraint function $c_j$ is defined as follows,
\begin{equation}
  c_j(\vec{x}) = \sum_{e \in \zeta_j} V_e(\vec{x}) - V_e^0,
\end{equation}
where $V_e^0$ is the reference volume of element $e$, which belongs to zone $j$ with element index set $\zeta_j$.

Imposing this constraint for each zone gives us a new constrained minimization problem:
\begin{align}
  \argmin_{\vec{x}}\  & \int_{\Omega} \Psi(\vec{x})       \\
  \text{s.t.}\     & c_j(\vec{x}) = 0 \quad \forall j.
  \label{eq:constrained-min}
\end{align}
As a special case we can preserve the total volume with a single \emph{global} constraint; by
contrast, classical incompressible neo-Hookean models require the volume of each and every element
to be preserved.

To illustrate the simplicity of this type of constraint we define the volume constraint for a
tetrahedral mesh. The volume of the tetrahedron $e$ is defined
(up to a constant scaling) as the triple scalar product
\begin{align*}
  V_e(\vec{x}) = \vec{v}_3 \cdot (\vec{v}_1 \times \vec{v}_2),
\end{align*}
where $\vec{v}_i = \vec{x}_i - \vec{x}_0$.  Then the Jacobian of the constraint function can be
computed as
\begin{align*}
  \frac{\partial c_j}{\partial \vec{x}} = \sum_{e} \frac{\partial V_e}{\partial \vec{x}},
\end{align*}
where the sparse vector $\frac{\partial V_e}{\partial \vec{x}} \in \R^{3n}$ is zero everywhere
except for the vertices of element $e$, where
\begin{align*}
  \left[\frac{\partial V_e}{\partial \vec{x}}\right]_0 =
  -(\vec{v}_2 \times \vec{v}_3 + \vec{v}_3 \times \vec{v}_1 + \vec{v}_1 \times
  \vec{v}_2)
\end{align*}\begin{align*}
  \left[\frac{\partial V_e}{\partial \vec{x}}\right]_1 = \vec{v}_2 \times \vec{v}_3, \quad
  \left[\frac{\partial V_e}{\partial \vec{x}}\right]_2 = \vec{v}_3 \times \vec{v}_1, \quad
  \left[\frac{\partial V_e}{\partial \vec{x}}\right]_3 = \vec{v}_1 \times \vec{v}_2.
\end{align*}

%\begin{figure}[t]
%  \centering
%  \begin{subfigure}{.49\linewidth}
%    \centering
%    \includegraphics[width=1.0\textwidth]{images/medpuck_049_50.png}
%    \caption*{(a)}
%    \label{sfig:medpuck_049_50}
%  \end{subfigure}%
%  \begin{subfigure}{.49\linewidth}
%    \centering
%    \includegraphics[width=1.0\textwidth]{images/medpuck_049_vcip_50.png}
%    \caption*{(b)}
%    \label{sfig:medpuck_049_vcip_50}
%  \end{subfigure}\par\medskip
%  \begin{subfigure}{.49\linewidth}
%    \centering
%    \includegraphics[width=1.0\textwidth]{images/puck_049_vcip.png}
%    \caption*{(b)}
%    \label{sfig:puck_049_vcip_50}
%  \end{subfigure}%
%  \begin{subfigure}{.49\linewidth}
%    \centering
%    \includegraphics[width=1.0\textwidth]{images/puck_0475_vcip_50.png}
%    \caption*{(d)}
%    \label{sfig:puck_0475_vcip_50}
%  \end{subfigure}%
%  \caption{\textbf{Reproducing High-Res Simulations}: (a) is a high-resolution puck with 180K
%    tetrahedrons with $\nu = 0.49$ without a zonal volume constraint. (b) is the result with a global
%    volume constraint with the same Poisson's ratio $\nu = 0.49$. (c) shows a low-res simulation
%    with global volume constraint with the Poisson's ratio $\nu = 0.49$, (d) is a lower resolution
%    22K tetrahedrons with our method with $\nu = 0.475$ that recreates the high-resolution results
%    (a) and (b) closely. \todo{remove example and replace with some other case where low-res recreates hi-res}}
%  \label{fig:res_compare}
%\end{figure}
%\todo{fig:res_compare parts c and d. not clear what you are trying to say}
Finally, the Hessian stencils for each $V_e$ will be simple linear skew-symmetric matrices.
% DKP: this needs elaboration that elements are linear in the
% vertex positions. For now, let's save that for another day, since we
% don't exploit it yet.
Thus, $c_{j}(\vec{x}) = 0$ is a one dimensional constraint with simple to implement sparse
derivatives, which gives true volume preservation.
% imcompressibility with respect to the surface of the solid
Note that this constraint can be further optimized by computing the volume of the entire zone by
iterating over zone boundary faces only.

Although uncomplicated, this constraint provides a powerful tool for emulating incompressible
elasticity. It allows users to achieve volume preservation without increasing Poisson's ratio,
which can cause instabilities and locking.
The constrained optimization problem can be solved with any non-linear optimizer which can deal with nonlinear constraints.
For most nonlinear solvers a few equality constraints should not be prohibitively expensive to solve, but for additional performance gain one may naturally use an Augmented Lagrangian method to solve the constraints.

The zone sizes are important in determining the level of local incompressibility. One global zone
for the entire mesh will essentially be a hydrostatic simulation, analogous to simulating a water
balloon. As the zone sizes decrease, there will be more local incompressibility around each element
which will result in a stiffer behavior. However, as long as the zones are at least as large as the
1-ring \cite{Irving:2007}, volumetric locking will not occur. Therefore, as we use a smaller zone
sizing, the results will become more similar to the results in \cite{Irving:2007}, but at a steeper performance cost.

Our method can be viewed as a simplification of the 2-field mixed formulation, where the pressure potential is given as
\begin{equation}
  \vec{p}^{T}\vec{c}(\vec{x}) = \sum_j \vec{p}_j c_j(\vec{x}) = \sum_j \vec{p}_j \left( \sum_{e \in G_j}
  V_e(\vec{x}) - V_e^0 \right),
  \label{eq:variational_lagrange}
\end{equation}
where the interpolated hydrostatic pressures $\vec{p}_j$ for zone $j$ are identified to be the
Lagrange multipliers for the $j$-th zonal volume constraint. If each element was assigned to a unique
zone, our method would recreate the mixed-element formulation for incompressible materials. However,
we use only a handful of zones, which keeps the problem size small and avoids locking and instability.

\subsection{Stabilization}
We apply the F-bar method \cite{neto:2005} to the energy density function to ensure stability.
\new{This method relaxes the near-incompressibility constraint enforced by the constitutive model by a modification of the deformation gradient,}
based on a multiplicative split of the deformation gradient $\vec{F}$ into a deviatoric and volumetric component. 
The deviatoric part is then computed as
\begin{equation}
  \bar{\vec{F}}= \alpha \vec{F}, \qquad \mbox{ where } \alpha = \frac{\bar{J}^{\frac{1}{3}}}{J^{\frac{1}{3}}},
  \label{eq:F-bar}
\end{equation}
and $\bar{J}$ is the average of $J$ computed over a set of local element stencils. In our case, the
local sets are the zones where the total volume is preserved, hence conveniently
$\bar{J} = 1$, and $\alpha = J^{-\frac{1}{3}}$. However, since as $J \rightarrow 0$ we have that $\alpha \rightarrow +\inf$, we instead apply a $C^2$ extension to $\alpha$ below a certain threshold $\epsilon$, similarly to Stomakhin et al. \shortcite{Stomakhin:2012}. Then the new extended deviatoric projector is given as
\begin{align}
  \tilde{\alpha} :=
  \begin{cases}
    \begin{array}{l} J^{-\frac{1}{3}} \end{array}  & \text{for } J > \epsilon    \\
    \begin{array}{l@{}l}
      \epsilon^{-\frac{1}{3}} & \,-\, \frac{1}{3} \epsilon^{-\frac{4}{3}} (J - \epsilon)
      \,+\, \frac{2}{9} \epsilon^{-\frac{7}{3}} (J - \epsilon)^2
    \end{array} & \text{for } J \leq \epsilon
  \end{cases}
  \label{eq:c2}
\end{align}
In practice, the choice of $\epsilon$ is not too important as long as it is small ($\sim 0.1$).

We then use $\bar{\vec{F}}(\vec{x})$ to compute the deviatoric part of the constitutive equation. For example,
the deviatoric part of neo-Hookean energy density function \eqref{eq:neohookean} will now be computed as
\begin{align}
  \bar{\Psi}_{\text{NH}}(\vec{\bar{\vec{F}}}; \lambda, \mu) = \frac{\mu}{2}(\tilde{\alpha}^2 I_C - 3).
  \label{eq:deviatoric_neohookean}
\end{align}
Note that this is similar to the form presented by Rivlin \shortcite{Rivlin:1948}, but extended below
$\epsilon$ to be continuously defined for $J \leq \epsilon$.

Using only the deviatoric component of deformation gradient for the elastic potential energy, we remove the contribution of the constitutive equation on the pressure. 
Hence, this allows the complete split of the total elastic stress, to the deviatoric stress from the elastic potential, and the volumetric stress from the constraint Lagrange multipliers and the volume penalty.

%%% Local Variables:
%%% mode: latex
%%% TeX-master: "sigconf"
%%% End:

\section{Local Compression Penalty}
\label{sec:penalty}
Our method ensures that volume is preserved within each zone, but without any element-wise compression
penalty the volume inside each zone can transfer between elements. This is a feature, as discussed in the Introduction, since
it reduces the cost and numerical challenges of enforcing volume preservation locally, while
ensuring good behavior globally.  Note that, unlike in the hydrostatic case, volume can not transfer
completely freely since elastic forces due to the shear modulus restrict large flows.

However, the zonal constraints by themselves model only the hydrostatic pressure in the coarse zones, hence we also need to model the finer-scale pressures in the individual elements.
We employ a more traditional approach to modeling element-wise pressure in the penalty method.
To model this local compression penalty function, we look at the volume penalty functions present in various neo-Hookean elasticity models.
In neo-Hookean models, the bulk modulus controls how much the material resists element-wise volumetric deformation.
The bulk modulus is represented in most neo-Hookean energy formulations in the first Lam\'e parameter $\lambda$, which is a combination of the shear and bulk modulus.
However, as discussed in Section~\ref{sec:back}, when $\lambda \rightarrow +\infty$ locking occurs, and the pressure computations become unstable.
But since we model the coarse-scale pressures as constraints, and we only need to model the finer-scale deviations in pressures, we are able to use a lower $\lambda$ and avoid locking.

\begin{figure}[t]
	\centering
	\begin{subfigure}{.3\linewidth}
		\centering \includegraphics[width=1.5in]{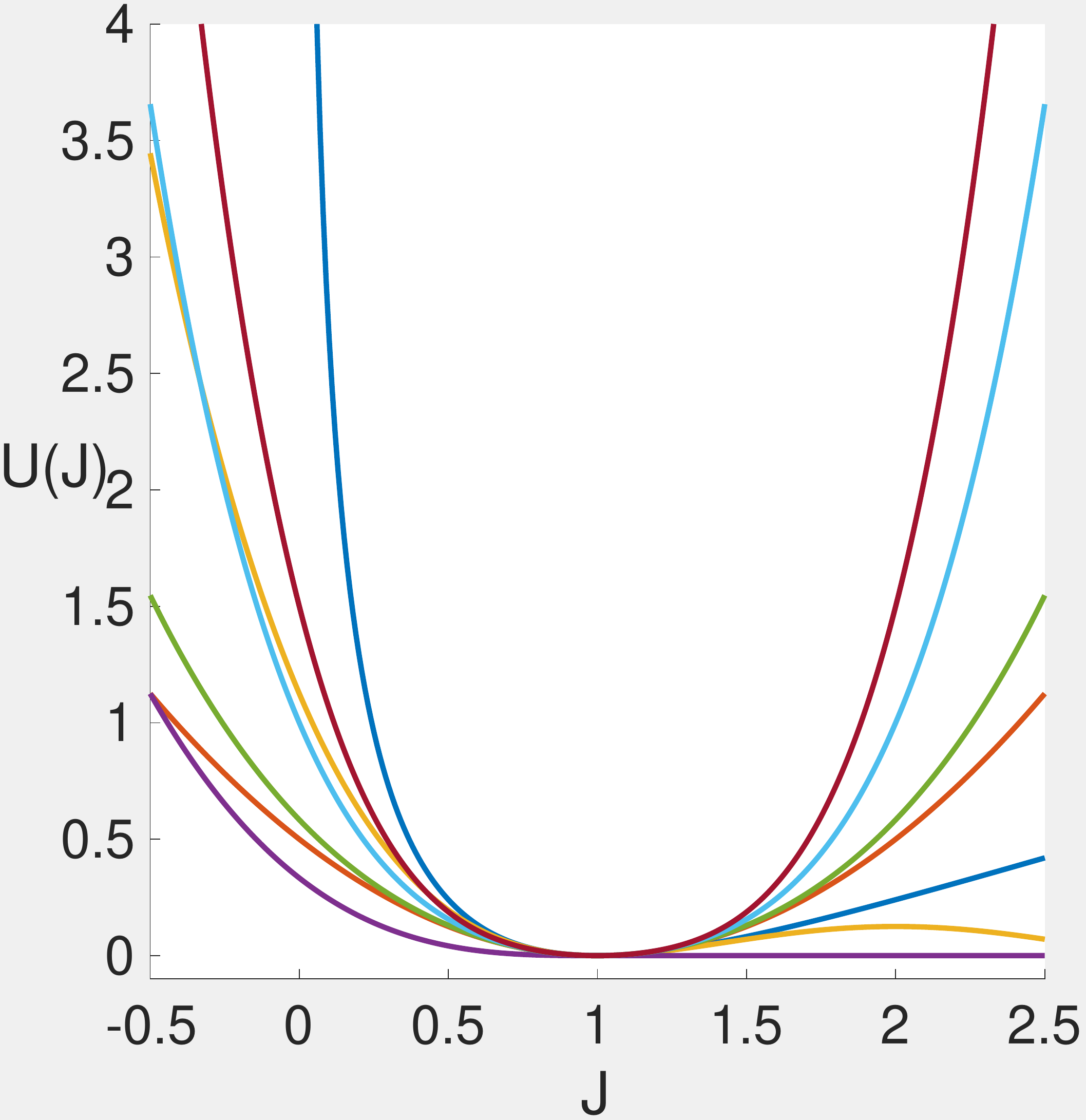}
		\caption*{(a)}
		\label{sfig:energy_plot}
	\end{subfigure}%
	\begin{subfigure}{.3\linewidth}
		\centering \includegraphics[width=1.5in]{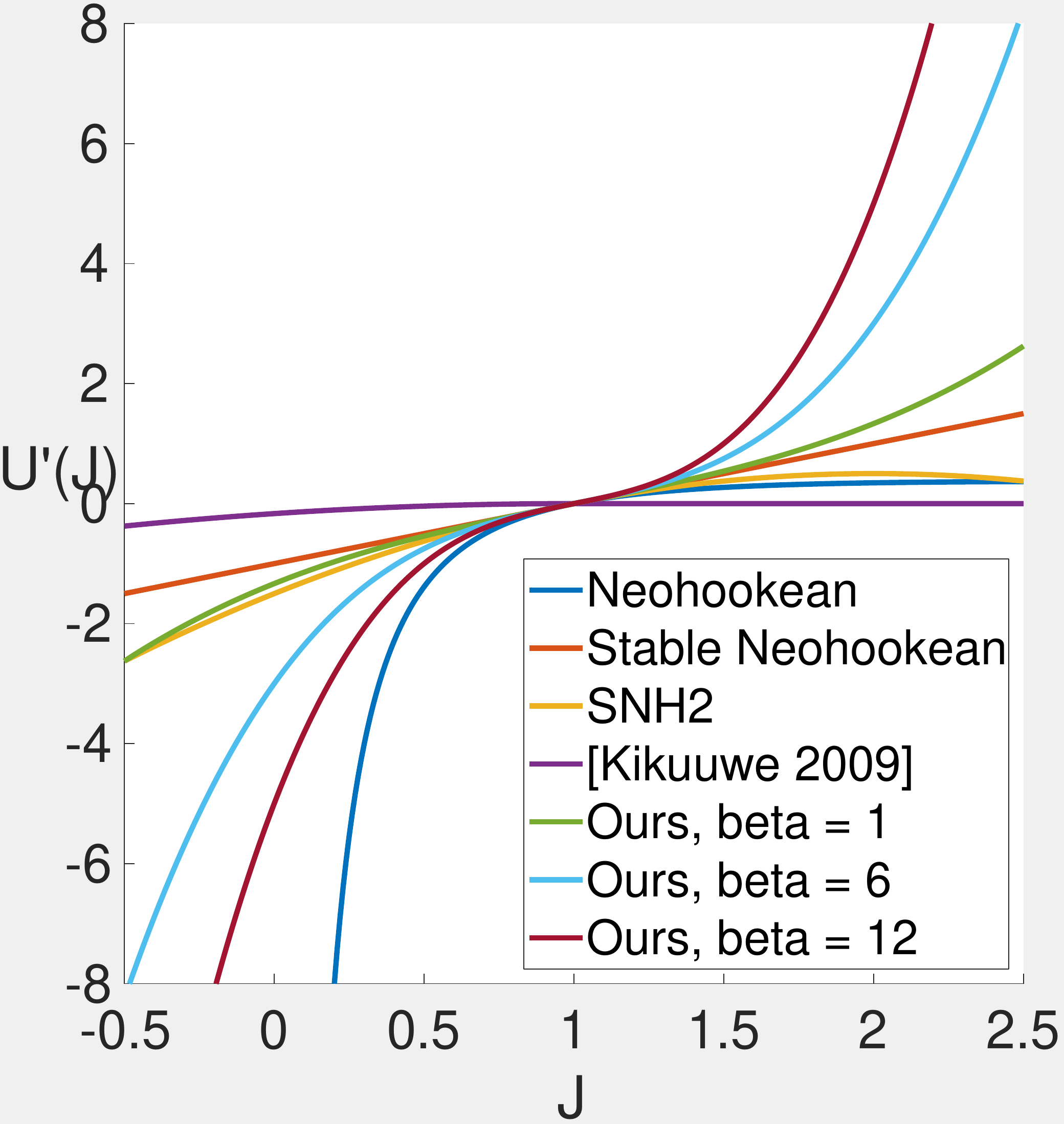}
		\caption*{(b)}
		\label{sfig:stress_plot}
	\end{subfigure}%
	\caption{\textbf{Penalty Plots}: A plot of the penalty terms $U(J)$ and their stresses $U'(J)$ from different
		energy formulations (neo-Hookean, Stable neo-Hookean \cite{Smith:2018}, the second-order expanded version of Stable neo-hookean (SNH2), \cite{kikuuwe:2009},
		ours with $\beta=1$, and ours with $\beta=6$) with $\lambda = 1$, in terms of the relative volume change $J$.
		The neo-Hookean volume term (blue) indicates a substantially larger penalty
		when compared to Stable neo-Hookean (orange)
		and \cite{kikuuwe:2009} (yellow), but the penalty term is undefined when $J \leq 0$ due to its log term.
		This is more evident to see in the stress plot during compression ($J < 1$), where the Stable
		neo-Hookean volumetric stress changes in a linear manner. Our method for both $\beta=1$ (purple) and $\beta=6$ (green) shows much more effective penalization under compression and stretch. Our stresses shows a nonlinear dependence on $J$
		similar to the neo-Hookean term, but also shows a effective growth not only during compression but also during stretch. }
	\label{fig:penalty_plots}
\end{figure}

If $\lambda$ is set too low the simulation is more susceptible to collapsing elements
and even equilibrium configurations with inverted elements for invertible energy models.
Consider the example in Figure~\ref{fig:pucks} of a cylindrical puck with a moving Dirichlet
boundary condition on a set of vertices on top of the puck. As the puck compresses, the tetrahedra
underneath the moving boundary are flattened to the point where subsequent steps cause boundary-adjacent tetrahedra to invert. At this point, incompressible energy models with a logarithmic volume
penalty ($\log J$) term will become undefined because $J \leq 0$. Other models, like co-rotated
elasticity, may permit inverted elements, but won't be able to recover from an inverted
configuration. This issue has motivated a number of solutions \cite{Irving:2004,Smith:2018} for
handling inverted elements, but we will focus on the recent work on the Stable neo-Hookean
model developed by Smith et al. While the proposed model attempts to solve many of the issues with
non-invertible energies and doesn't require additional parameters, as can be seen from the plot
of the volume change penalty term in terms of relative volume change $J$ in Figure~\ref{fig:penalty_plots},
the Stable neo-Hookean energy resists compression much more timidly as compared to
the standard neo-Hookean model defined in Equation~\ref{eq:neohookean}.

This results in the simulation possibly converging to an invalid configuration where inverted elements exist,
and a nonlinear optimization solver can struggle due to inverted elements being present in intermediate solutions which cause oscillations.
Especially, this oscillation can be aggravated when constraints are introduced, presenting major performance issues when one tries to use volume constraints.
We solve this by formulating a new volume penalty term that is both invertible and still resists compression effectively.

Let us write such penalty term as $U(J)$, controlled linearly by parameter $\lambda$.
To design such a penalty term, we first take a look at what conditions the function $U(J)$ must
meet.
A detailed study of various neo-Hookean compression penalty terms and explanations for each of the
conditions can be found in \cite{hartmann:2003}.

\begin{enumerate}[label=\alph*)]
	\item The function must evaluate to 0 at rest ($J = 1$).
	\item The gradient of the function, i.e., the volumetric stress, must also evaluate to 0 at rest.
	\item For the $\lambda$ of the penalty term to  correspond to the Lam\'e parameter in linear elasticity, $\frac{\partial^2 U(1)}{\partial J^2} = 1$ must hold.
	\item The function must be defined for all real numbers $(-\infty, +\infty)$.
	\item $\frac{\partial^2 U(J)}{\partial J^2} \geq 0, \,\forall J \in \R$ for the penalty to both penalize compression and stretch.
\end{enumerate}

Consider the following function,
\begin{equation}
	U(J; \beta) := \frac{1}{12} (J-1)^2 \left[ \beta (J-1)^2 + 6 \right],
	\label{eq:penalty_function}
\end{equation}
where the parameter $\beta \in [0, +\infty)$ controls how steeply the penalty function will penalize change in $J$.
The first and second derivatives of the function are
\begin{align}
	\frac{\partial U(J; \beta)}{\partial J}     & = \frac{1}{3} (J-1) \left[ \beta (J-1)^2 + 3 \right], \text{ and} \\
	\frac{\partial^2 U(J; \beta)}{\partial J^2} & = \beta (J-1)^2 + 1.
	\label{eq:penalty_derivatives}
\end{align}
Therefore, the function satisfies all of the conditions listed above.
Note that for $\beta = 0$, $U(J;0) = U_{\text{SNH}}(J) = \frac{1}{2} (J-1)^2$.
As $\beta$ increases, the penalty function penalizes compression and stretch more effectively than the Stable
neo-Hookean penalty term, while still being fully invertible.
Therefore, this is a suitable choice for our compression penalty term.
Plots comparing different penalty terms $U(J)$ and stresses $\frac{\partial U(J)}{\partial J}$ are shown in Figure~3.
Experimentally, $\beta = 1$ was sufficient for most realistic examples governed by external force, but for examples where inversions were more likely due to contact or boundary conditions,
% we could easily find
higher values of $\beta$ resolved all inversions.

The additional nonlinearity introduced in the gradient (Equation~\ref{eq:penalty_derivatives}) of our penalty function compared to a standard Stable neo-Hookean penalty is the main reason for the inversion-robustness in our model.
It is possible to formulate models with even higher nonlinearity than what we propose here, but in our
experiments we found that such energy models provide no significant benefit in resolving inversions
compared to \eqref{eq:penalty_function} and only increase the number of nonlinear solver iterations
until convergence.
The plot in Figure~\ref{fig:penalty_plots} demonstrates how effectively our penalty resists compression compared to other invertible methods.
%The details of the penalty term is presented in Section~2 of the Supplemental Document.

%% TODO with new energy model

We demonstrate that by this simple addition to the energy potential, we can obtain results similar to
that of using mixed finite elements as in \cite{Irving:2007}, but with very few global constraints (or even one constraint). The results of Figure~\ref{fig:suspended-cubes} demonstrate
that with a local compression penalty equivalent to $\nu = 0.45$ and with just one global zone, the deformations are close to
using a 1-ring constraint around each vertex.
Also, we found that simply adding this additional nonlinearity to the energy resulted in faster performances in most examples 
when volume constraints were used, and even in many cases where there were no constraints.
In Table~\ref{tab:performance}, we compare the performance results of using $\beta = 0$ (equivalent to SNH) and higher $\beta$.

%%e Local Variables:
%%% mode: latex
%%% TeX-master: "sigconf"
%%% End:

\section{Volumetric Zoning}
\label{sec:zoning}

\begin{figure}[ht] 
  \centering
  \adjustbox{trim={.25\width} {.0\height} {.25\width} {.0\height},clip}%
    {\includegraphics[width=4.0in]{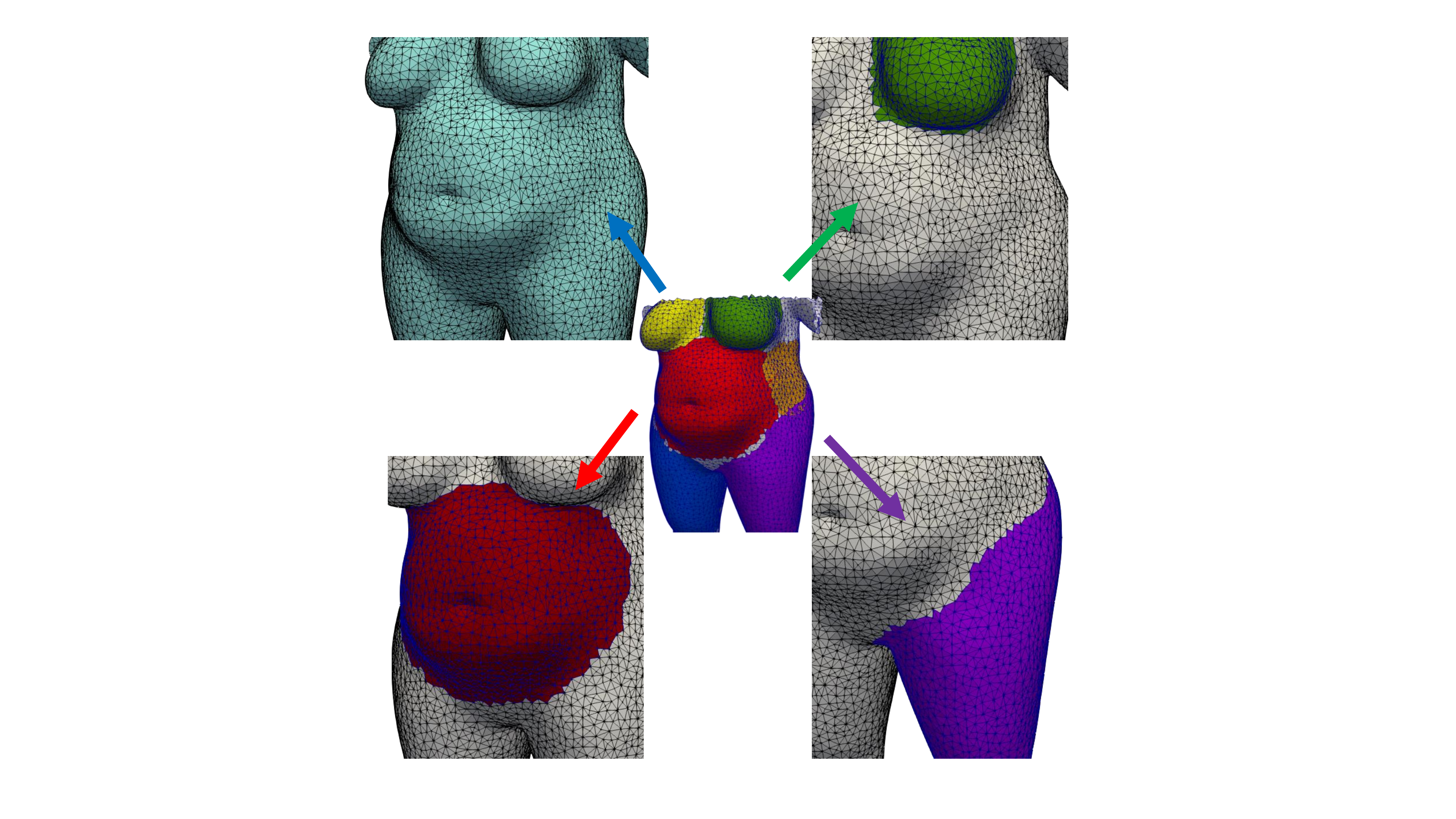}}
    \caption{\textbf{Body Zones}: Volume preserving zones on the body mesh used for simulation in
      Figure~\ref{fig:teaser}. Six significant anatomical compartments (belly (red), side and back waist (orange), 2 legs (blue and purple), 2 breasts (yellow and green)) are selected as zones. Additionally, another zone is added including the full body (cyan), so that the entire body would be volume preserving. Note that the full body zone is not necessary, but can be added to ensure the volume preservation of not only each compartments but also the entire body. The zones are drawn on the mesh surface with a texture painting tool in a visual effects software, and then projected to the inner body using our projection method. }
    \label{fig:body_zones}
  \end{figure}
Human soft tissues are not homogenous but naturally segmented into anatomical compartments. Our zonal volume preservation method provides additional flexibility for modelers to align volume preservation with such anatomical compartments (see Figure~\ref{fig:body_zones} for example).
For a more complete description of one possible workflow for defining such zones, we propose a simple method of obtaining
volumetric zones on a tetrahedral mesh from surface vertex annotation.

Manual vertex painting is a very common part of
many character animation pipelines, allowing users to assign attributes such as skinning weights.
Alternatively, there are methods to automatically compute transformation weights from
skeleton meshes \cite{Rohmer:2009, Baran:2007, Weber:2007}, from sparse subsets of degrees of freedom
\cite{Jacobson:2012}, or from animation data \cite{James:2005}. 

Using either of the aforementioned methods, we end up with a set of weights defining possibly
overlapping zones on the surface of our FEM mesh. We then transfer this data onto the surface
triangles. Naturally these surface zones should be simply connected in order for the volumetric
zones to follow suit.

In order to transfer zone information to the rest of the tetrahedral mesh, we first construct a
smooth potential field around the mesh surface using Hermite Radial Basis Functions
\cite{Pai:2018, Vaillant:2013, Macedo:2009, Wendland:2004}, although any signed distance field will
do.  We then project each tetrahedron centroid along the potential gradient onto the surface
triangles. The triangle zone information is then copied from the triangles, back to the
source tetrahedra.

Albeit simple, this projection is an effective method to map internal tetrahedral mesh elements to
surface triangles. This way, we allow the users to define volume preserving zones by simply
painting surface vertices with any existing tool. We demonstrate
this approach with an sample female body simulation mesh on Figure~\ref{fig:body_zones}, where the
surface zones are chosen manually to capture the anatomical volume-preserving regions. 

\begin{figure*}
	\centering
	\begin{subfigure}{.24\linewidth}
		\centering
		\includegraphics[width=1.0\textwidth]{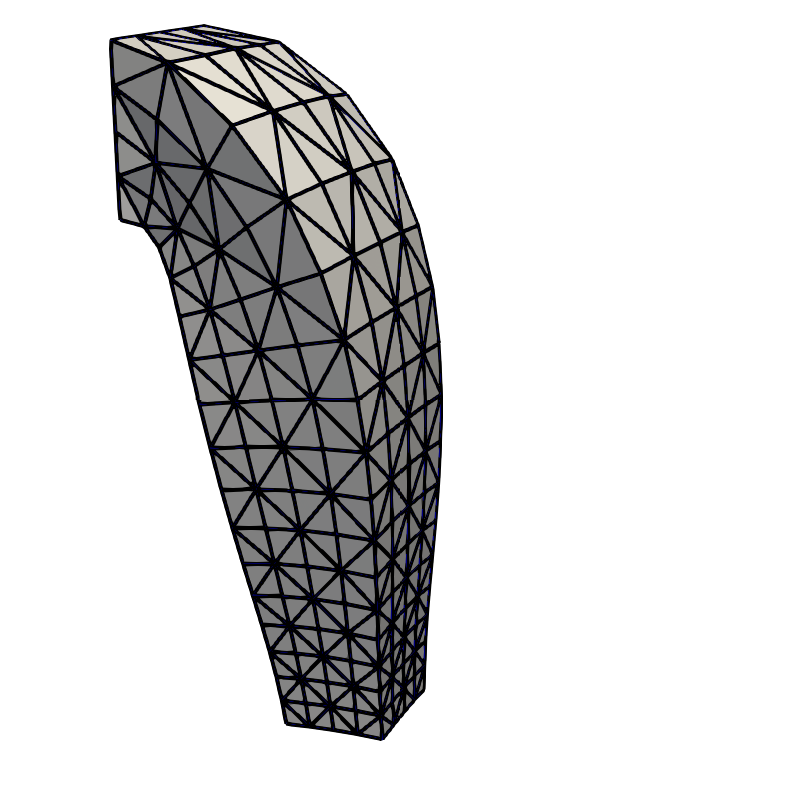}
		\caption*{(a) Hydrostatic}
		\label{sfig:haning_cuboid-hydrostatic}
	\end{subfigure}%
	\begin{subfigure}{.24\linewidth}
		\centering
		\includegraphics[width=1.0\textwidth]{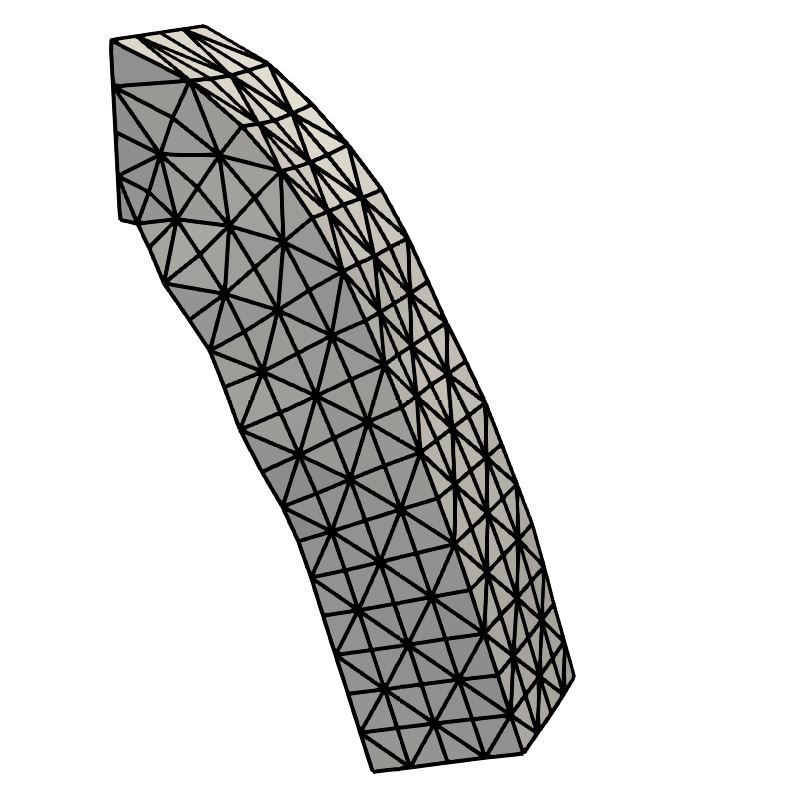}
		\caption*{(b) 4-ring}
		\label{sfig:haning_cuboid-hydrostatic}
	\end{subfigure}%
	\begin{subfigure}{.24\linewidth}
		\centering
		\includegraphics[width=1.0\textwidth]{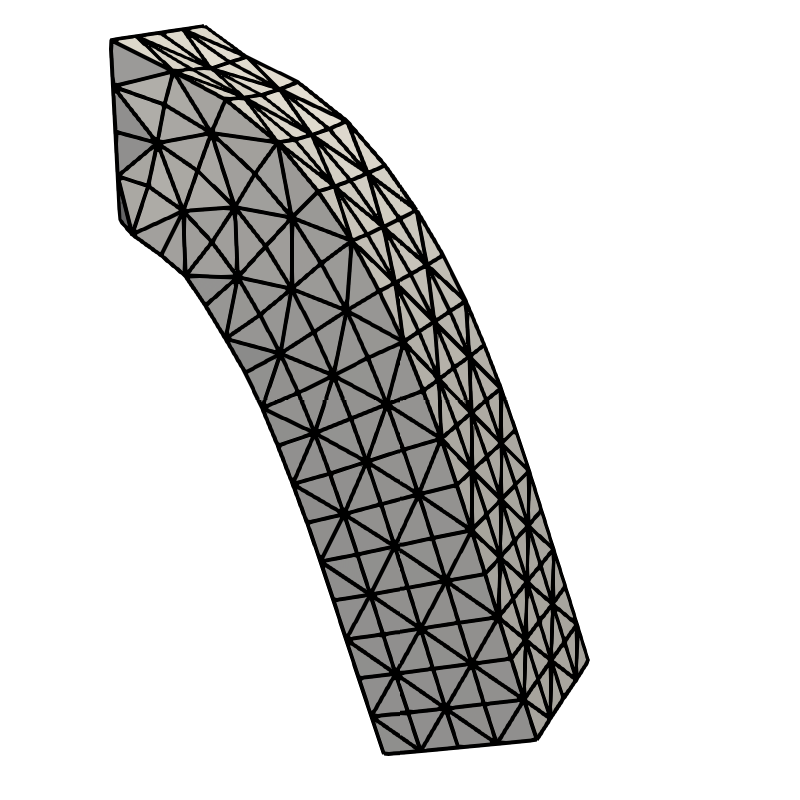}
		\caption*{(c) 2-ring}
		\label{sfig:haning_cuboid-hydrostatic}
	\end{subfigure} \par \medskip
	\begin{subfigure}{.24\linewidth}
		\centering
		\includegraphics[width=1.0\textwidth]{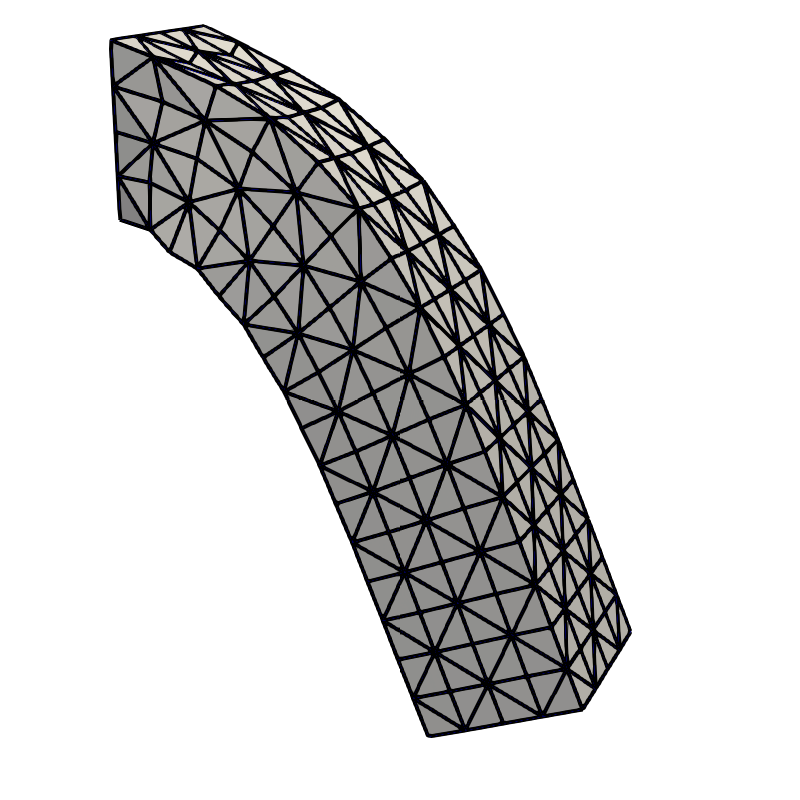}
		\caption*{(d) 1-ring}
		\label{sfig:haning_cuboid-hydrostatic}
	\end{subfigure}%
	\begin{subfigure}{.24\linewidth}
		\centering
		\includegraphics[width=1.0\textwidth]{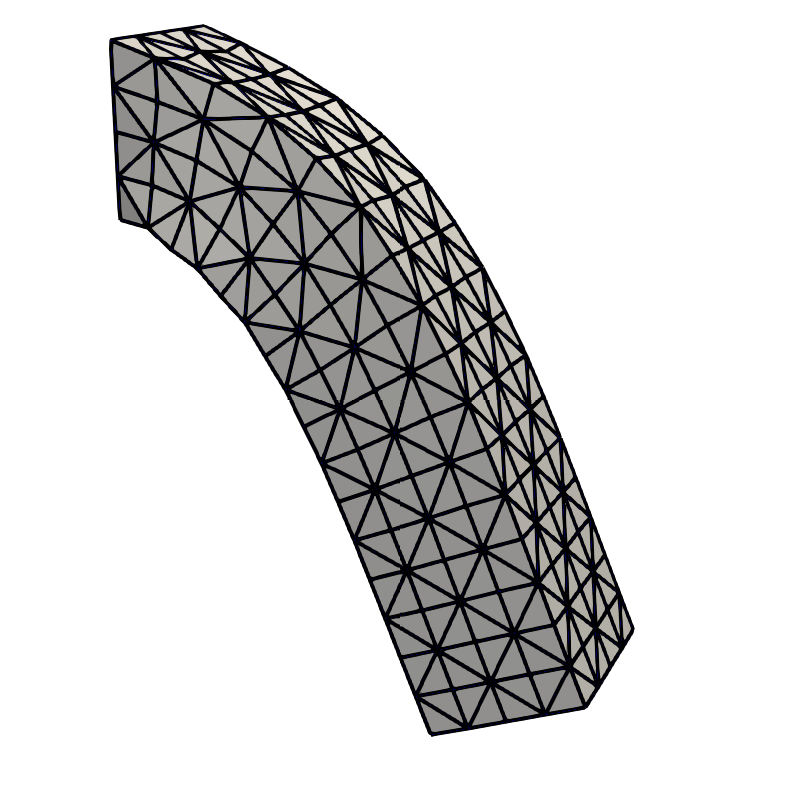}
		\caption*{(e) Hydrostatic + Penalty}
		\label{sfig:haning_cuboid-hydrostatic}
	\end{subfigure}%
	\begin{subfigure}{.24\linewidth}
		\centering
		\includegraphics[width=1.0\textwidth]{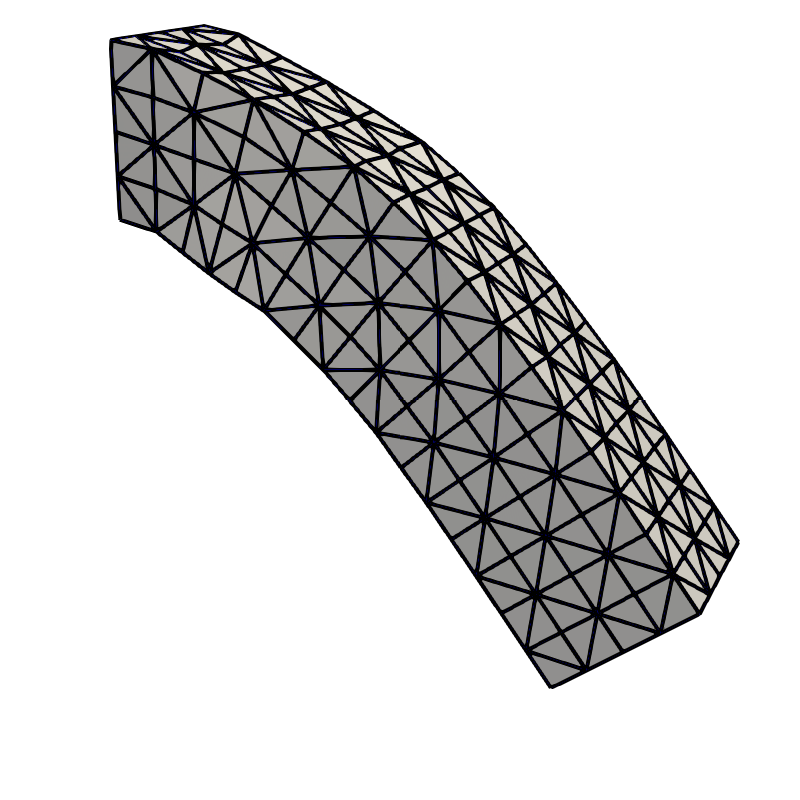}
		\caption*{(f) $\nu = 0.499$}
		\label{sfig:haning_cuboid-hydrostatic}
	\end{subfigure}%
	\caption{\new {\textbf{Zoning Tests}: A cantilever beam with $16\times4\times4$ vertices is suspended under gravity with the left surface fixed with Dirichlet boundary conditions, (a, b, c, d) demonstrates the results of using different choices of volumetric zones without any extra penalty. (a) is the hydrostatic case, where the entire mesh is one zone; (b, c, d) respectively are the results of using a 4-ring, 2-ring, 1-ring zone \cite{Irving:2007}. (e) shows the result of our method with the hydrostatic zone with penalty, which closely resembles the 1-ring result. (f) is the case where a high Poisson's ratio was used, which is essentially a zone per element, where locking prevents the beam from bending further. } }
	\label{fig:hanging_cuboids}
\end{figure*}

\new{
	The sizes of the volumetric zones determine the extent of the volume flow: smaller zones mean the volume flow is more restricted.
	This leads to the object appearing stiffer as the zones get smaller, and when each zone is a single element locking would occur.
	As the zones become smaller, the zonal volume constraints approximate the element-wise volume constraint more closely.
	Therefore the solution becomes more accurate, so long as stiffness is sufficiently low (see Section~\ref{subsec:IncompLocking}).
	However, volumetric stiffness inside zones can also be controlled through the penalty, which approximates the fine-scale pressures.
	This allows the user to use a much coarser choice of zones, with additional volume penalty to get an accurate result similar to using finer zones.
	Figure~\ref{fig:hanging_cuboids} demonstrates the results of a simple simulation of a cantilever beam, where different sized volumetric zones were used.
	Our method with a hydrostatic zone and a local penalty achieves a similar result to the one-ring constraint result, with 11.15 times faster performance.
}

%%% Local Variables:
%%% mode: latex
%%% TeX-master: "sigconf"
%%% End:

\section{Epidermis Model}

The human body is covered by skin, and the  epidermis constitutes the outer-most layers of the skin. It is a complex layered structure whose layers (especially the stratum corneum) are much stiffer than the underlying
hypodermis, fat and muscle tissues. Multilayer skin models have be previously proposed in computer graphics and biomechanics \cite{magnenat2002,flynn2009,li2014}. The layered structure is fundamental to the quality of deformation and formation
of wrinkles \cite{cerda2003geometry}.  We propose a simple model for the stiffness of the epidermis as an area-preserving potential of the surface of the mesh,
which is simply added to the total energy potential to be minimized. Our results show that this
simple extension contributes heavily to the appearance of realistic tissue deformation.

% The outermost layer of human tissue consists of the epidermis --- a very thin layer of stiff tissue
% \todo{(citation \& more details on epidermis needed)}.
% It is this stiffness that causes fine wrinkling of the skin. Further below the epidermis, is a
% thicker layer called the dermis, which is responsible for deeper wrinkles, especially with aging.
% In this work we primarily focus on a larger scale of human body simulation where these types of
% wrinkles are not captured by the simulation mesh. However, the disparity between the skin and fascia
% bulk moduli can drastically affect the deformation of human tissue.
% To capture the effects of skin on tissue deformation, we add a bulk elasticity component of the
% surface of the tetrahedral mesh to our total energy potential.

For this, we can use the 2D version of the penalty function we formulated in
the previous section. The energy density function has the same form as the 3D
version of the penalty, but with a 2D reduced
deformation gradient \cite{Li:2013} $\tilde{\vec{F}} \in \R^{3 \times 2}$.

\begin{equation}
  \Psi_e(\tilde{\vec{F}}; \gamma; \lambda_e) :=
  \frac{\lambda_e}{12} (\tilde{J}-1)^2 \left[ \gamma (\tilde{J}-1)^2 + 6 \right],
  \label{eq:epidermis}
\end{equation}

where $\tilde{J} = \det \tilde{\vec{F}}$, and $\gamma$ controls how steeply severe area change is penalized. In practice, $\gamma = 1.0$ seemed to be a good choice in almost all cases we tested for.
This energy is defined for the surface triangles of the 3D mesh, and controlled by the first Lam\'e
parameter $\lambda_e$, which is defined separately from the volumetric material parameters. For
convenience we compute $\lambda_e$ from the more familiar Young's modulus $E_e$ and Poisson's ratio
$\nu_e$ using Equation~\ref{eq:poisson2lame}.

We use a high Poisson's Ratio ($\nu \in [0.4, 0.49]$) for the epidermis energy density
functional~\eqref{eq:epidermis}. We then add this energy potential to the original variational
problem~\eqref{eq:energy_min} to find the local minima of elastic potential such that it minimizes
surface area change.

\begin{figure}[ht]
	\centering
	\begin{subfigure}{.24\linewidth}
		\centering
		\includegraphics[width=1.0\textwidth]{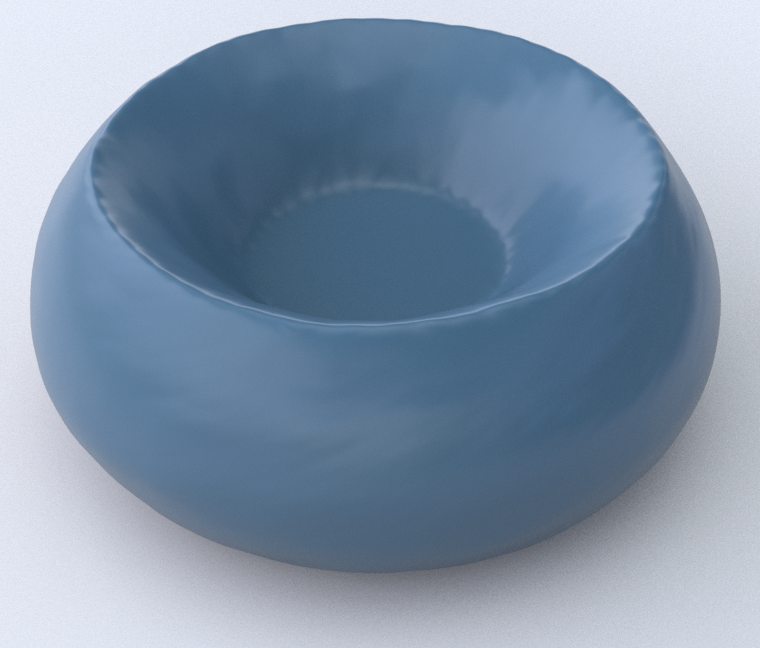}
		\caption*{(a) UNH, $\nu = 0.495$}
		\label{sfig:epi_unconstrained}
	\end{subfigure}%
	\begin{subfigure}{.24\linewidth}
		\centering
		\includegraphics[width=1.0\textwidth]{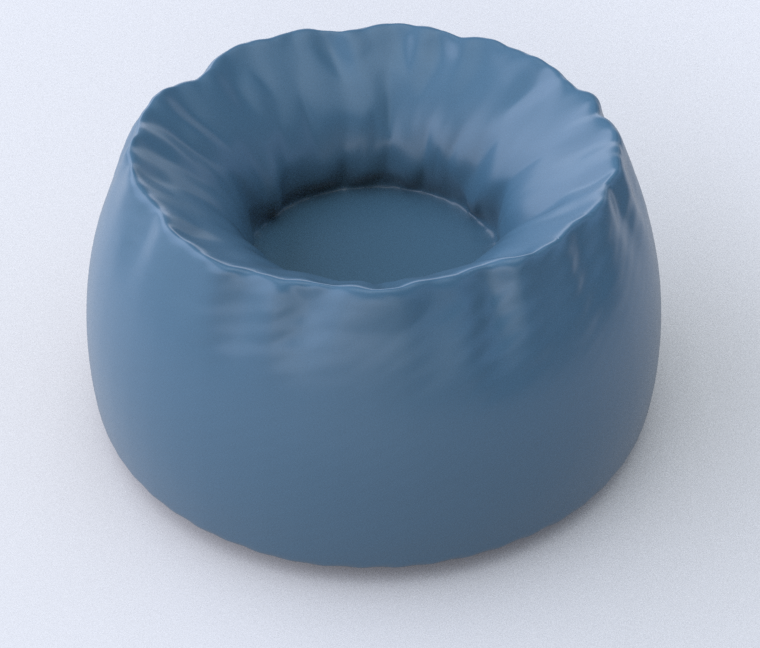}
		\caption*{(b) UNH + Epidermis}
		\label{sfig:epi_unconstrained_epi}
	\end{subfigure}%\par\medskip
	\begin{subfigure}{.24\linewidth}
		\centering
		\includegraphics[width=1.0\textwidth]{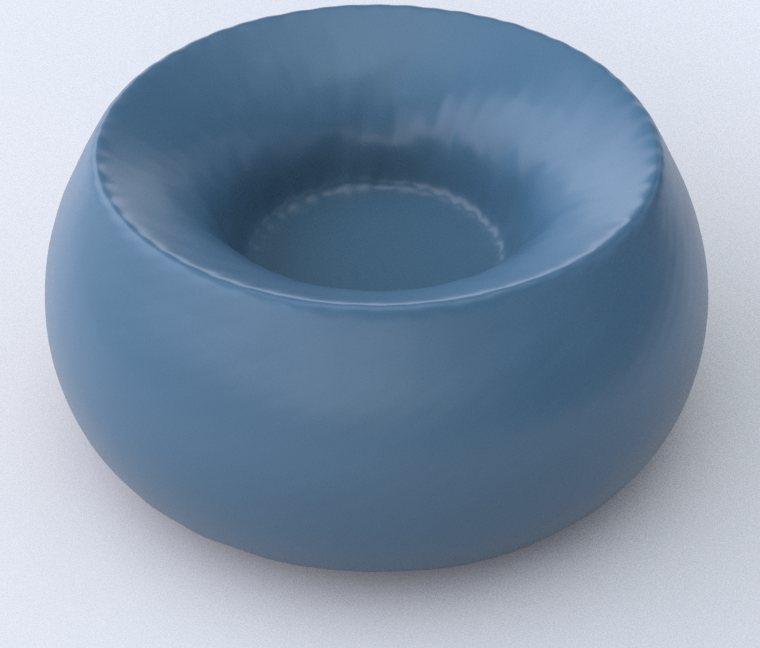}
		\caption*{(c) CNH, $\lambda = 40$}
		\label{sfig:epi_unconstrained_vc}
	\end{subfigure}%
	\begin{subfigure}{.24\linewidth}
		\centering
		\includegraphics[width=1.0\textwidth]{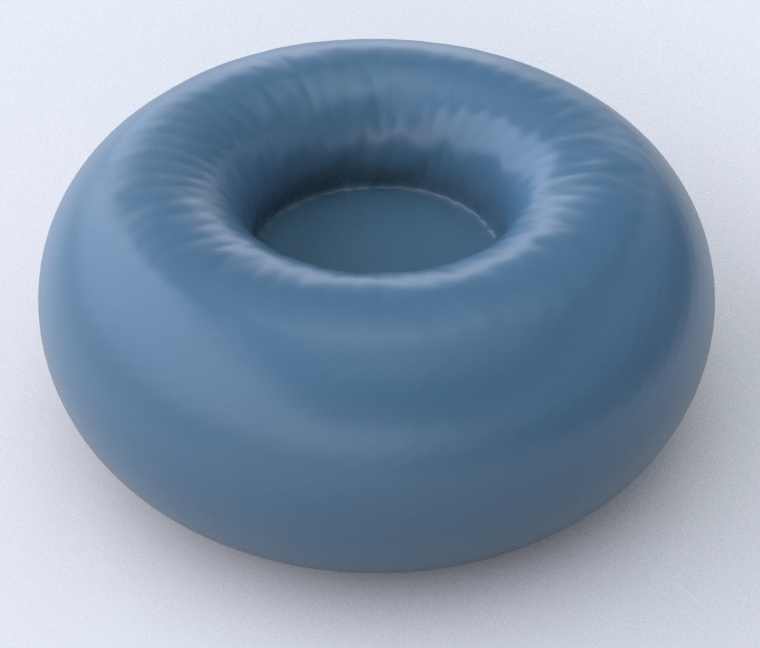}
		\caption*{(d) CNH + Epidermis}
		\label{sfig:epi_unconstrained_vc_epi}
	\end{subfigure}%
	\caption{\textbf{Epidermis Simulation}: Cylindrical pucks with unconstrained sides were compressed with Dirichlet boundary conditions in the center. (a) shows the result of the unconstrained neo-Hookean (UNH) model, without the epidermis model, volume constraints, or penalty. (b) shows UNH with just the epidermis energy with $\lambda_e = 40$ applied to (a); the surface deformation becomes more regular and wrinkles appear as a result. (c) and (d) show the corresponding results with our constrained neo-Hookean (CNH) model; (c) shows the effect of just volume preservation and compression penalty while (d) shows the effect of adding the epidermis energy to (c). }
	\label{fig:epidermis}
\end{figure}

To demonstrate the effectiveness of this simple modification, we indent
the cylindrical puck example again with the epidermis model as shown in
Figure~\ref{fig:epidermis}.
This simulation produces a more organic surface without unnaturally sharp edges.

%%% Local Variables:
%%% mode: latex
%%% TeX-master: "sigconf"
%%% End:

\section{Results}

The following results demonstrate the versatility of our method.  We use a tetrahedral mesh discretization for
all our simulations.
Our implementation relied on the Ipopt non-linear optimization package \cite{Wachter:2006} to solve the constrained optimization problem proposed in Equation~\ref{eq:constrained-min}.
In our experiments, we found that excessive parameter tuning was not required to use our method with Ipopt: only when used with additional nonlinear constraints, we  occasionally tuned the \texttt{nlp_scaling_max_gradient} parameter.
For all our examples, we set the Young's Modulus to $\mu = 16.0$ KPa, and we started with $\beta = 1.0$ and tuned it to improve performance.
However, for just resolving inversions and numerical instability, $\beta = 1.0$ works well for all of these examples.

\subsection{Two Tetrahedra}

As the simplest proof of concept example for the volume constraint, we constructed a mesh with two
tetrahedra with equal volume joined together by a face as shown in Figure~\ref{fig:twotets}. We
then compressed one of the tetrahedra with a Dirichlet boundary condition to a plane, to see the
effects of the volume constraint. The energy model used is neo-Hookean with $\nu = 0.495$.

As expected, the example without the global volume constraint loses around $50\%$ of the volume,
since the deformation of the compressed tetrahedron only affects its neighbor due to the change in shape
of their shared face.
When a volume constraint is applied, the unconstrained tetrahedron inflates to twice the original
size, keeping the total volume constant.

\subsection{Pressure Distribution}

\begin{figure*}
	\centering
	\begin{subfigure}{.32\linewidth}
		\centering
		\adjustbox{trim={.15\width} {.00\height} {.15\width} {.00\height},clip}%
		{\includegraphics[width=1.0\textwidth]{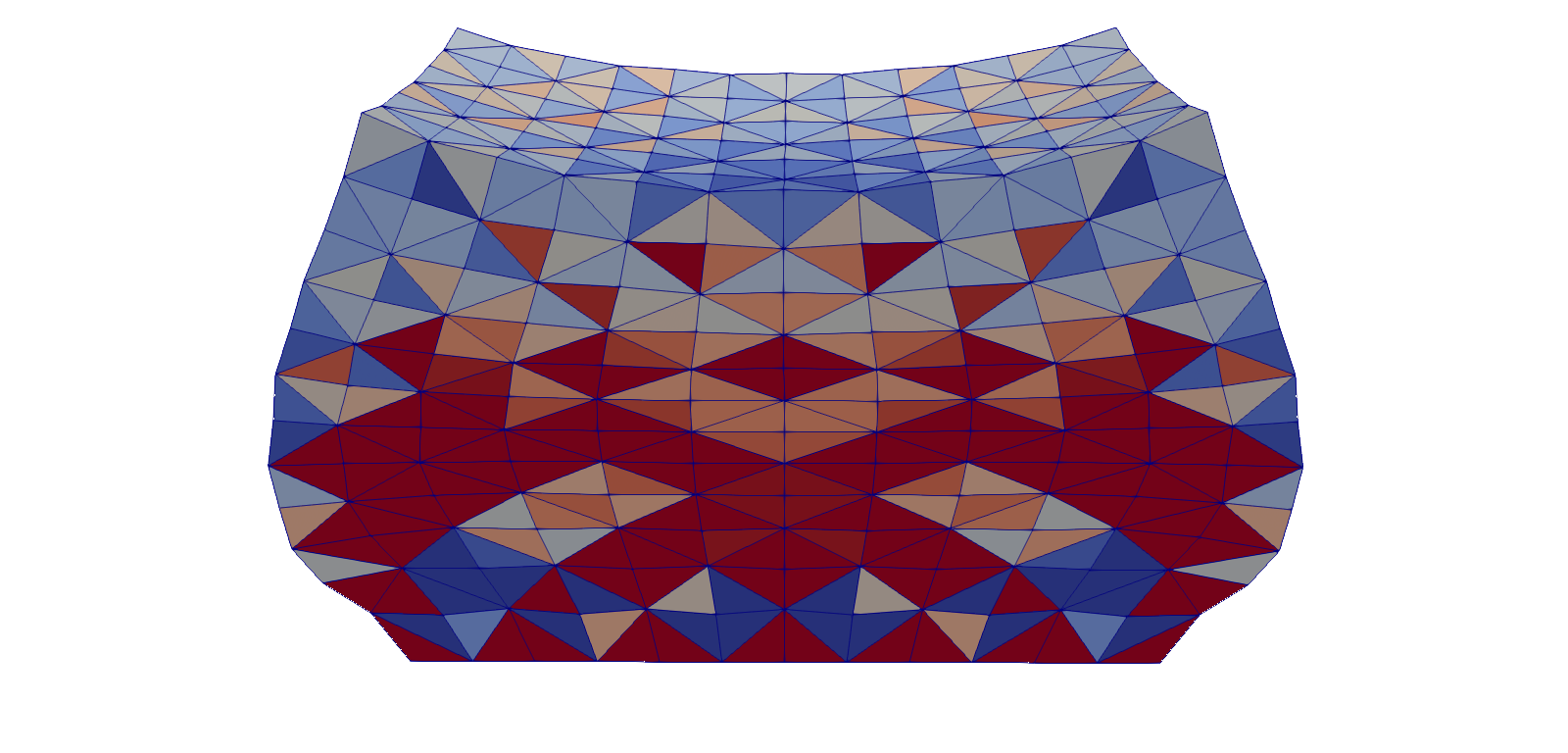}}
		\caption*{(a) UNH, $\nu = 0.495$}
		\label{sfig:suspended-cube-pr-0495}
	\end{subfigure}%
	\begin{subfigure}{.32\linewidth}
		\centering
		\adjustbox{trim={.15\width} {.00\height} {.15\width} {.00\height},clip}%
		{\includegraphics[width=1.0\textwidth]{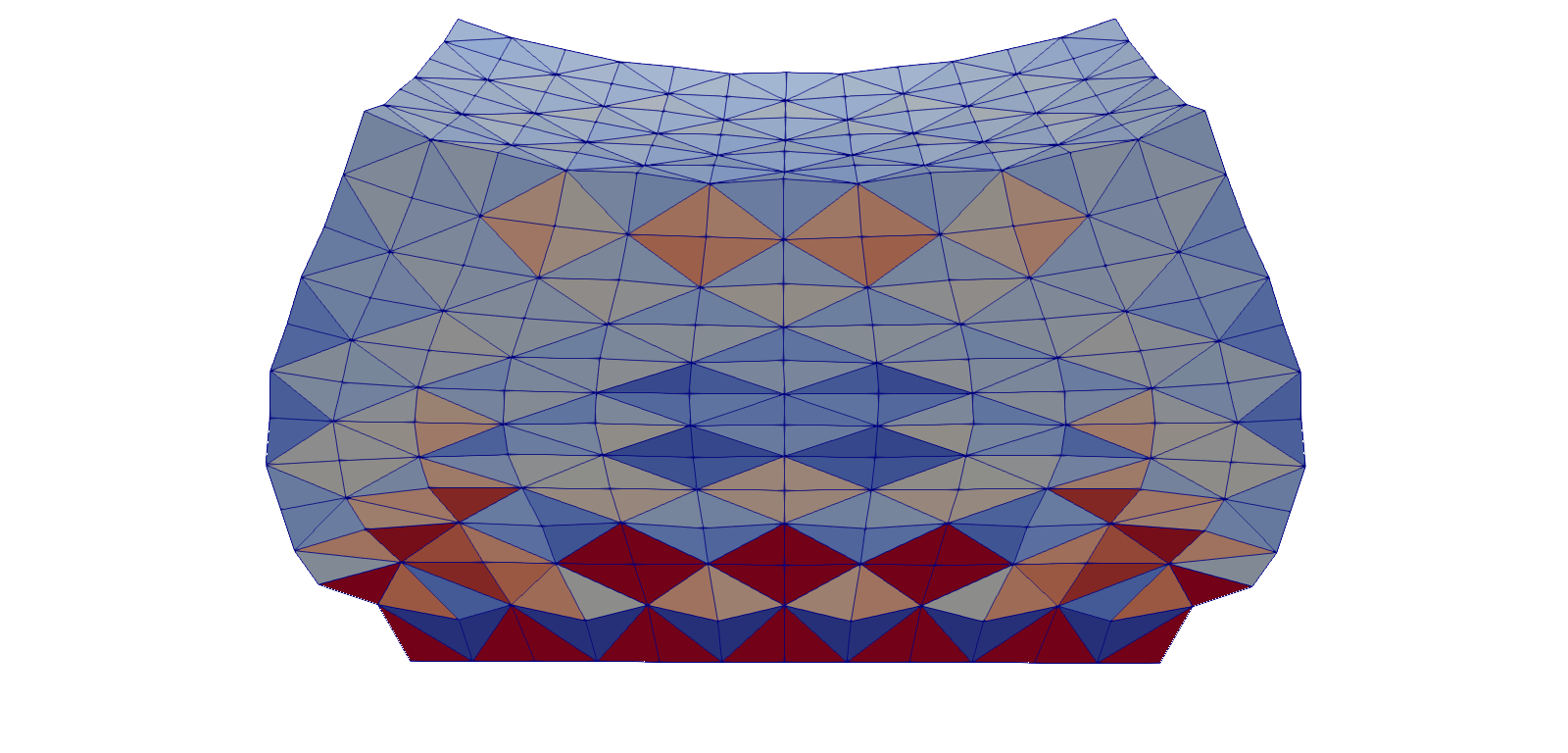}}
		\caption*{(b) One-ring}
		\label{sfig:suspended-cube-onering}
	\end{subfigure}%
	\begin{subfigure}{.32\linewidth}
		\centering
		\adjustbox{trim={.15\width} {.00\height} {.15\width} {.00\height},clip}%
		{\includegraphics[width=1.0\textwidth]{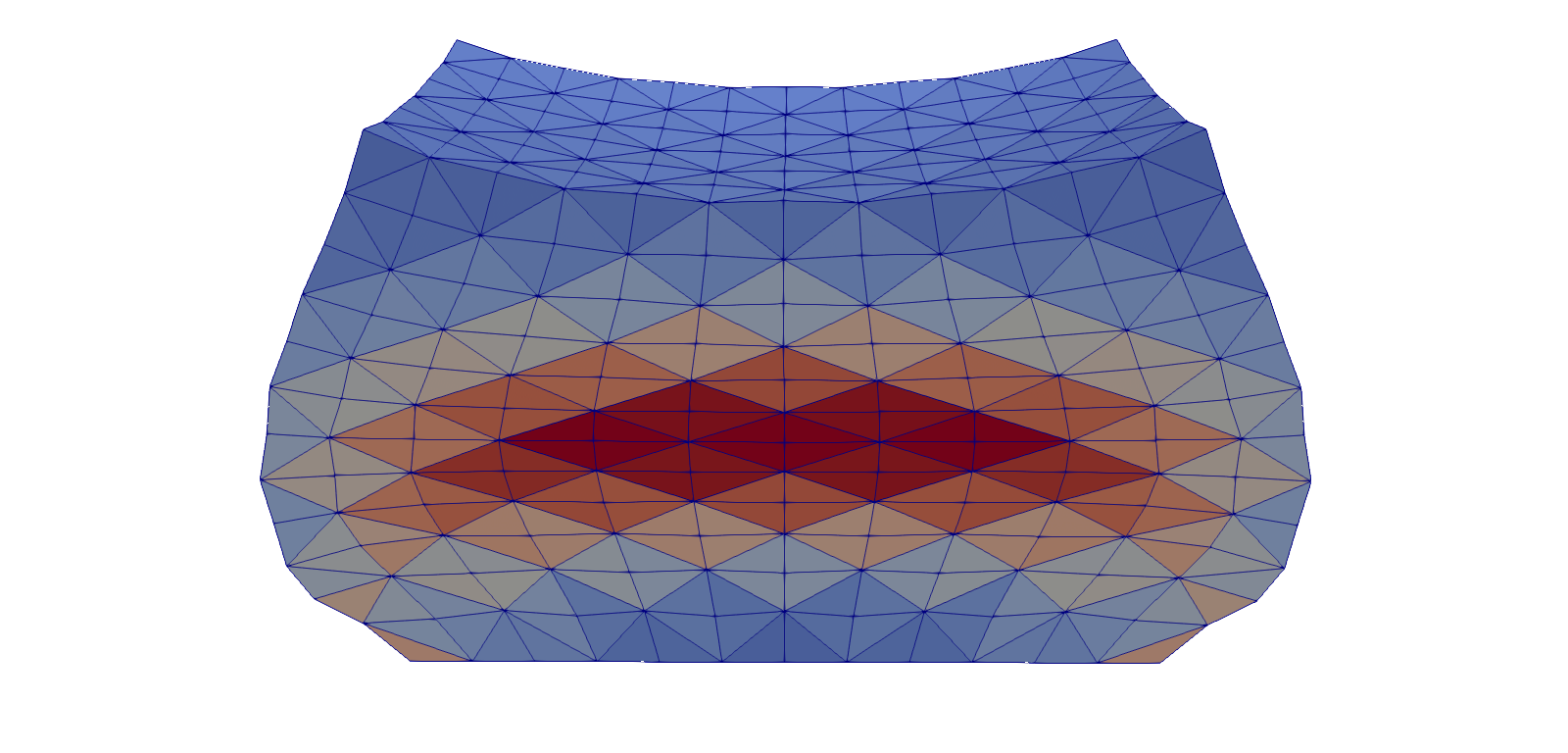}}
		\caption*{(c) CNH, $\lambda = 100, \beta = 1$}
		\label{sfig:suspended-cube-vc-100}
	\end{subfigure}%
	\caption{\textbf{Pressure Distribution}: A cube with $12^3$ vertices is supported under gravity with the bottom surface fixed with Dirichlet boundary conditions, visualized with the per-tet pressure distribution in the inner body. (a) shows the per-tet Poisson's ratio approach with $\nu = 0.495$, where the effects of volumetric locking results in highly irregular pressure distribution. (b) is the result where a one-ring volume constraint is used, where a checkerboard pattern appears in the pressure distribution due to a lack of stability. Finally, (c) is our method with $\lambda = 100, \beta = 1$, where the pressure distribution is regular and realistic. }
	\label{fig:suspended-cubes}
\end{figure*}

Following a classic volumetric locking example, we test our methods with a standing cube simulation. A neo-Hookean cube with shear modulus $\mu = 10.0$ is supported under gravity with the bottom surface fixed with Dirichlet boundary conditions, results shown in Figure~\ref{fig:suspended-cubes}. This example demonstrates the effects of external force on the pressure distribution on the body surface and interior with regards to different approaches of simulating incompressibility. Irregular pressure distribution may not visually affect the results much, but when simulating frictional contact or fracture, they might result in unrealistic solutions. The high Poisson's ratio ($\nu = 0.495$) approach results in obvious irregularities in the pressure distribution, a manifestation of volumetric locking. Per-tet hydrostatic pressures in this case are computed as the the volumetric components of the stress tensors, that is, the negative divergence of the Cauchy stress. The high bulk modulus per each element causes pressure computation from displacement variables to be unreliable. The one-ring volume constraint approach \cite{Irving:2007}, which is equivalent to the Average Nodal Pressure element \cite{bonet:1998}, shows a more regular pressure distribution, but also shows checkerboard patterns. In this case, the pressures are the Lagrange multipliers for the volume constraints, scaled to be in the same units as the volumetric stress then mapped back to the cells. The checkerboard pattern is an artifact of the instability of the one-ring constraint approach, where the averaging of the pressure variables on the nodes allows solutions with such checkerboarding to occur. With our method with one global zone and a local compression penalty of $\lambda = 100, \beta = 1$, we compute the per-tet pressure as a sum of the average zonal pressure (the Lagrange multiplier of the zonal volume constraint) and the fine-scale pressures computed from the volumetric stress component from the penalty.
The bulk modulus is much smaller, compared to $\nu = 0.495$ (corresponding to $\lambda = 400$) in the Poisson's ratio approach. Therefore volumetric locking is avoided and the volumetric stress components are more regular. Since the volume preserving zone is global and the average pressure is constant throughout the zone, the checkerboard artifact in the Lagrange multipliers is eliminated.

\subsection{Dynamic Stability}

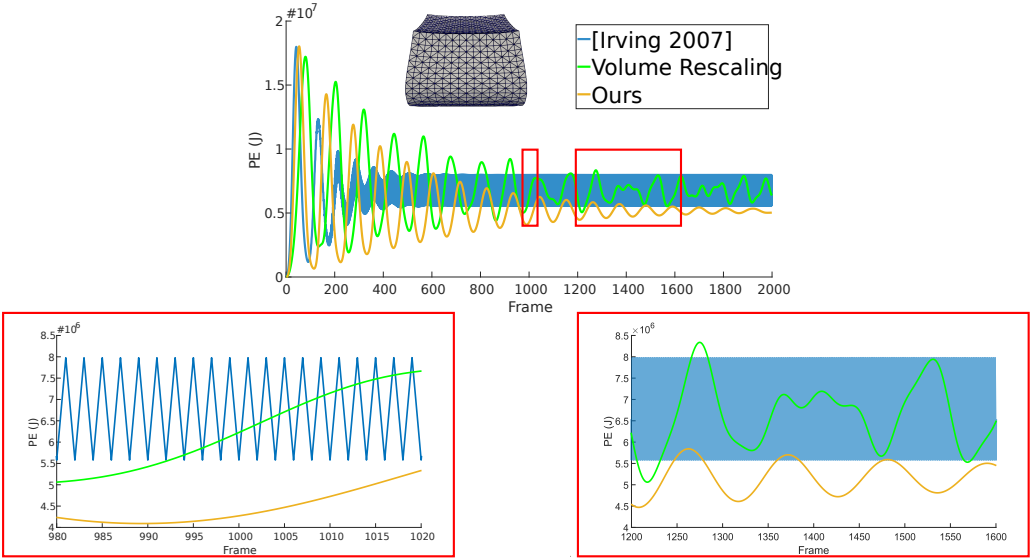
\begin{figure}
	\centering
	\begin{overpic}[width=0.6\columnwidth]{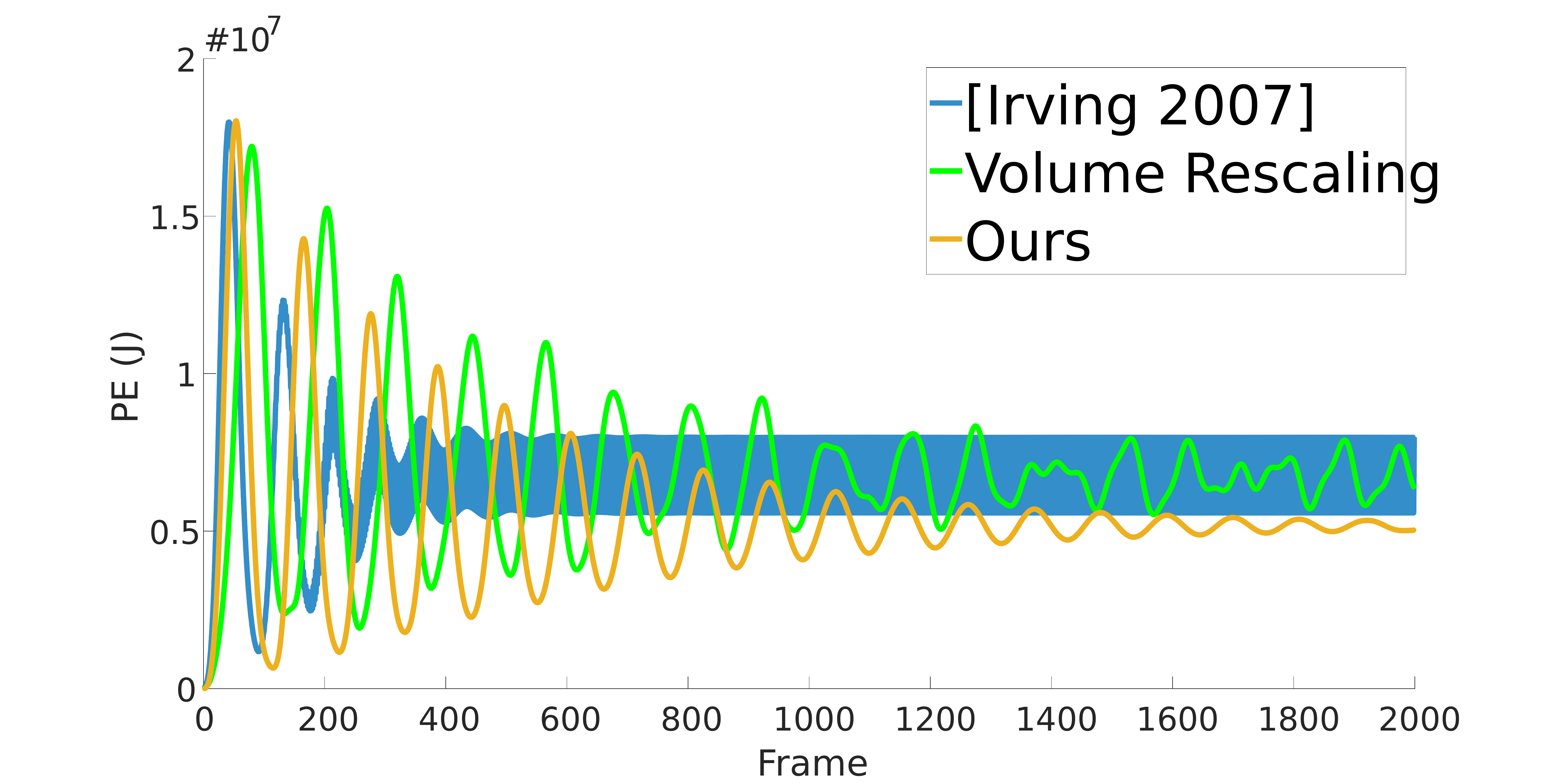}
		\put(25,30){\includegraphics[width=0.2\columnwidth]{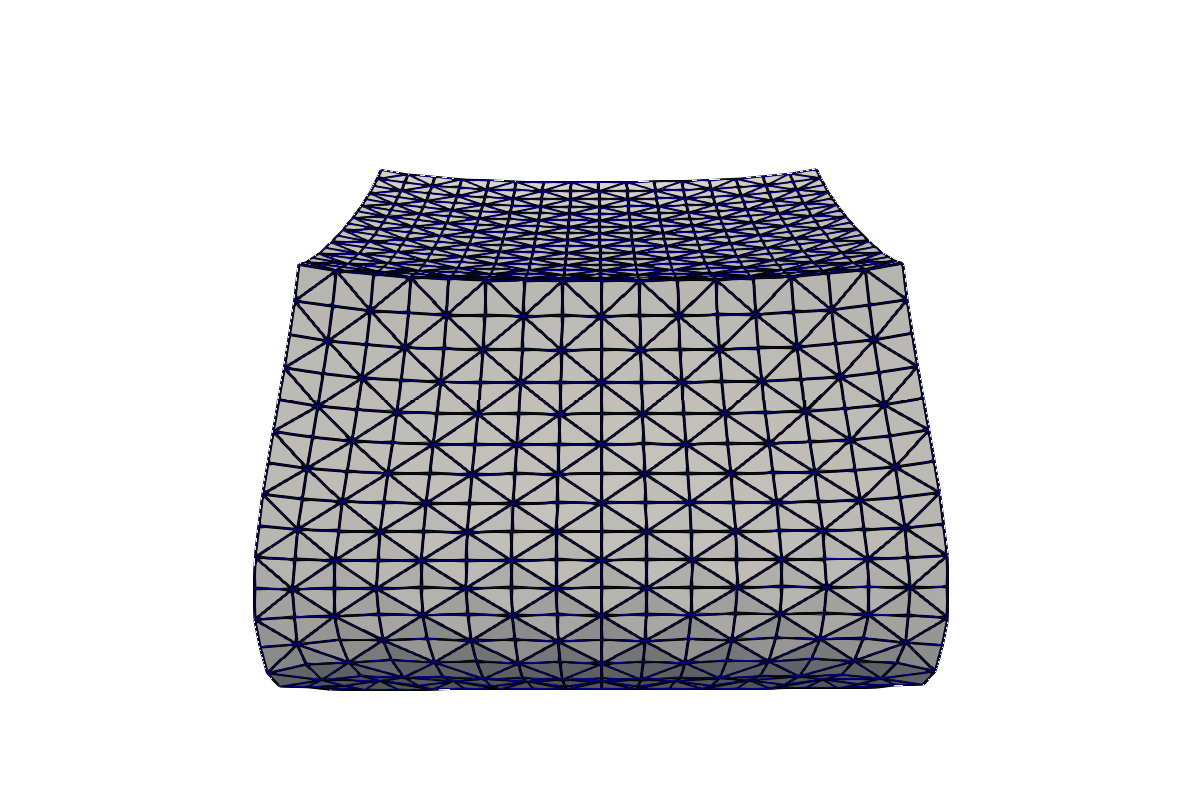}}
		\put(49.5,14){
			\begin{tikzpicture}
				\draw[thick,draw=red] (0,0) rectangle ++(0.2,1.0);
			\end{tikzpicture}
		}
		\put(58,14){
			\begin{tikzpicture}
				\draw[thick,draw=red] (0,0) rectangle ++(1.4,1.0);
			\end{tikzpicture}
		}
	\end{overpic} \hfill
	\begin{overpic}[width=0.45\columnwidth]{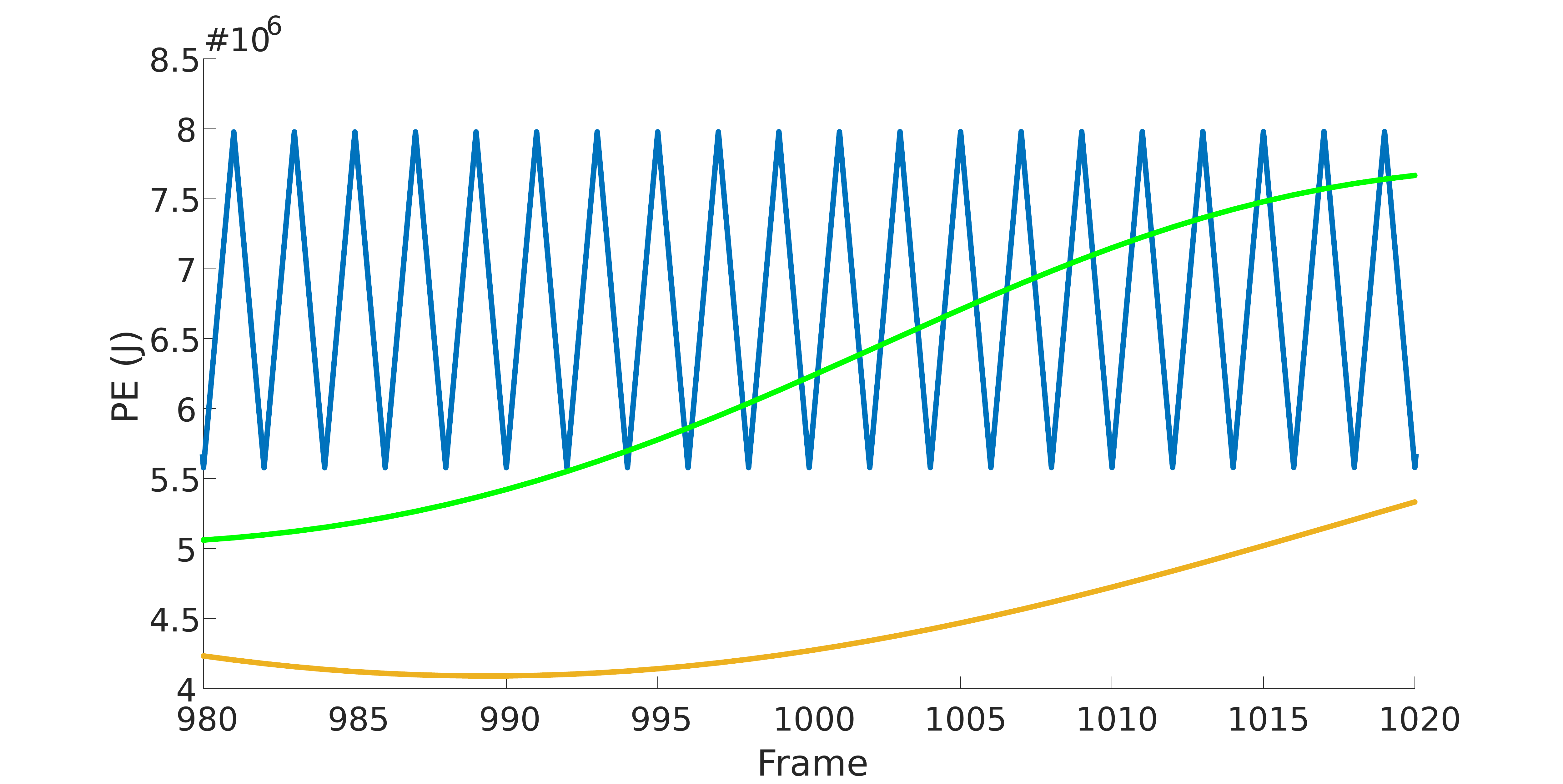}
		\put(0,-0.2){
			\begin{tikzpicture}
				\draw[thick,draw=red] (0,0) rectangle ++(6.0,3.2);
			\end{tikzpicture}
		}
	\end{overpic} \hfill
	\begin{overpic}[width=0.45\columnwidth]{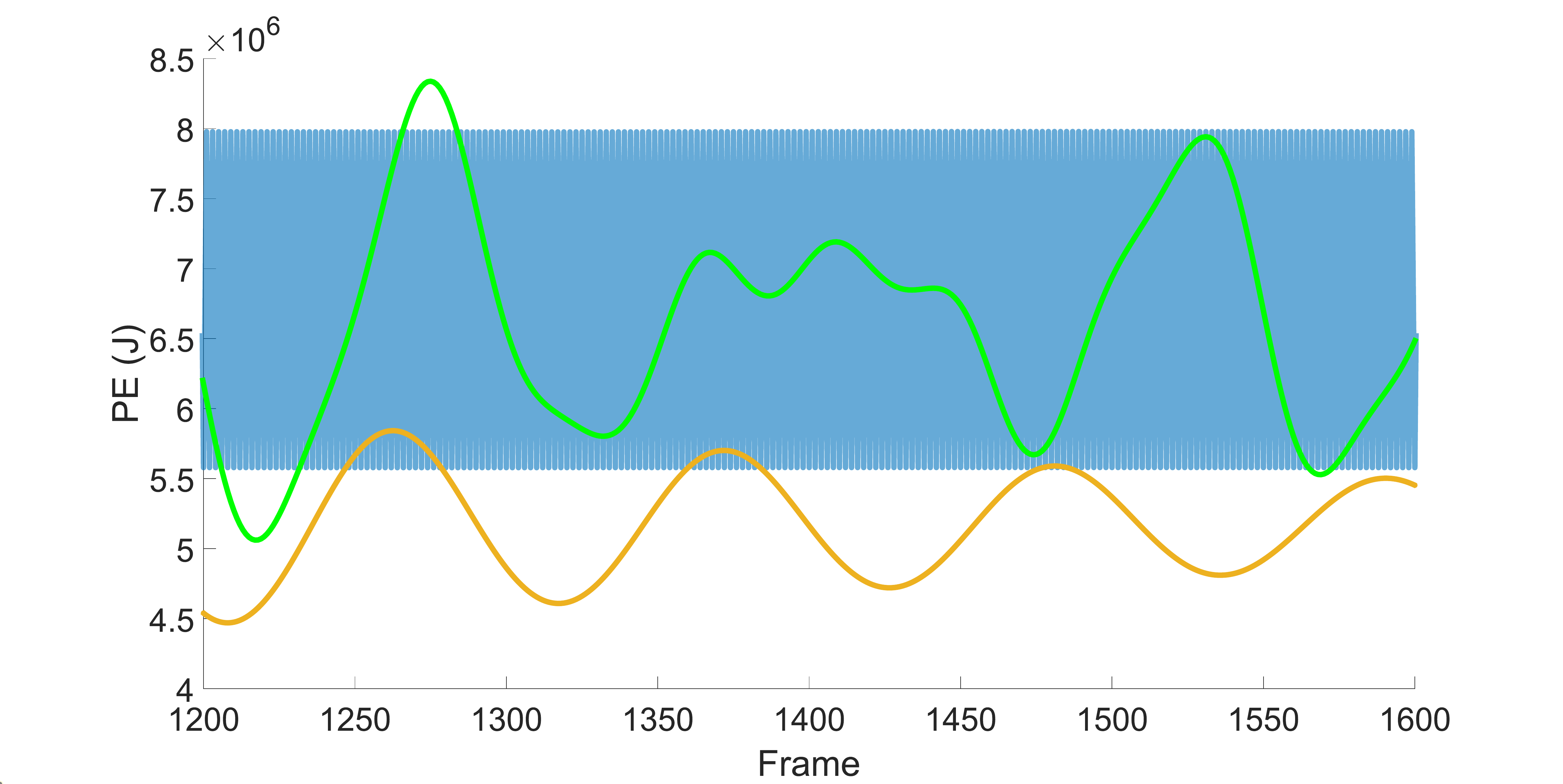}
		\put(0,-0.2){
			\begin{tikzpicture}
				\draw[thick,draw=red] (0,0) rectangle ++(6.0,3.2);
			\end{tikzpicture}
		}
	\end{overpic}
	\caption{\textbf{Dynamic Stability}: Energy plot from the simulation of a soft $16^3$ cell cube subjected to gravity, with a moderately large timestep of 8ms. We compare three different algorithms for simulating incompressibility, where the blue curve is the energy plot for the one-ring pressure algorithm of \cite{Irving:2007}, the green is a na\"ive global volume rescaling algorithm, and yellow is ours. \cite{Irving:2007} (blue) shows severe oscillation even with an aggressive clamping of the volume recovery. Note how due to the extreme oscillations in potential energy, the energy plot for the method appears as a thick line in the top plot. The bottom plots show a magnified plot to demonstrate how the energy behaves in small timeframes. The volume rescaling algorithm (green) gains and loses energy arbitrarily due to a unrealistic simulation of the volumetric stress, which can be seen from the irregularities in the plot. Our method (yellow) displays realistic and stable energy behavior, and converges to a stable configuration. \label{fig:stability}}
\end{figure}

\begin{figure*}[ht!]
	\centering
	\begin{subfigure}{.24\linewidth}
		\centering
		\includegraphics[width=1.0\textwidth]{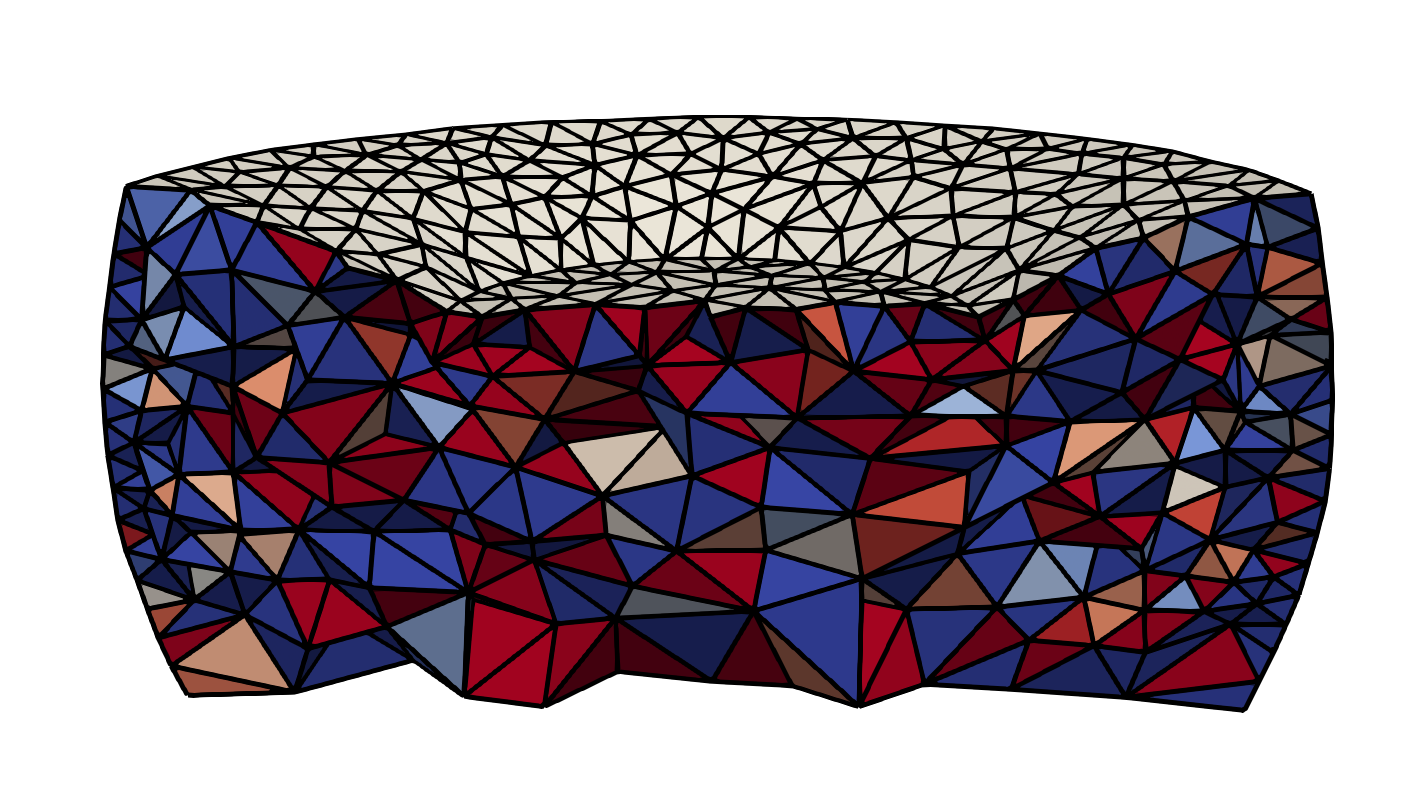}
		\includegraphics[width=1.0\textwidth]{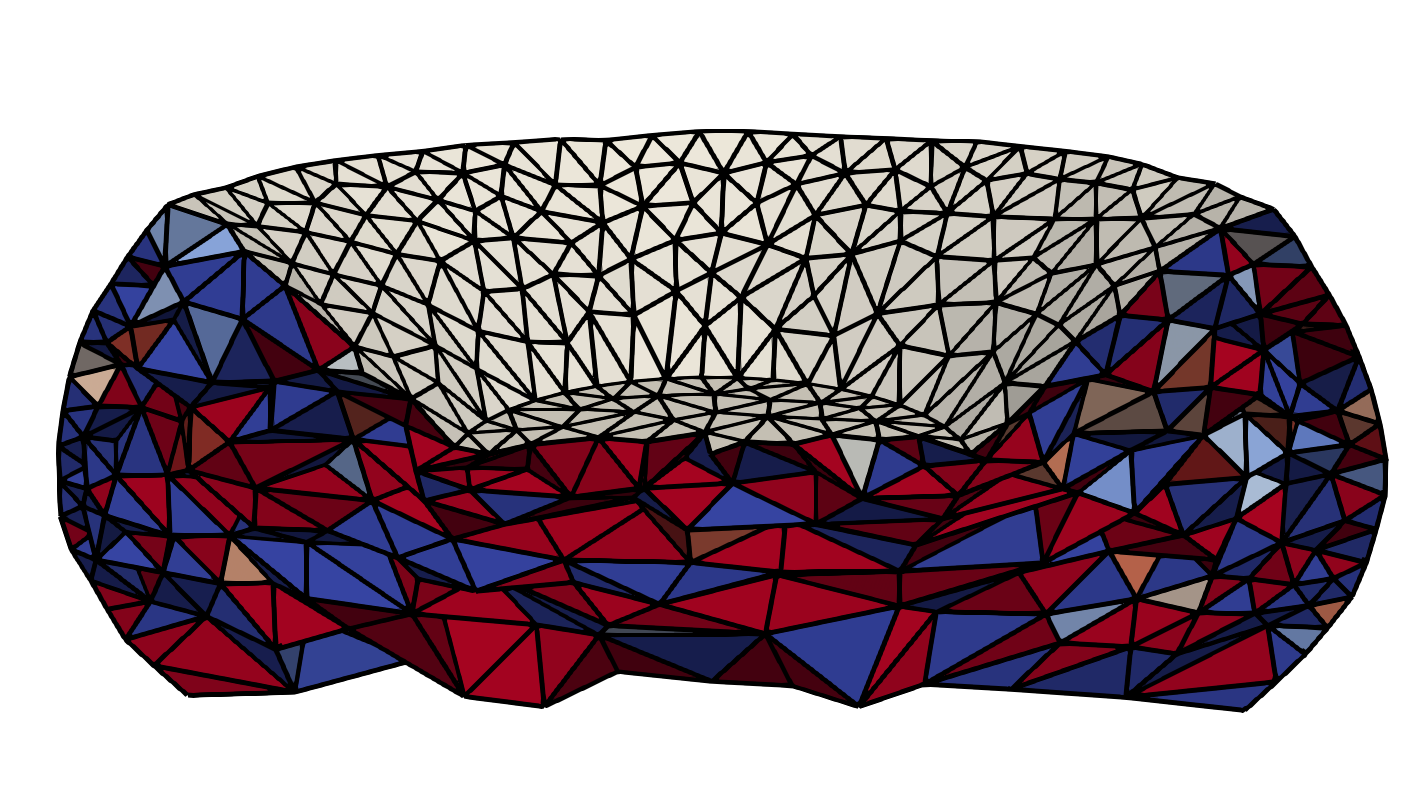}
		\adjustbox{trim={.15\width} {.0\height} {.15\width} {.1\height},clip}%
		{\includegraphics[width=1.4\textwidth]{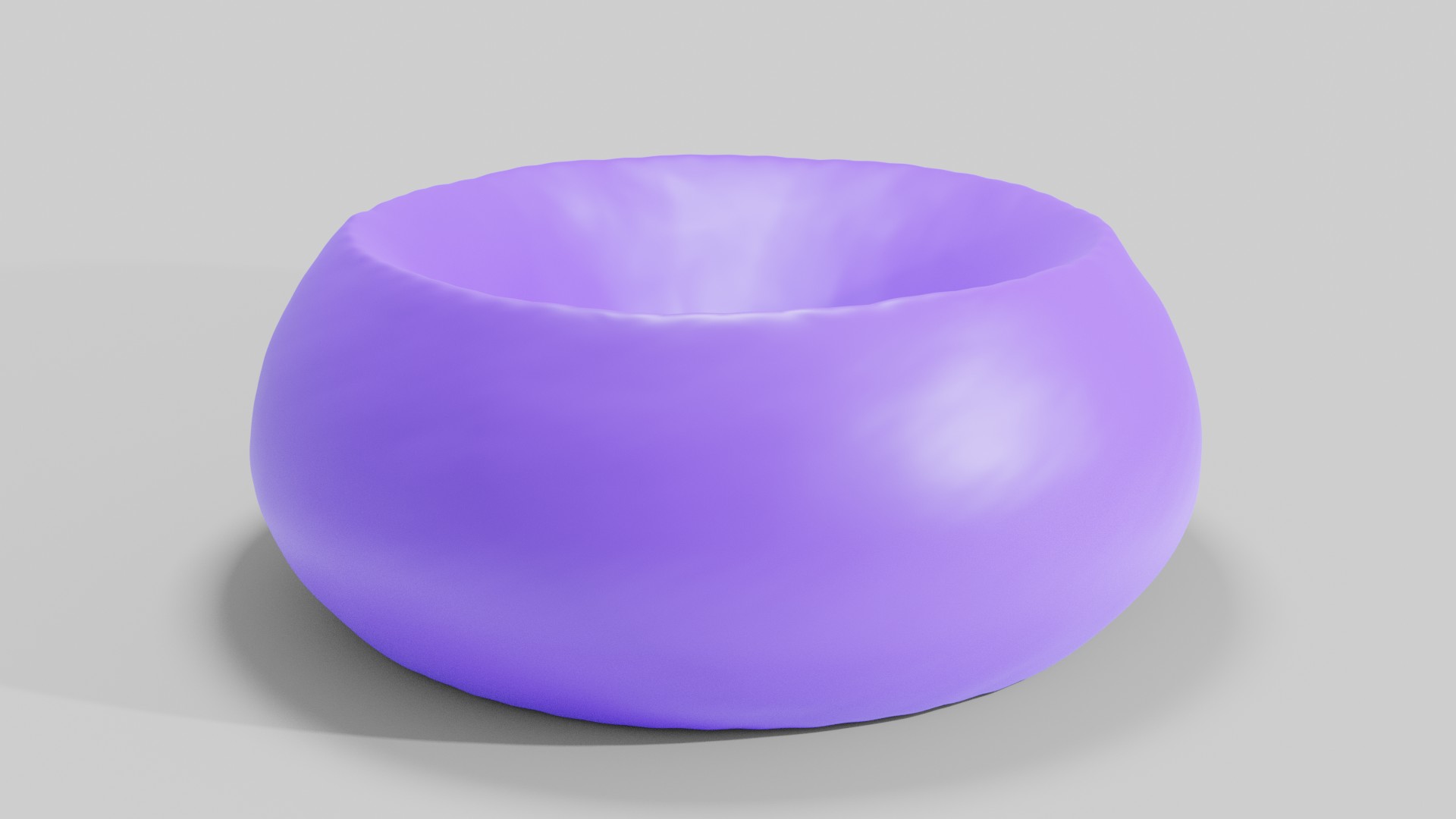}}
		\caption*{(a) UNH, $\nu = 0.495$}
		\label{sfig:puck-pr0495}
	\end{subfigure}%
	\begin{subfigure}{.24\linewidth}
		\centering
		\includegraphics[width=1.0\textwidth]{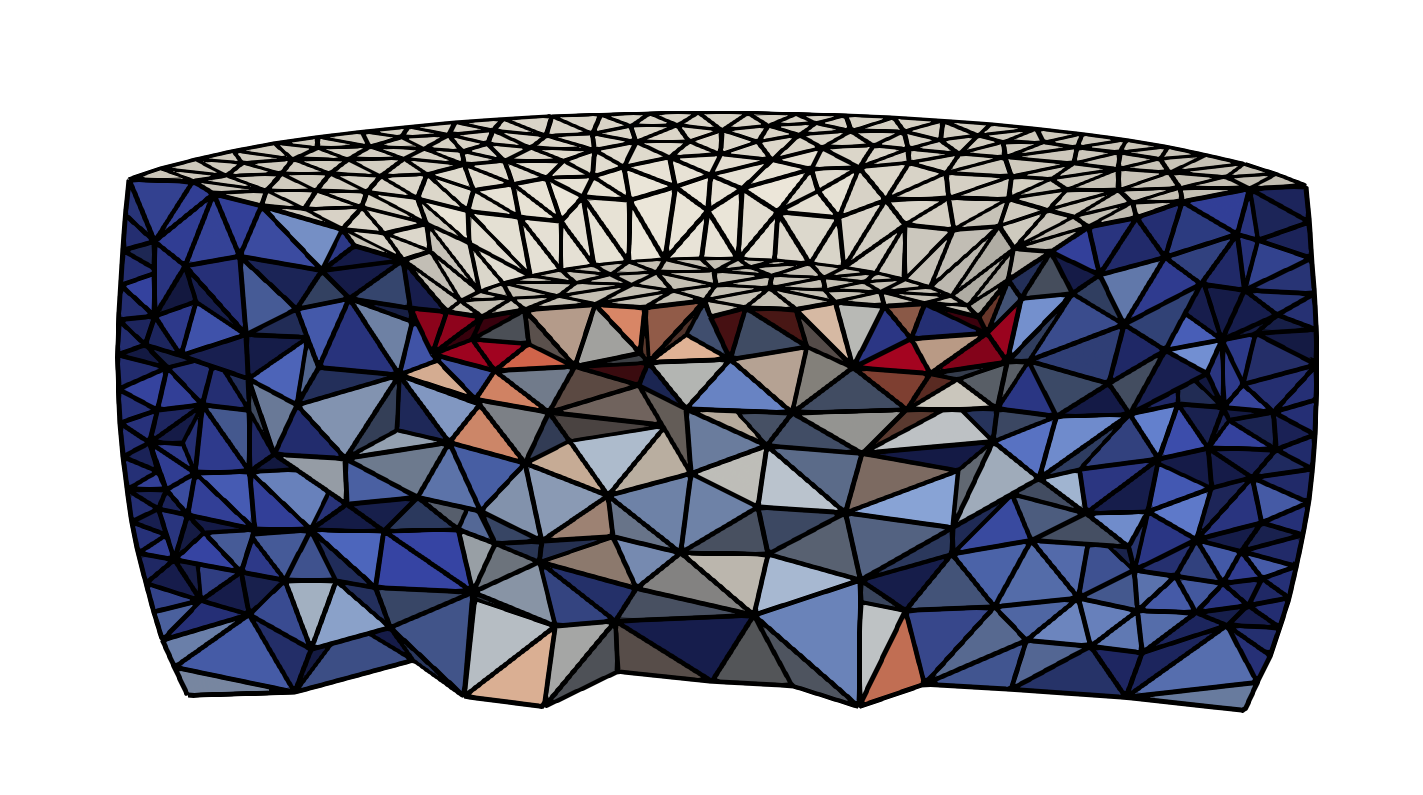}
		\includegraphics[width=1.0\textwidth]{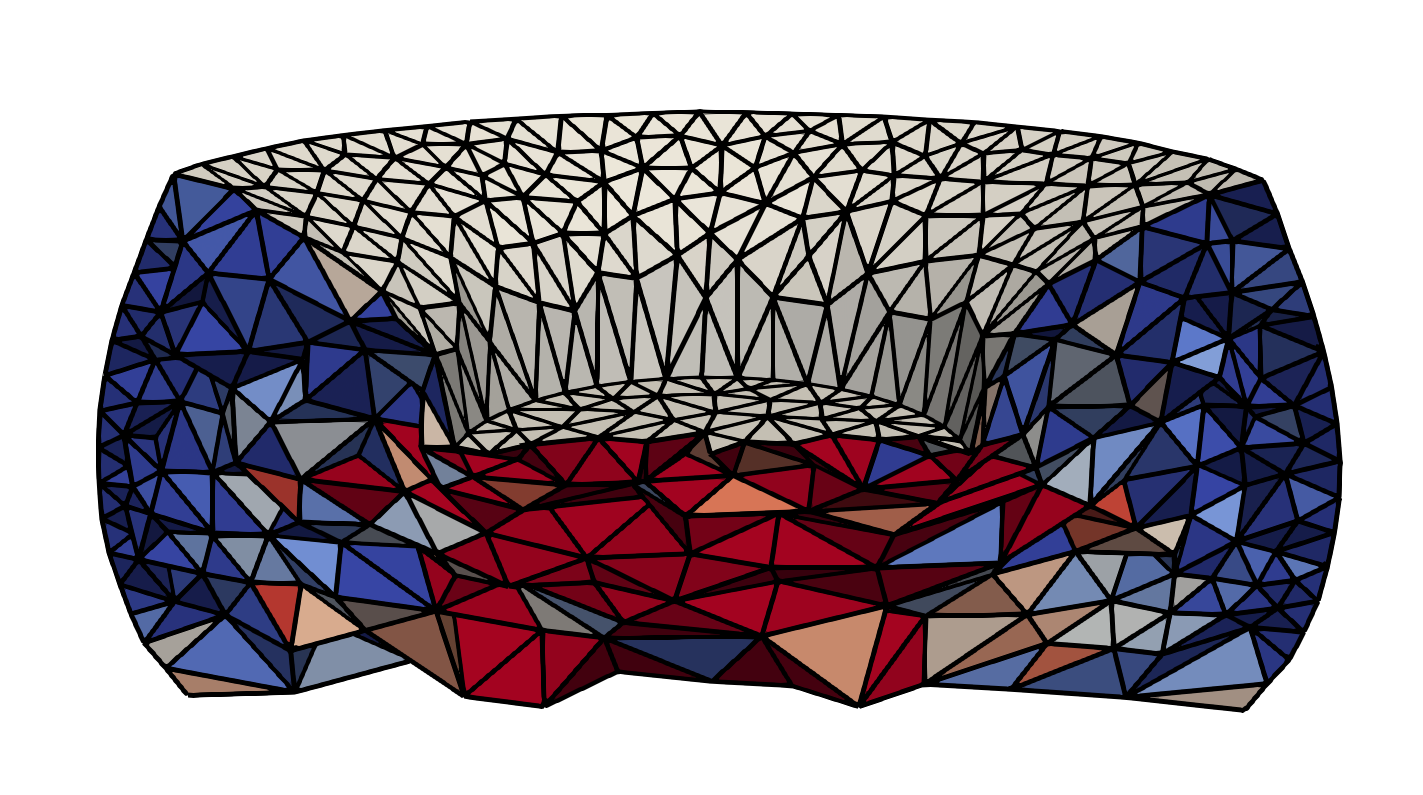}
		\adjustbox{trim={.15\width} {.0\height} {.15\width} {.1\height},clip}%
		{\includegraphics[width=1.4\textwidth]{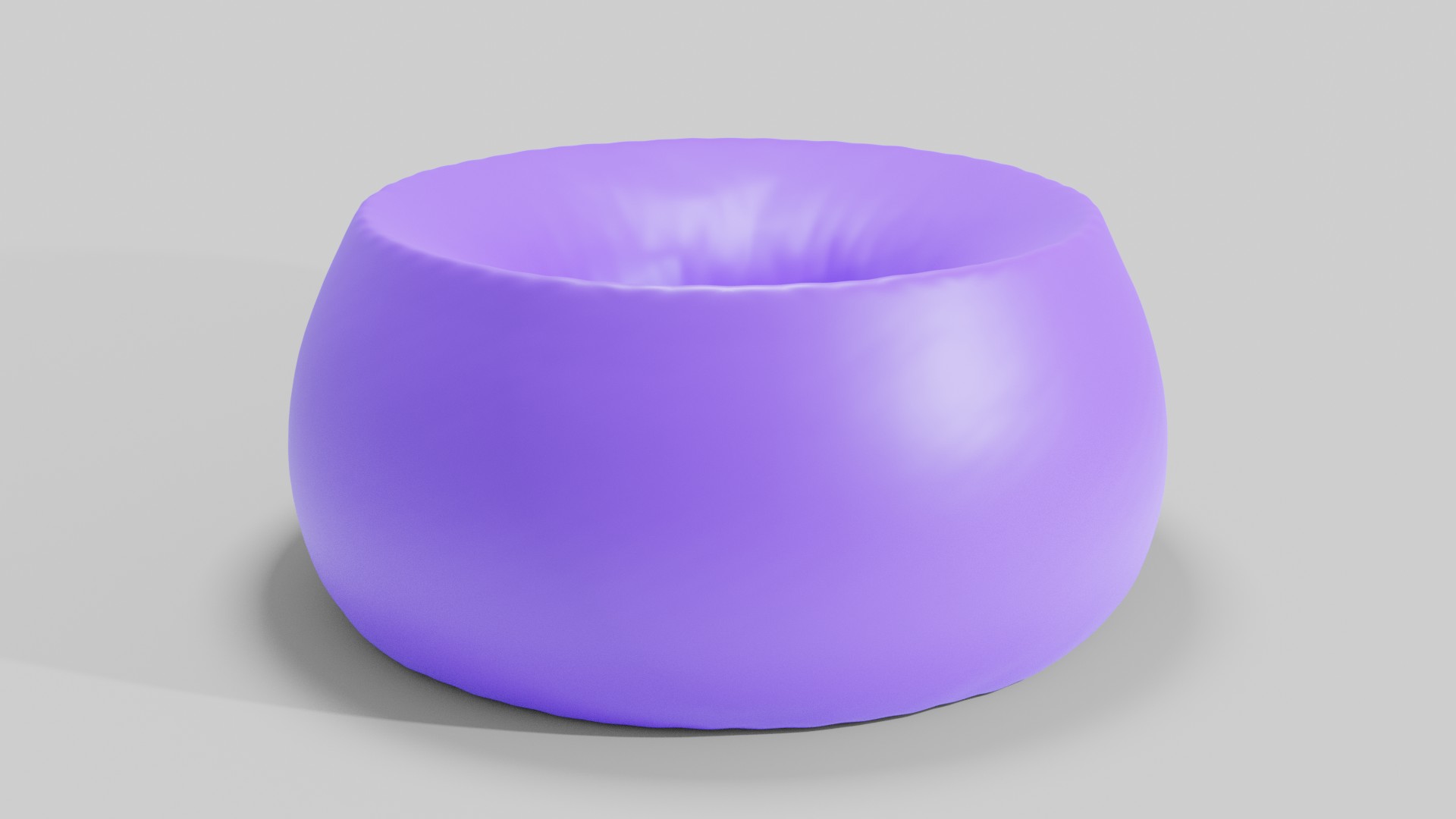}}
		\caption*{(b) UNH, $\nu = 0.4545$}
		\label{sfig:puck-pr04545}
	\end{subfigure}%
	\begin{subfigure}{.24\linewidth}
		\centering
		\includegraphics[width=1.0\textwidth]{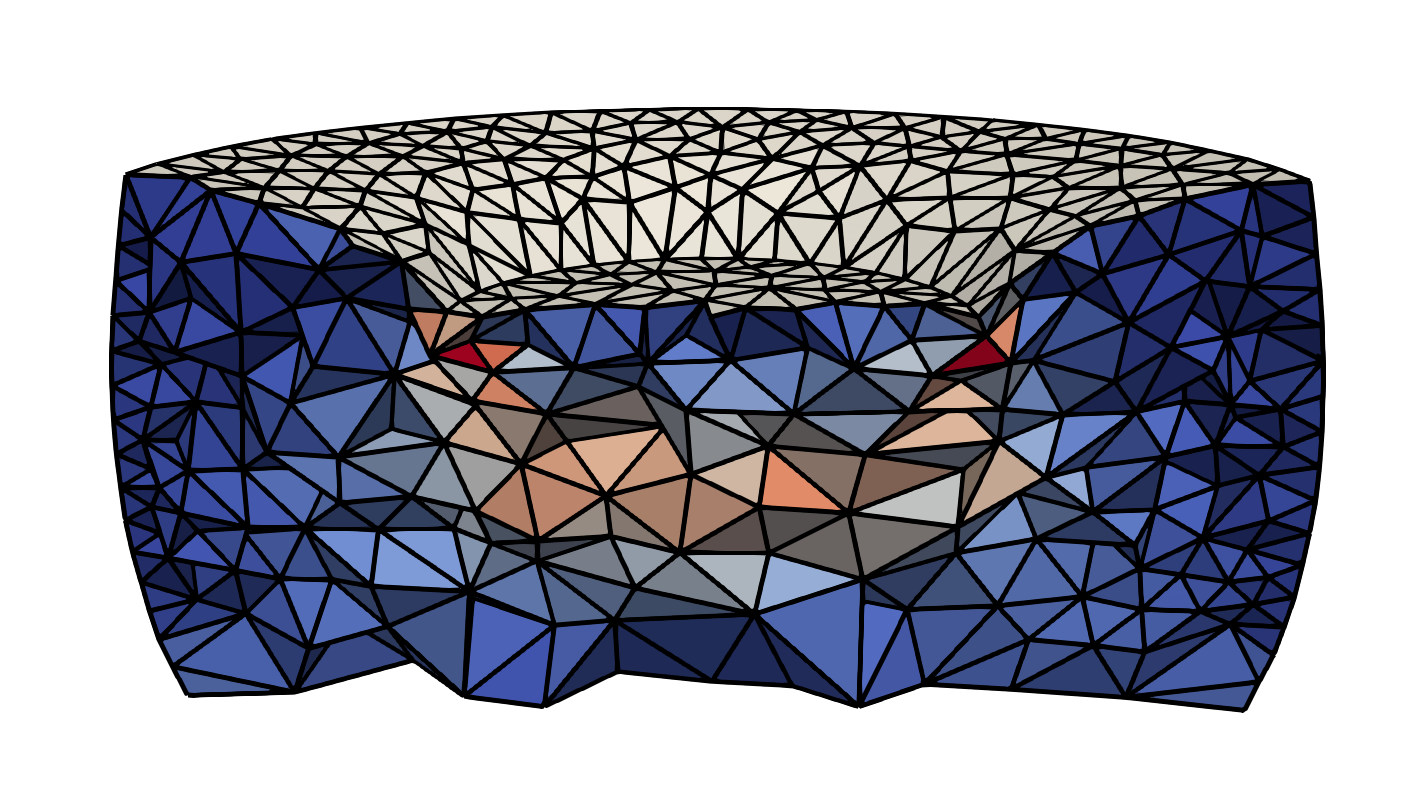}
		\includegraphics[width=1.0\textwidth]{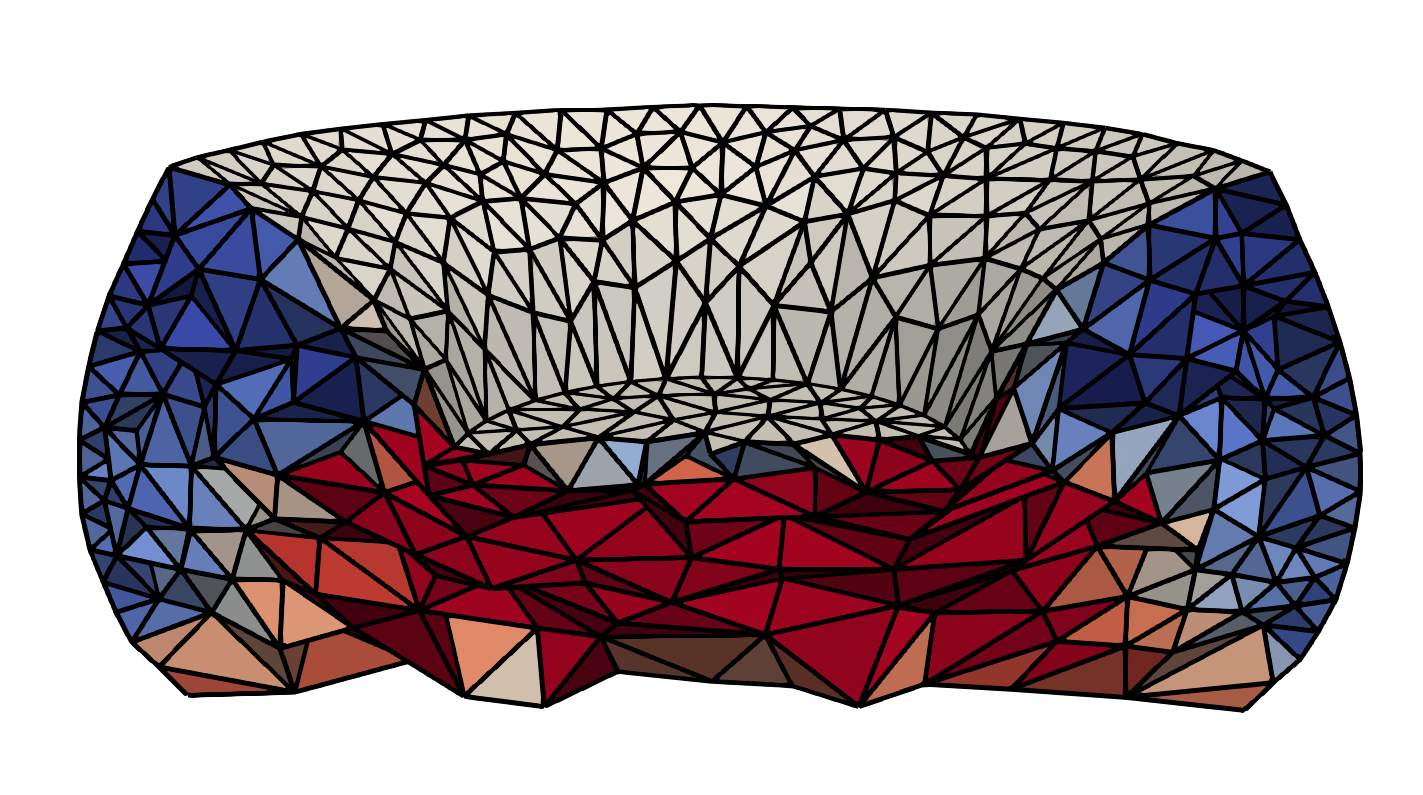}
		\adjustbox{trim={.15\width} {.0\height} {.15\width} {.1\height},clip}%
		{\includegraphics[width=1.4\textwidth]{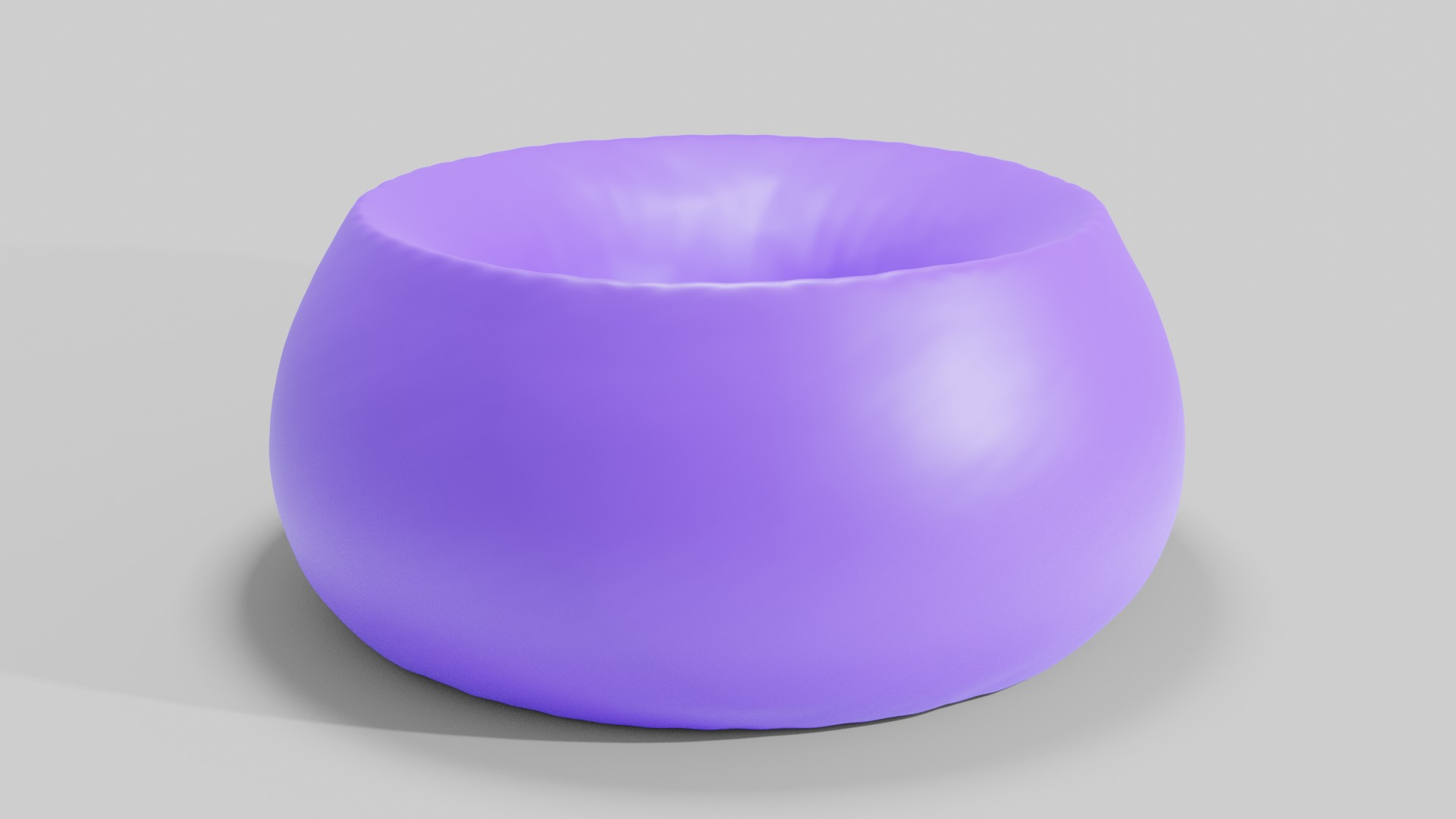}}
		\caption*{(c) CNH}
		\label{sfig:puck-vc}
	\end{subfigure}%
	\begin{subfigure}{.24\linewidth}
		\centering
		\includegraphics[width=1.0\textwidth]{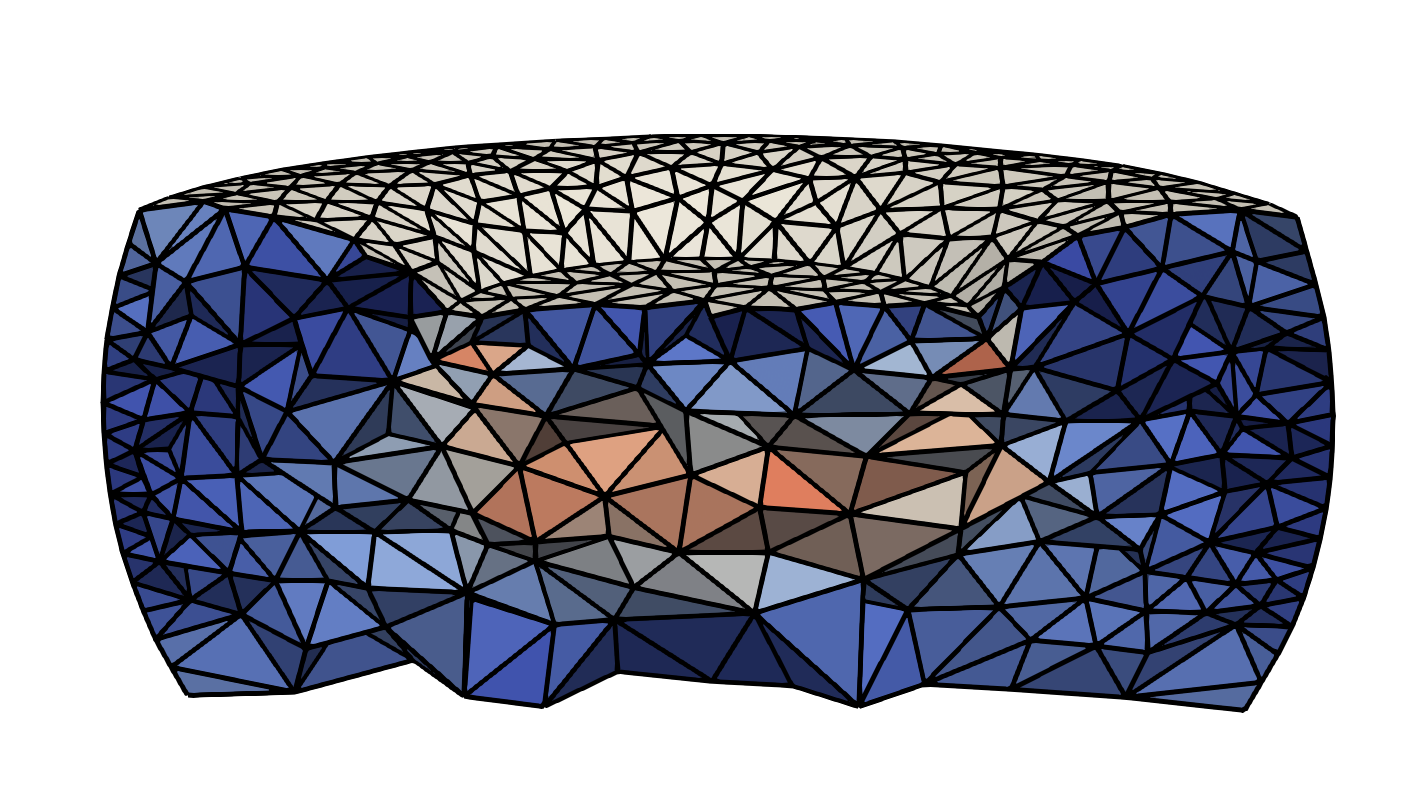}
		\includegraphics[width=1.0\textwidth]{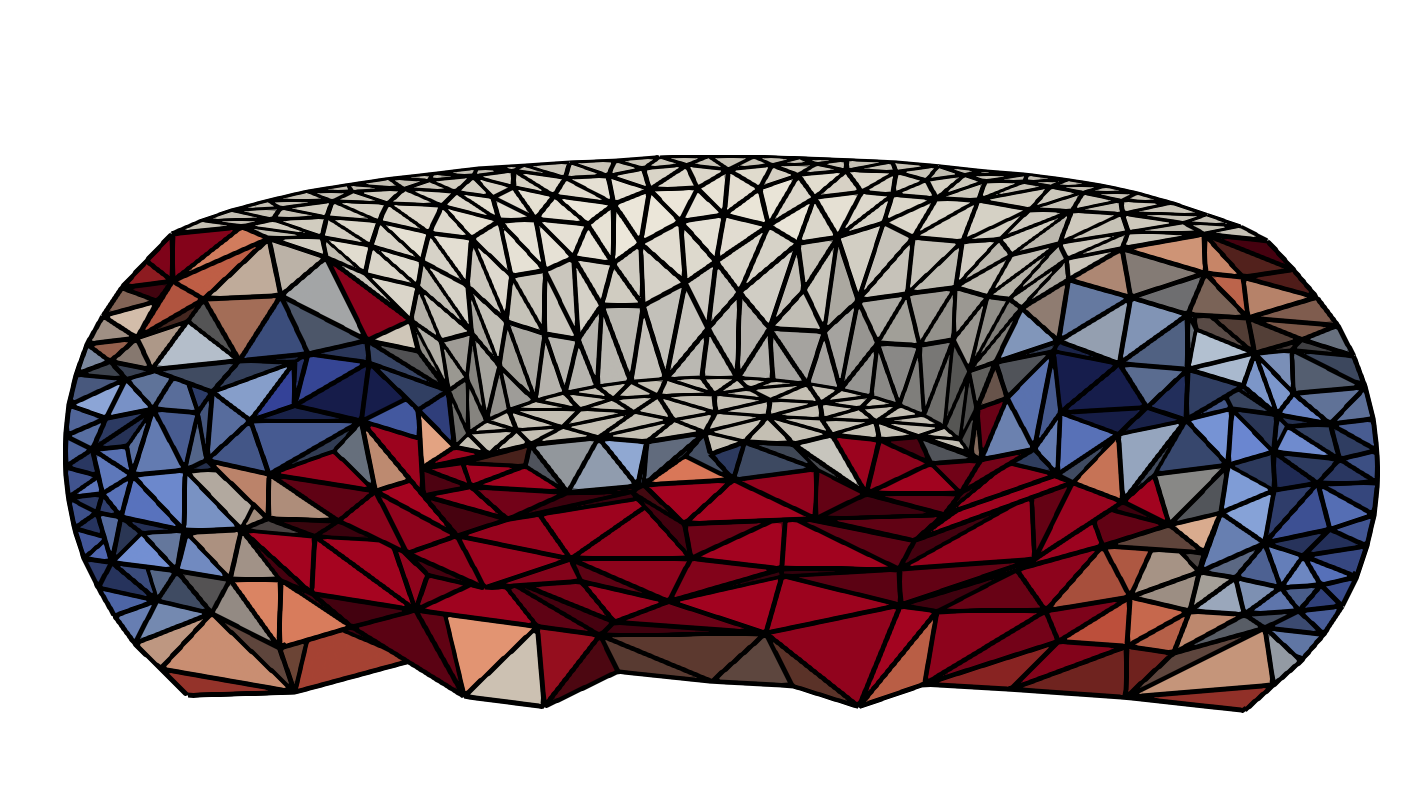}
		\adjustbox{trim={.15\width} {.0\height} {.15\width} {.1\height},clip}%
		{\includegraphics[width=1.4\textwidth]{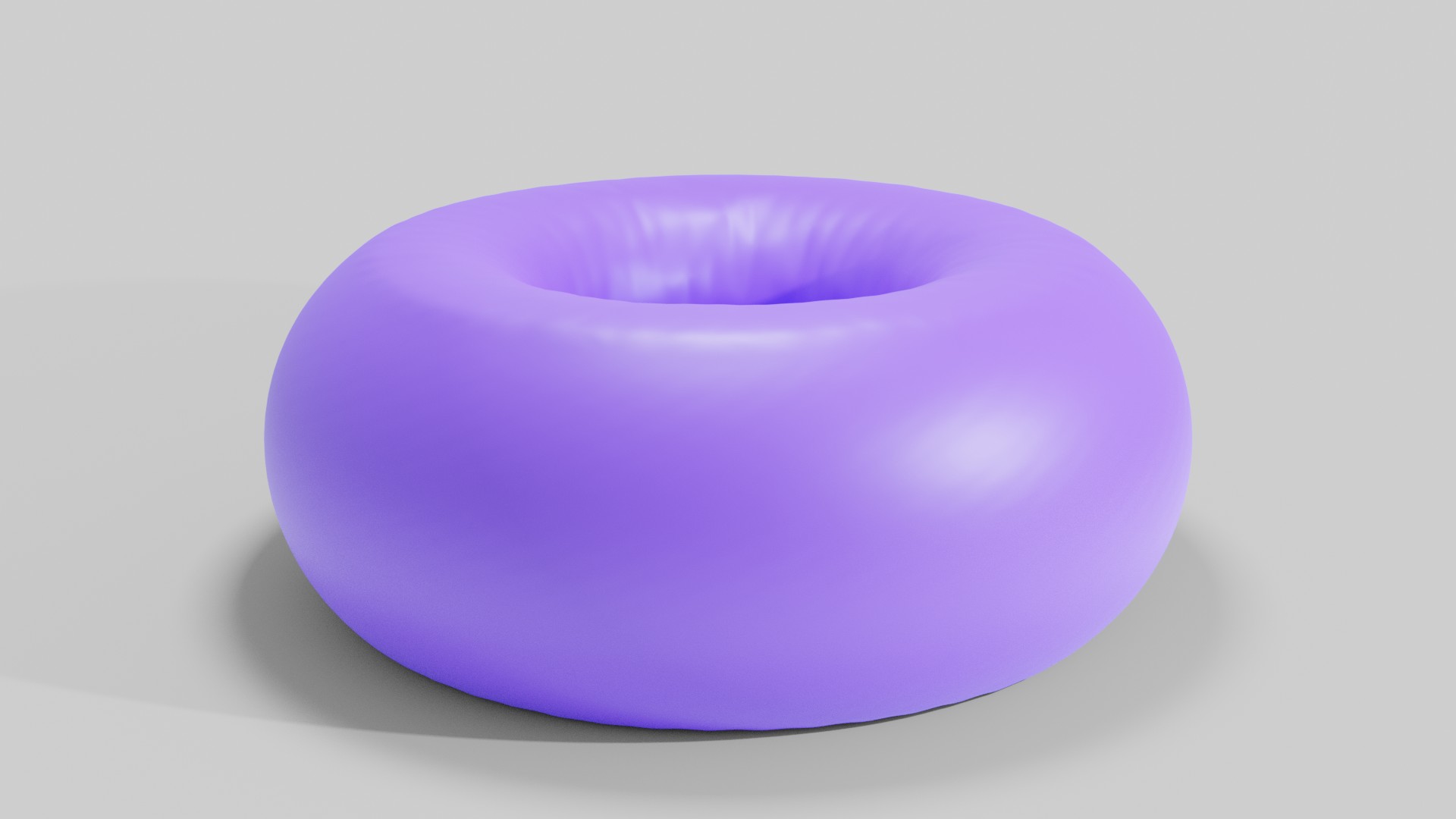}}
		\caption*{(d) CNH + Epidermis}
		\label{sfig:puck-vc-epi}
	\end{subfigure}%
	\caption{\textbf{Bulge Test}: A cross-section of a simulated puck is shown. The puck is indented at $25\%$ and $50\%$ of its height. (a) is the result without volume constraint and $\nu = 0.495$. These base results show visible volume loss of $0.6\%$ and $1.8\%$ and artificial stiffness from locking, resulting in no visible bulging at the cylinder top. Also, the pressure distribution visualized by the color scheme shows severe irregularities. (b) is the simulation with $\nu = 0.4545$, which loses $2.1\%$ and $5.1\%$ of the total volume respectively. (c) shows the result with volume constraint and $\lambda = 100$ and $\beta = 9$, showing a realistic bulging due to successful volume preservation. (d) shows the result where an epidermis of $\lambda = 100$ was added, where the surface demonstrates a more organic deformation due to surface area preservation. }
	\label{fig:pucks}
\end{figure*}

We test the stability of our method in dynamic simulations compared to \cite{Irving:2007} and a na\"ive volume rescaling method that readers may be tempted to use. We simulate a cube consisting of $16^3$ cells with $\mu = 100.0$, under gravity. We tested the simulation with a timestep of 8ms for a total of 1000 frames.
We found that the volume projection in the Irving algorithm can be unstable when used with large timesteps, and clamping of volume preservation forces must be applied to make it stable. However, we found that the clamping threshold to make this specific example not blow up was quite low, resulting in a volume error of ~2.4\% at worst. Even with such aggressive clamping, we noticed visible oscillations on the top surface of the cube.
We also tested a simple volume rescaling algorithm, where the mesh was projected at every time step based on its center of mass, such that the global volume is conserved. We found that this method is highly unstable and virtually impossible with Semi-Implicit Euler integration (where only one Newton step is used) without increasing Poisson Ratio to greater than 0.495, which leads to locking. With full Backward Euler integration, we could get the simulation to not blow up, but it still required a quite high Poisson Ratio of 0.48. However, an unrealistically exaggerated oscillation of the entire mesh was present even until the final frames.
Compared to the other methods, we found our method to be very stable, and the runtimes were comparable (around 2\% faster) to Irving's, which is a semi-implicit method where ours is fully implicit. With our method, this simulation is stable at much higher timesteps, i.e. 33ms.

The plot of the potential energy for this simulation can be seen Figure~\ref{fig:stability}. Note the extreme oscillations present in the potential energy plot for the one-ring nodal pressure algorithm \cite{Irving:2007} and the unrealistic irregularities in the plot for the volume rescaling algorithm.
For a visual comparison of the instabilities of the other two methods compared to ours, please refer to our accompanied video.

\subsection{Bulge Test}

A natural example of an incompressible material is human soft tissue, so to test the effects of our method, we test on the ``skin puck'' model from \cite{Pai:2018}, where the vertices on the bottom are fixed with Dirichlet boundary conditions.
When simulating biological tissue, bulging is a crucial visual characteristic that depicts the incompressibility of the underlying material.
Therefore, it is important that this simulation shows visually significant bulging under compression.
We use a 22K tetrahedron simulation mesh for the puck.  To produce substantial compression, we animate a set of vertices on the top of the puck with a Dirichlet boundary condition moving these vertices down by a fixed amount per time step. We performed a quasi-static simulation where at each step, the animated surface is indented by 1\% of the height. We test the displacement until 50\% of the total height of the puck, which produces extreme compression.

Without the volume constraint, a low Poisson's Ratio $\nu \in [0.0, 0.45]$ will result in little noticeable bulging at the top surface due to volume loss, and a higher $\nu$ will result in unnaturally stiff visual results and irregular pressure distribution due to volumetric locking. Also, even at $\nu = 0.495$, there was a $1.8\%$ volume loss at an indentation of 50\% of the puck height.

Adding the volume constraint allows a completely incompressible simulation with realistic pressure solutions for this example, without being a heavy burden on performance in most cases. At $50\%$ indentation, the amount of volume loss can be made arbitrarily low\footnote{Up to machine precision.} with the global volume constraint using any type of energy model and parameters. By contrast, a standard simulation without the constraint produces approximately 22\% volume loss with $\nu = 0.4$, 5.1\% loss with $\nu = 0.45$, and 1.8\% loss with $\nu = 0.495$. However, without any local compression penalty the simulation converges to an infeasible state with many inverted tets around the border of the Dirichet boundaries.

Adding our local penalty term with $\lambda = 100$ allows the simulation to be completely free of inverted tets. Although even with $\beta = 0$ the solution does not converge to an infeasible state, the lack of a sufficient resistance to volumetric deformation causes numerical instability and results in a very slow convergence of the nonlinear optimizer during the timesteps with more extreme deformations (after the 25\% indentation). By using $\beta = 9$ we were able to achieve better numerical stability, resulting in a 14.84\% faster runtime on total, and 21.17\% faster runtime when only considering the frames after the 25\% indentation where the moving Dirichlet boundary starts to invert elements. Finally, with an epidermis model of $\lambda_e = 100, \beta_e = 1$ added, we are able to generate a more regular surface deformation and achieve an visually organic deformation overall.

\subsection{Dynamic Impact}

\begin{figure*}[t!]
	\centering
	\begin{subfigure}{.15\linewidth}
		\rotatebox[origin=c]{0}{\scriptsize{(a) UNH, $\nu=0.45$}}
	\end{subfigure}%
	\begin{subfigure}{.14\linewidth}
		\centering
		\adjustbox{trim={.25\width} {.0\height} {.25\width} {.4\height},clip}%
		{\includegraphics[width=2.0\textwidth]{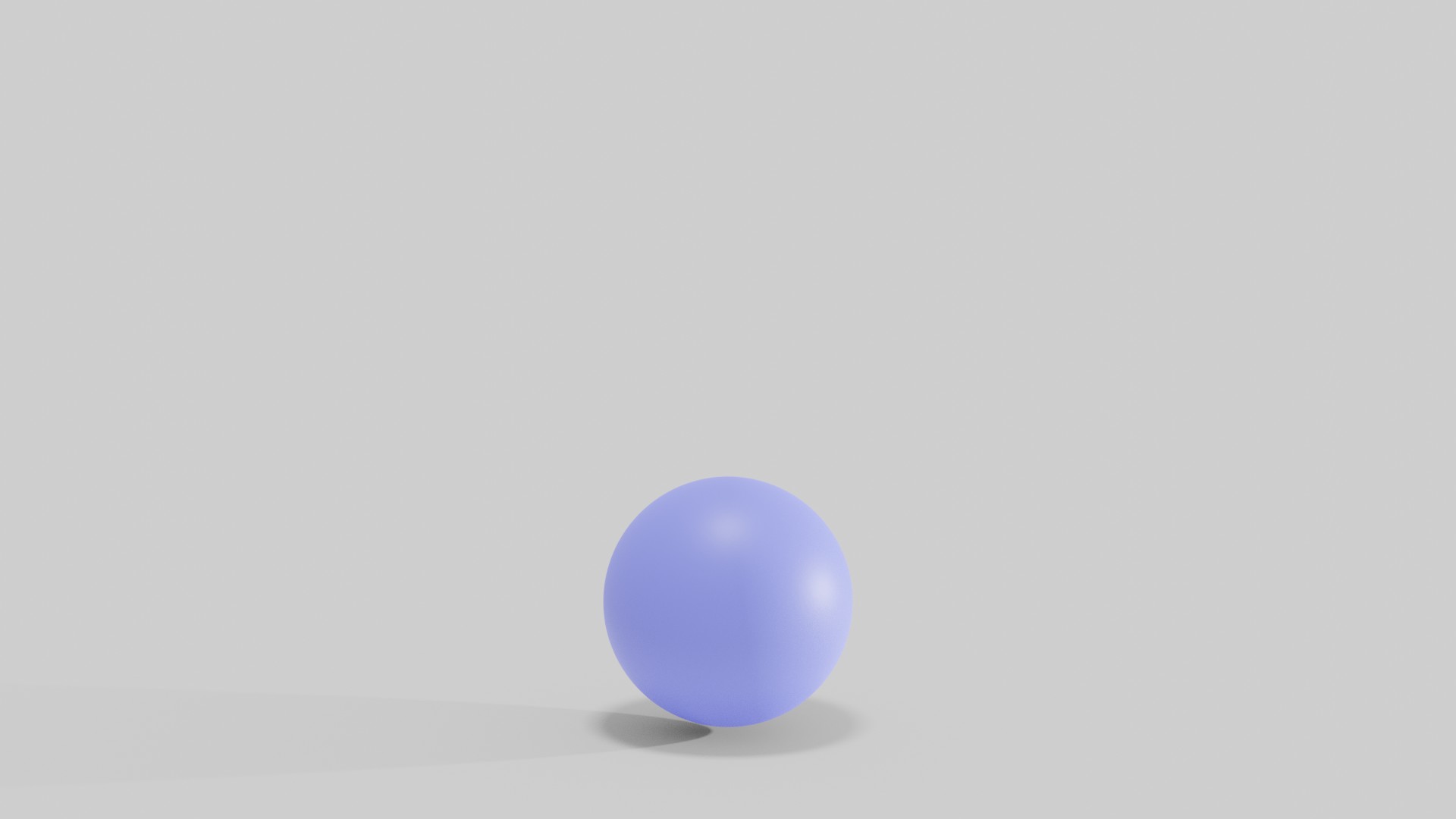}}
		%\caption*{(a1)}
		\label{sfig:ball-045-1}
	\end{subfigure}%
	\begin{subfigure}{.14\linewidth}
		\centering
		\adjustbox{trim={.25\width} {.0\height} {.25\width} {.4\height},clip}%
		{\includegraphics[width=2.0\textwidth]{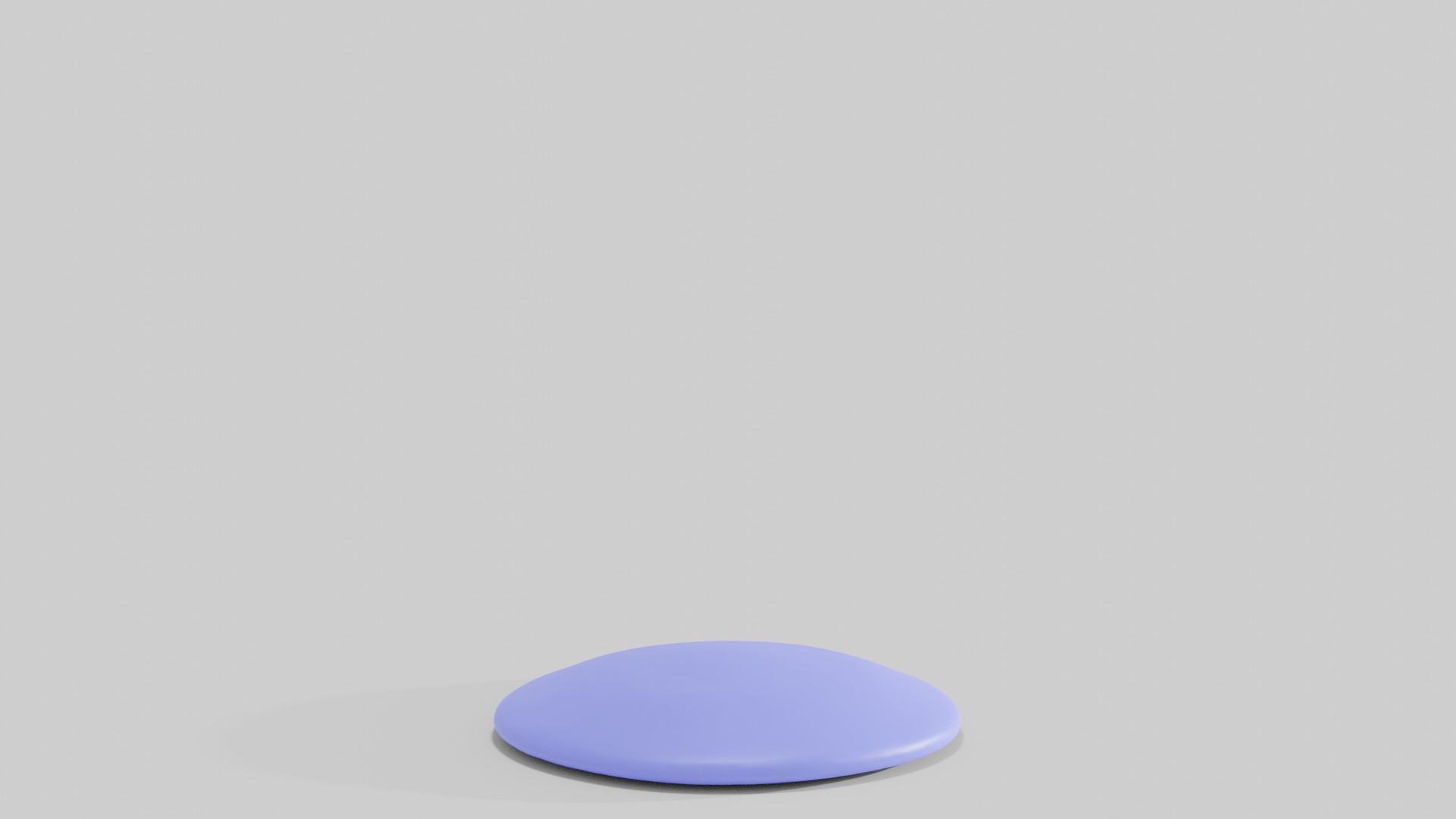}}
		%\caption*{(a2)}
		\label{sfig:ball-045-2}
	\end{subfigure}%
	\begin{subfigure}{.14\linewidth}
		\centering
		\adjustbox{trim={.25\width} {.0\height} {.25\width} {.4\height},clip}%
		{\includegraphics[width=2.0\textwidth]{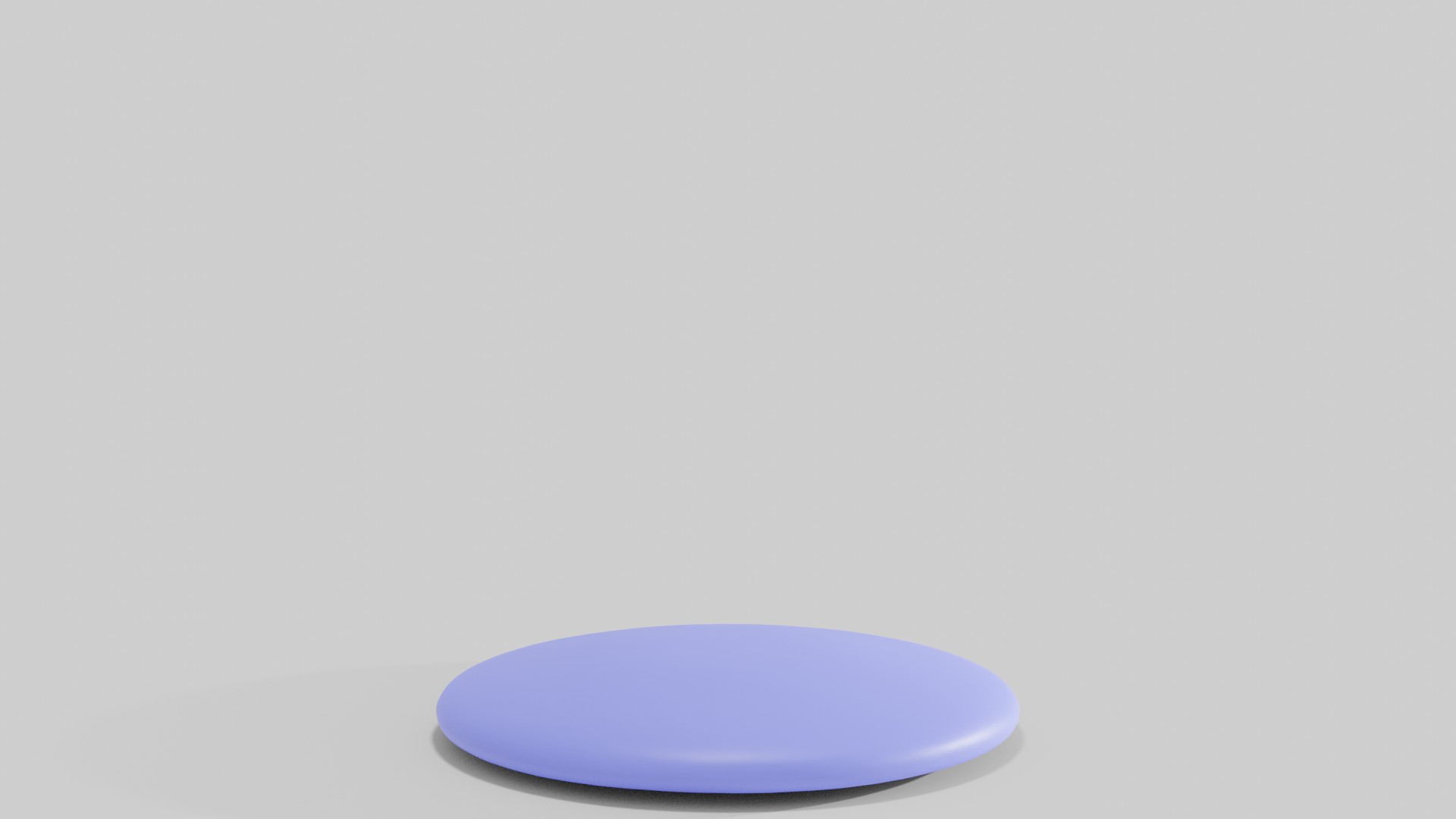}}
		%\caption*{(a3)}
		\label{sfig:ball-045-3}
	\end{subfigure}%
	\begin{subfigure}{.14\linewidth}
		\centering
		\adjustbox{trim={.25\width} {.0\height} {.25\width} {.4\height},clip}%
		{\includegraphics[width=2.0\textwidth]{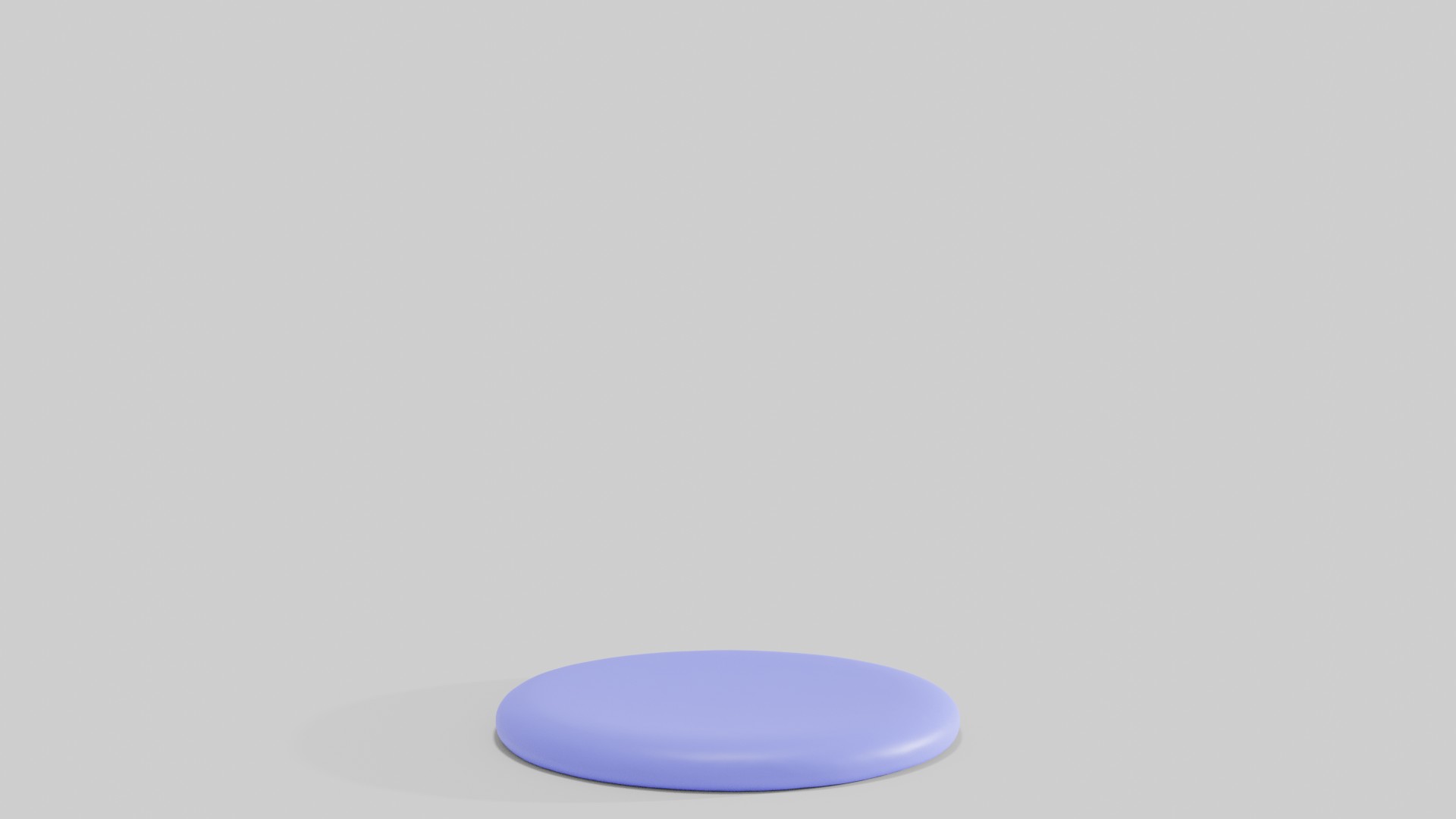}}
		%\caption*{(a4)}
		\label{sfig:ball-045-4}
	\end{subfigure}%
	\begin{subfigure}{.14\linewidth}
		\centering
		\adjustbox{trim={.25\width} {.0\height} {.25\width} {.4\height},clip}%
		{\includegraphics[width=2.0\textwidth]{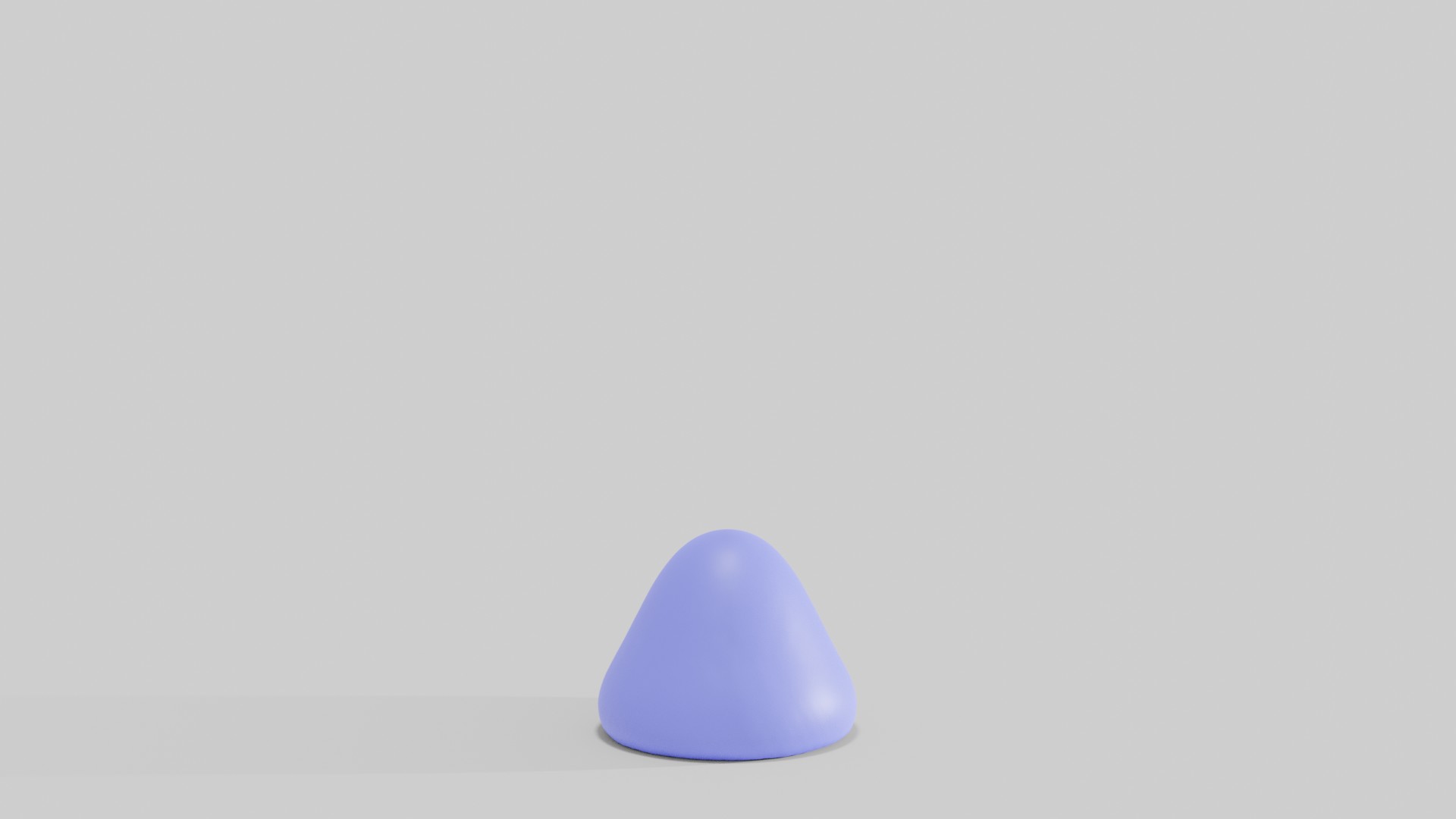}}
		%\caption*{(a5)}
		\label{sfig:ball-045-5}
	\end{subfigure}%
	\begin{subfigure}{.14\linewidth}
		\centering
		\adjustbox{trim={.25\width} {.0\height} {.25\width} {.4\height},clip}%
		{\includegraphics[width=2.0\textwidth]{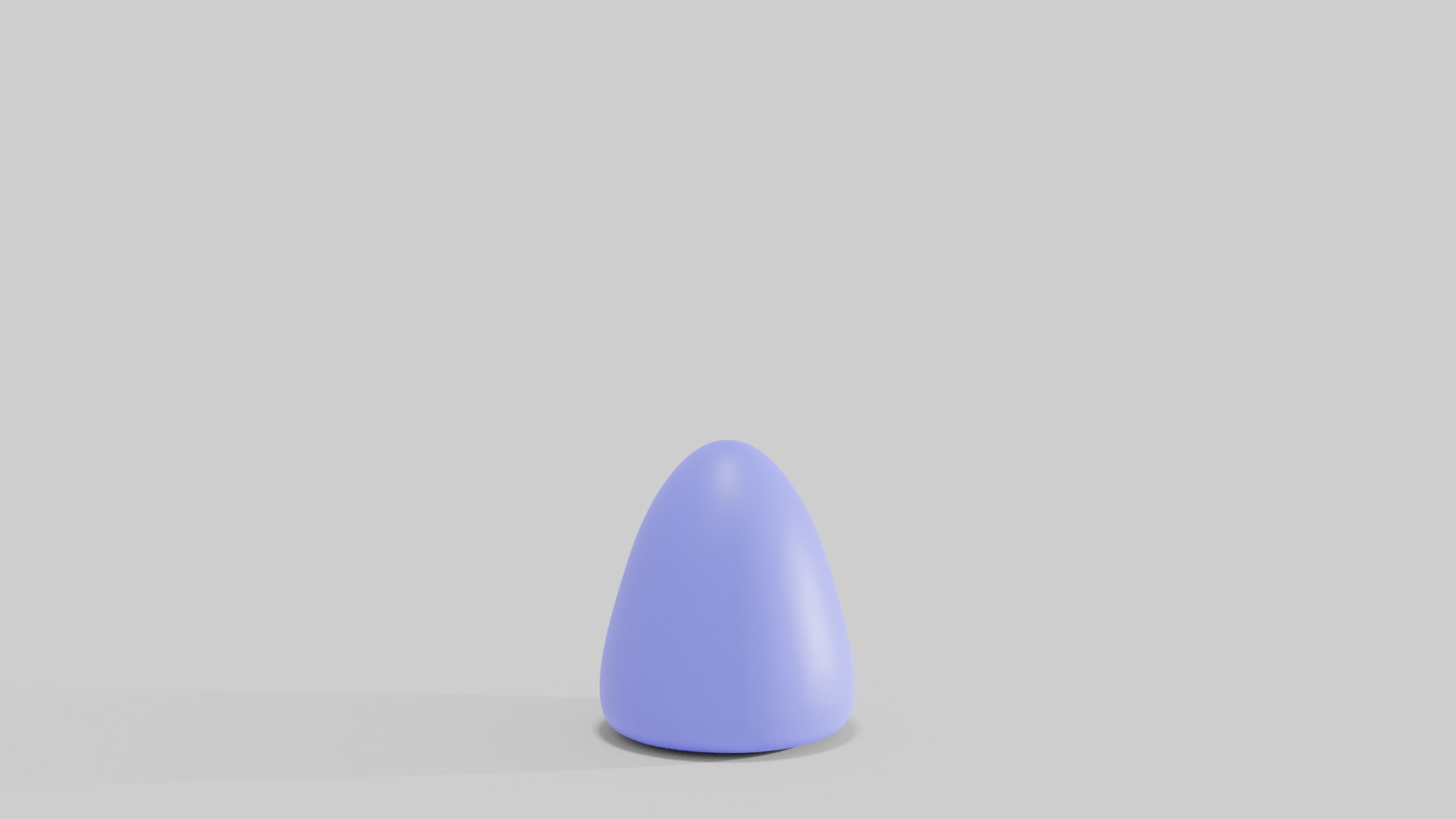}}
		%\caption*{(a6)}
		\label{sfig:ball-045-6}
	\end{subfigure}\hfill
	\begin{subfigure}{.15\linewidth}
		\rotatebox[origin=c]{0}{\scriptsize{(b) UNH, $\nu=0.495$}}
	\end{subfigure}%
	\begin{subfigure}{.14\linewidth}
		\centering
		\adjustbox{trim={.25\width} {.0\height} {.25\width} {.4\height},clip}%
		{\includegraphics[width=2.0\textwidth]{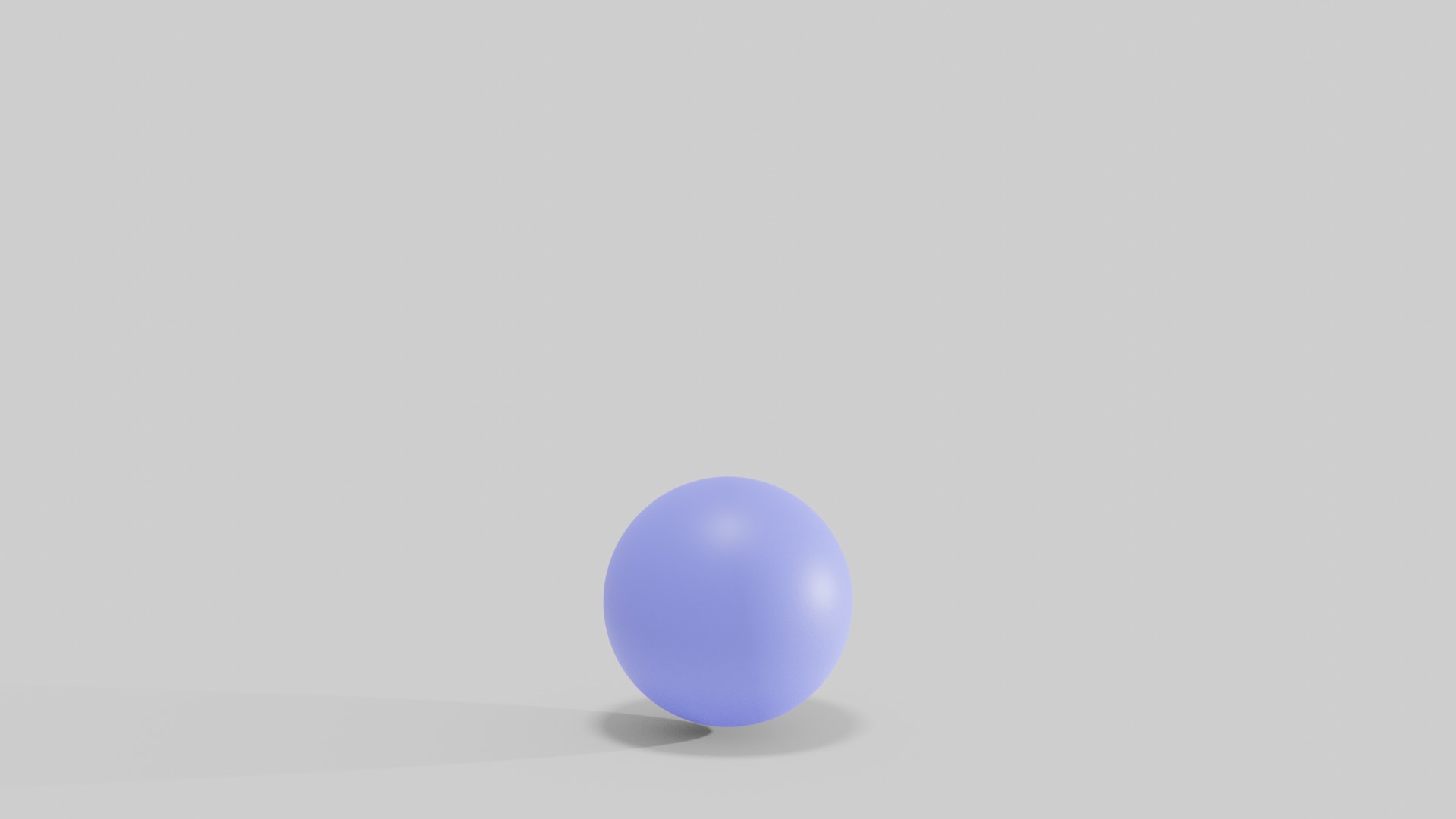}}
		%\caption*{(a1)}
		\label{sfig:ball-0495-1}
	\end{subfigure}%
	\begin{subfigure}{.14\linewidth}
		\centering
		\adjustbox{trim={.25\width} {.0\height} {.25\width} {.4\height},clip}%
		{\includegraphics[width=2.0\textwidth]{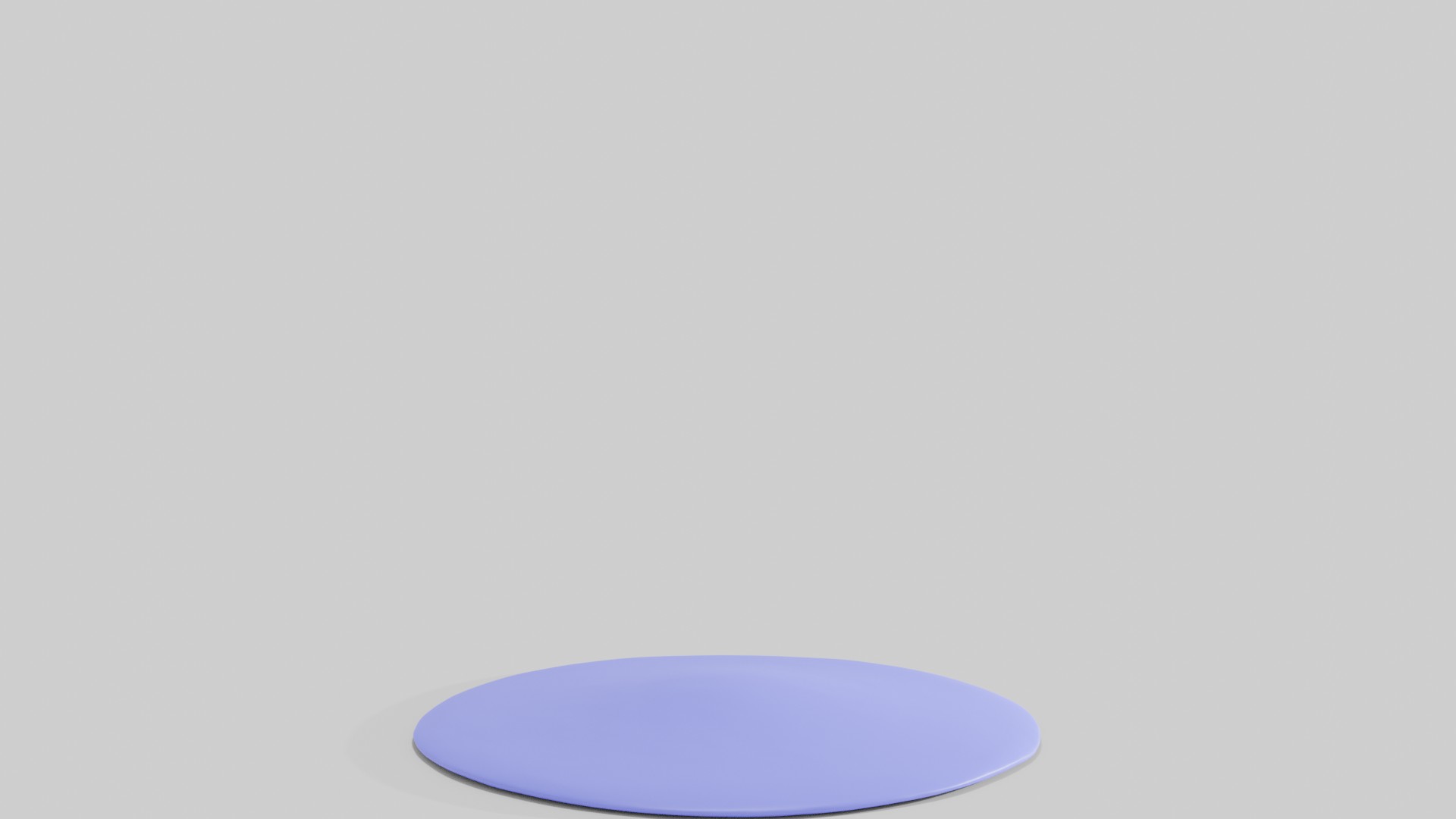}}
		%\caption*{(a2)}
		\label{sfig:ball-0495-2}
	\end{subfigure}%
	\begin{subfigure}{.14\linewidth}
		\centering
		\adjustbox{trim={.25\width} {.0\height} {.25\width} {.4\height},clip}%
		{\includegraphics[width=2.0\textwidth]{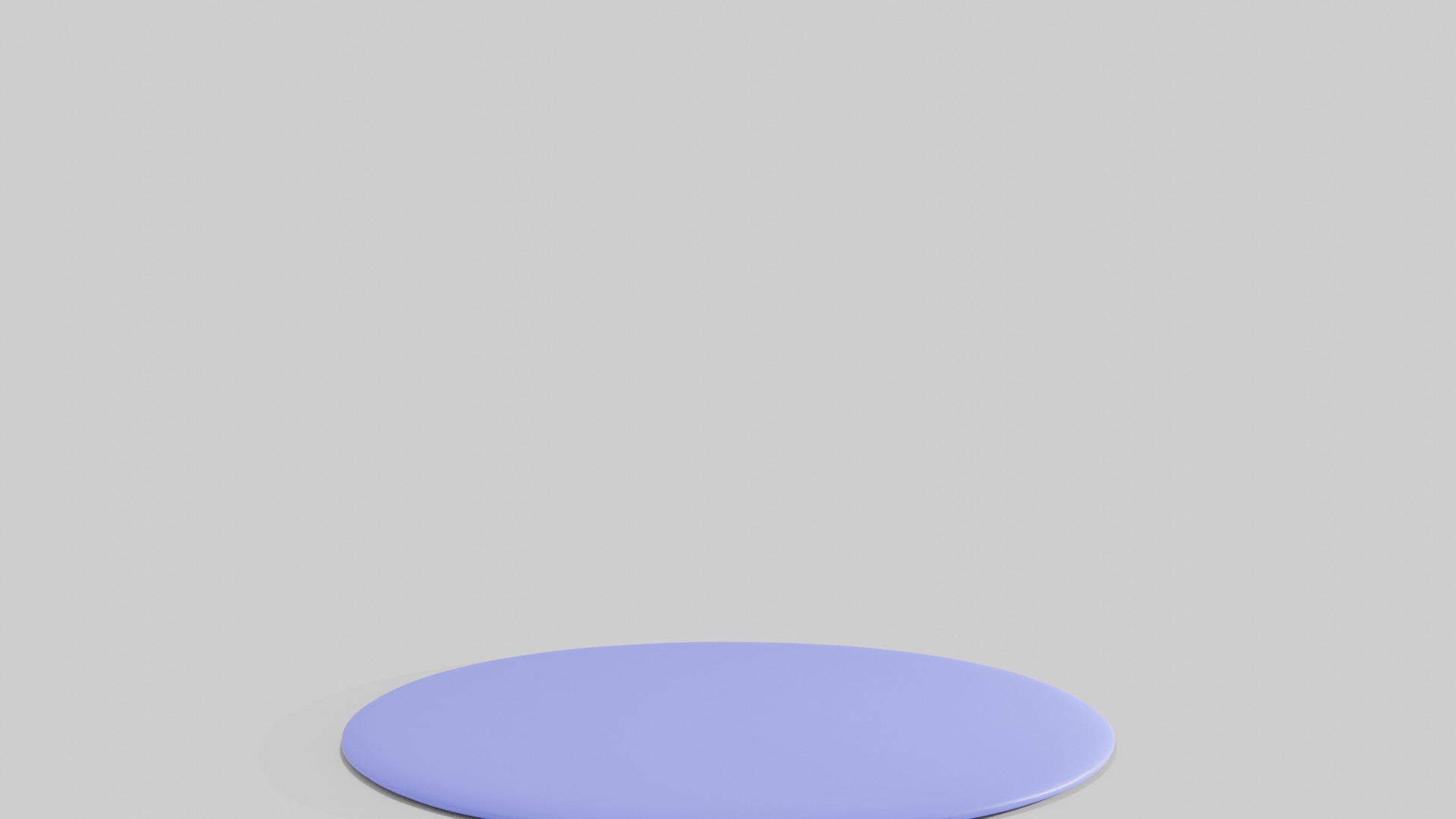}}
		%\caption*{(a3)}
		\label{sfig:ball-0495-3}
	\end{subfigure}%
	\begin{subfigure}{.14\linewidth}
		\centering
		\adjustbox{trim={.25\width} {.0\height} {.25\width} {.4\height},clip}%
		{\includegraphics[width=2.0\textwidth]{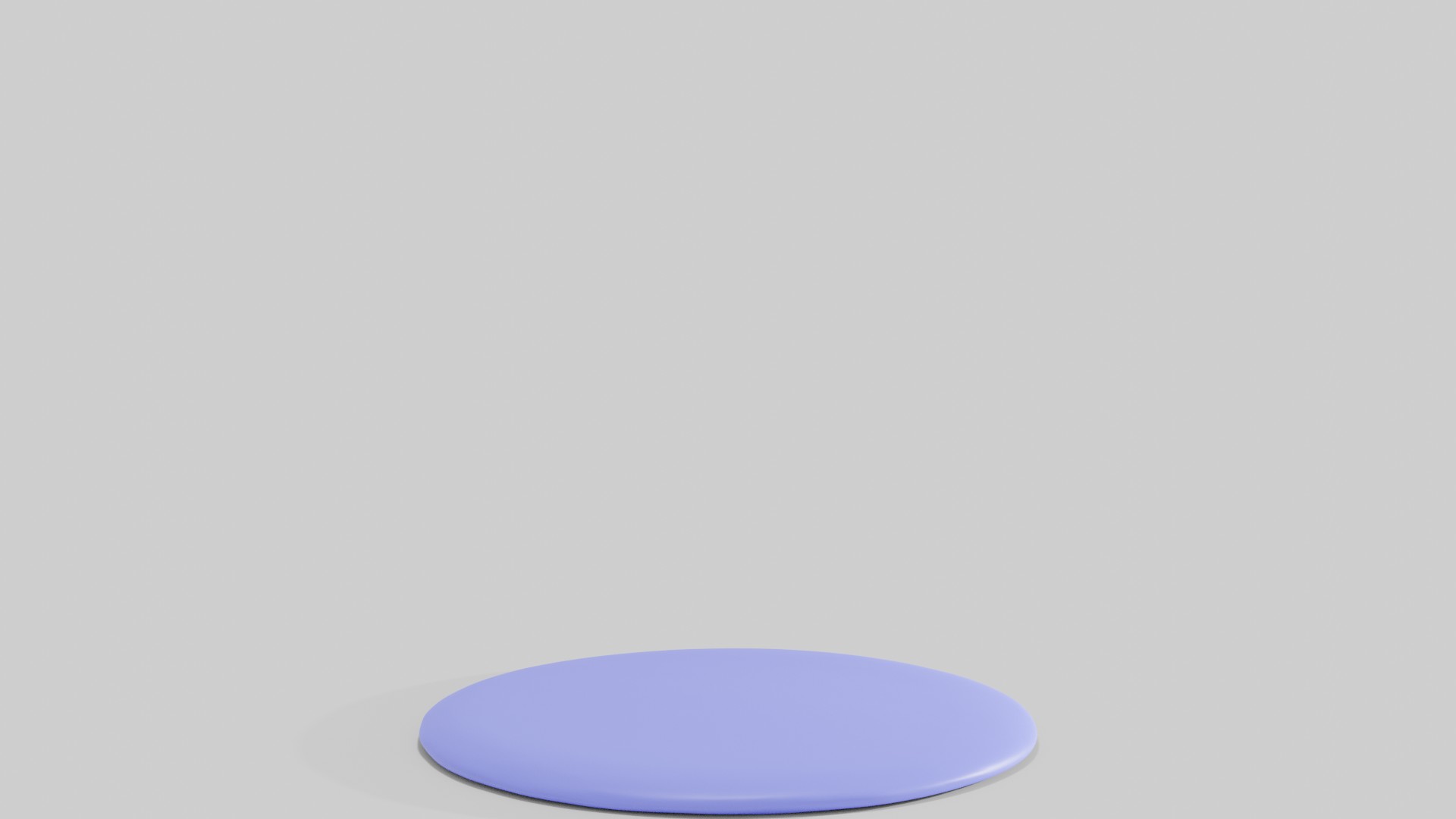}}
		%\caption*{(a4)}
		\label{sfig:ball-0495-4}
	\end{subfigure}%
	\begin{subfigure}{.14\linewidth}
		\centering
		\adjustbox{trim={.25\width} {.0\height} {.25\width} {.4\height},clip}%
		{\includegraphics[width=2.0\textwidth]{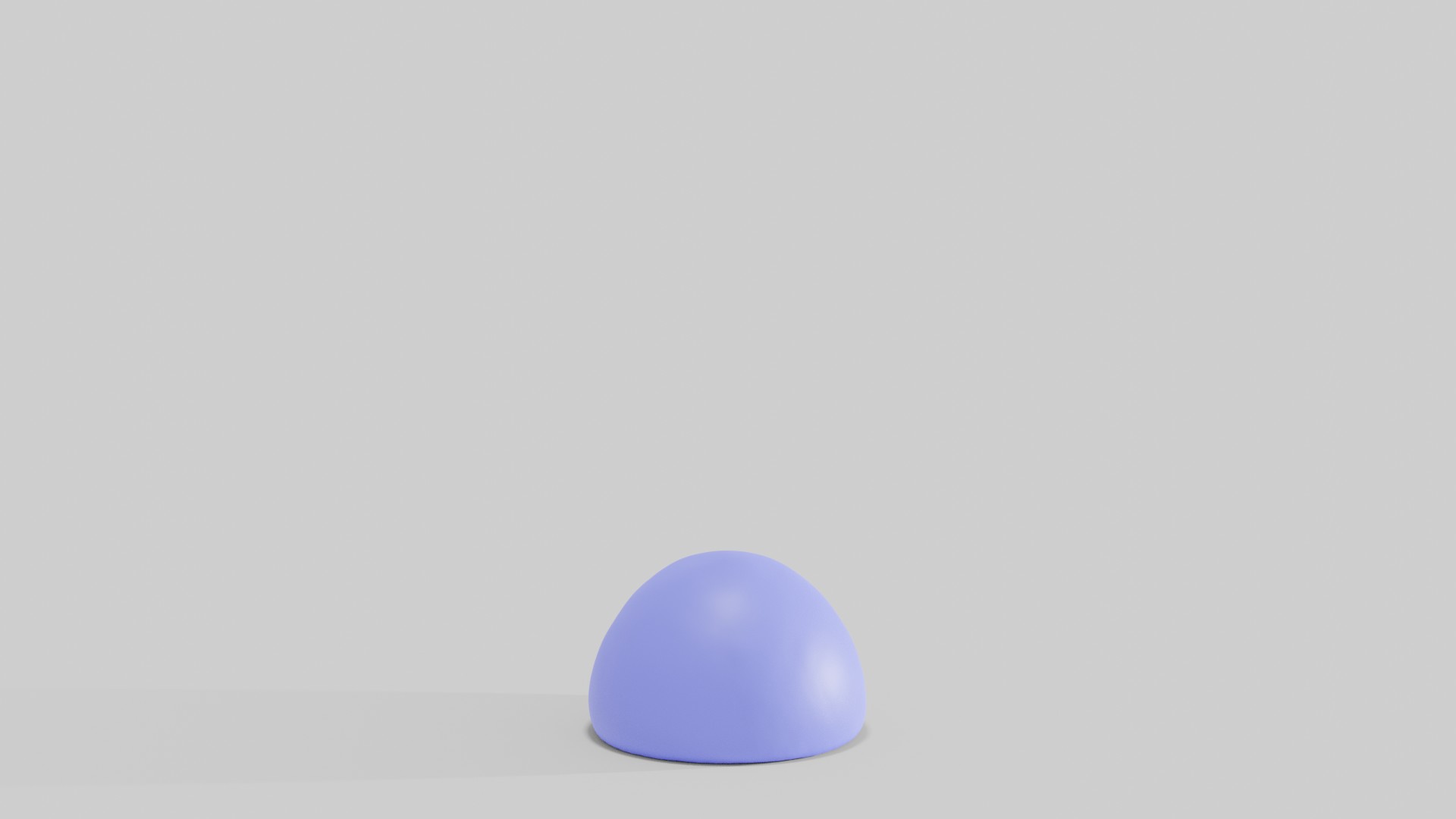}}
		%\caption*{(a5)}
		\label{sfig:ball-0495-5}
	\end{subfigure}%
	\begin{subfigure}{.14\linewidth}
		\centering
		\adjustbox{trim={.25\width} {.0\height} {.25\width} {.4\height},clip}%
		{\includegraphics[width=2.0\textwidth]{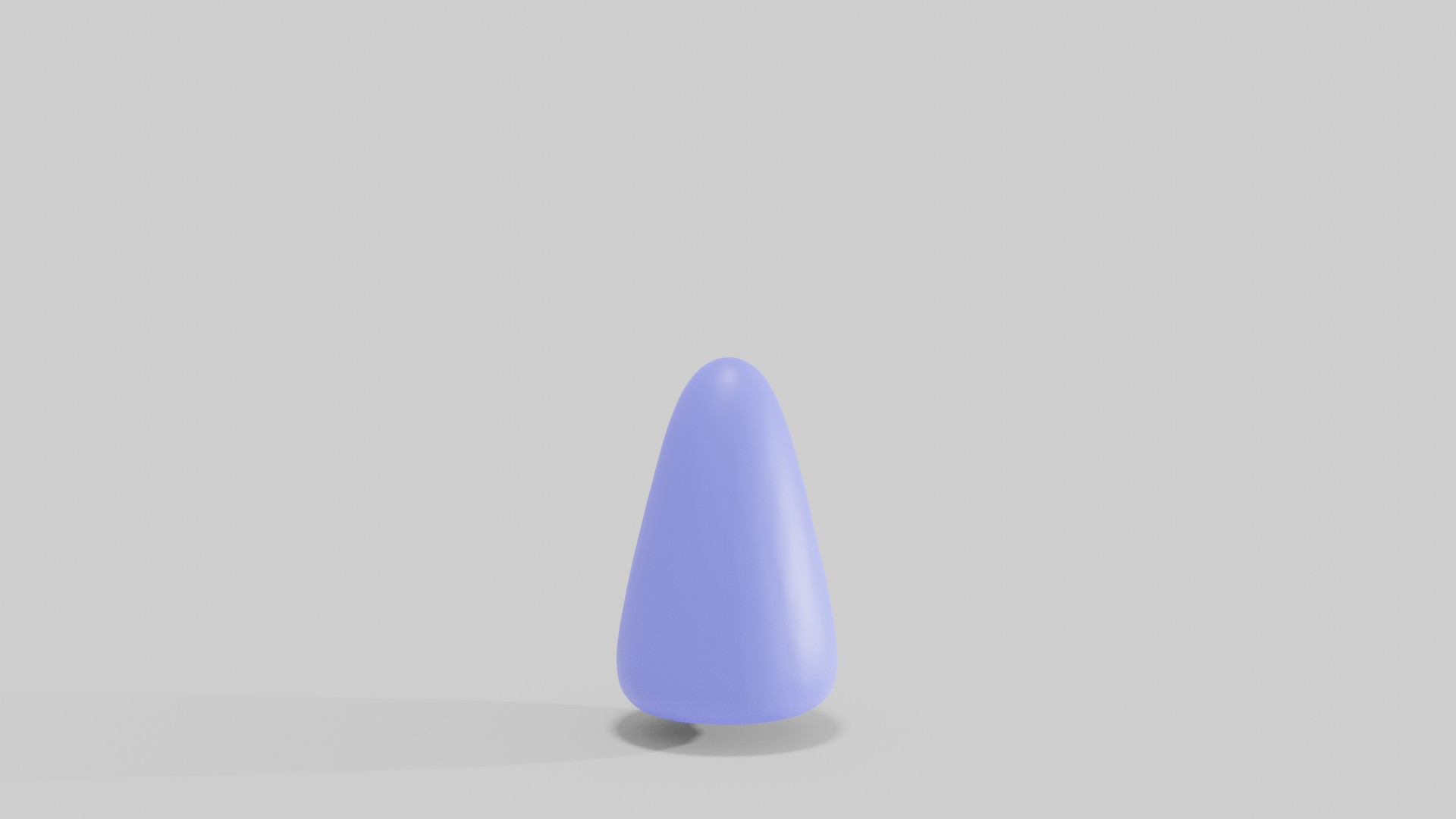}}
		%\caption*{(a6)}
		\label{sfig:ball-0495-6}
	\end{subfigure}\hfill
	\begin{subfigure}{.15\linewidth}
		\rotatebox[origin=c]{0}{\scriptsize{(c) Ours}}
	\end{subfigure}%
	\begin{subfigure}{.14\linewidth}
		\centering
		\adjustbox{trim={.25\width} {.0\height} {.25\width} {.4\height},clip}%
		{\includegraphics[width=2.0\textwidth]{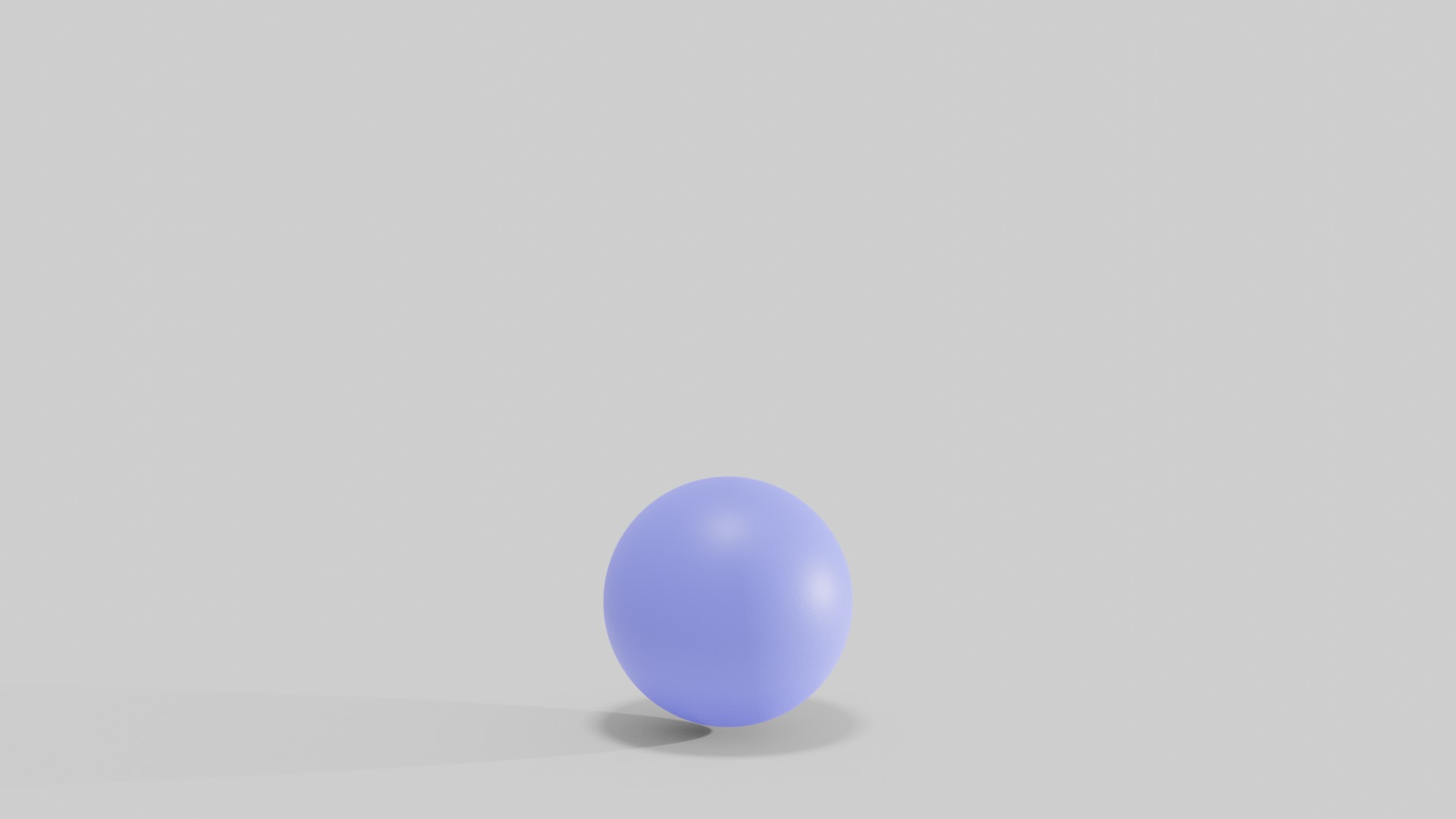}}
		%\caption*{(a1)}
		\label{sfig:ball-vc-1}
	\end{subfigure}%
	\begin{subfigure}{.14\linewidth}
		\centering
		\adjustbox{trim={.25\width} {.0\height} {.25\width} {.4\height},clip}%
		{\includegraphics[width=2.0\textwidth]{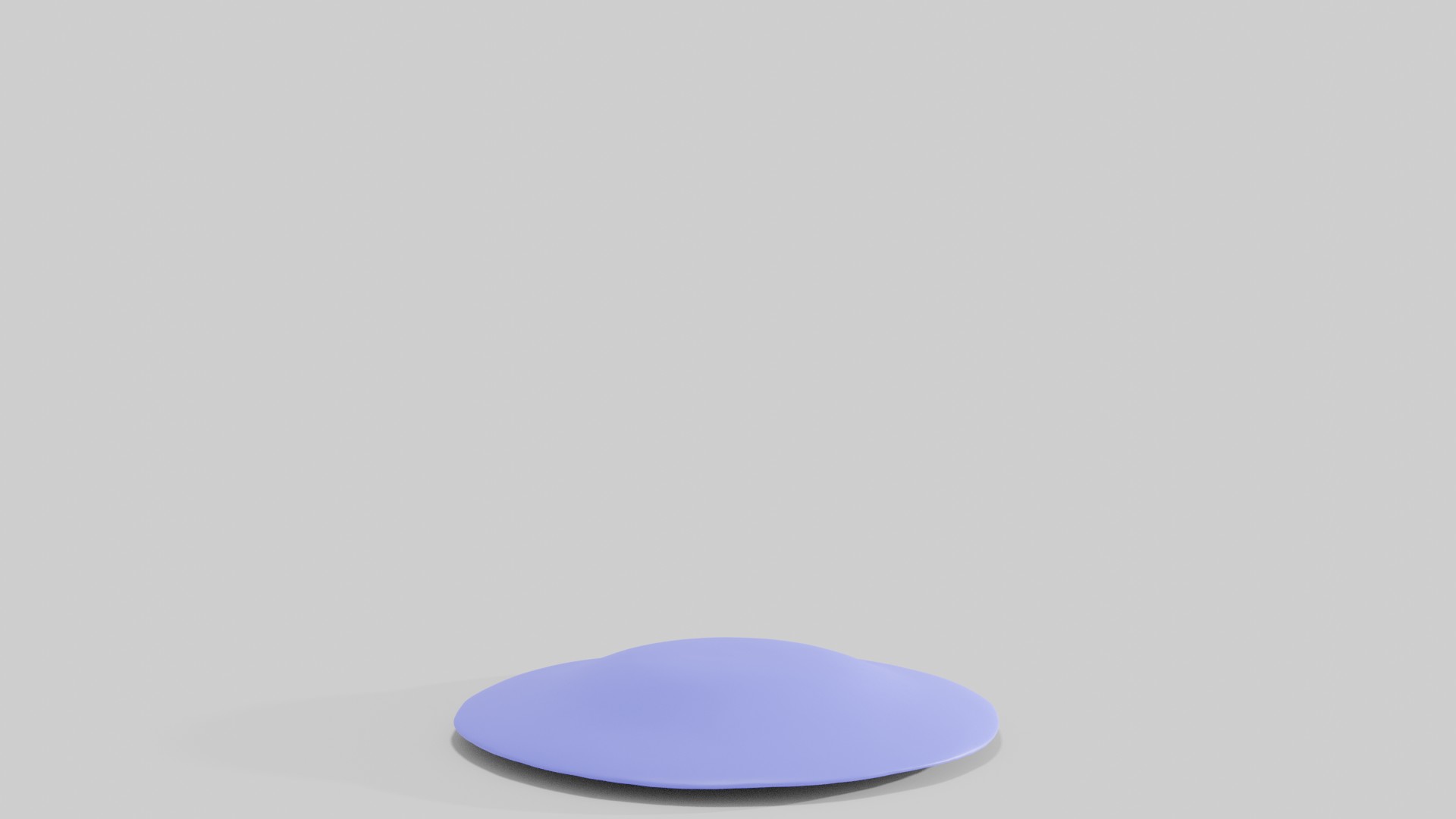}}
		%\caption*{(a2)}
		\label{sfig:ball-vc-2}
	\end{subfigure}%
	\begin{subfigure}{.14\linewidth}
		\centering
		\adjustbox{trim={.25\width} {.0\height} {.25\width} {.4\height},clip}%
		{\includegraphics[width=2.0\textwidth]{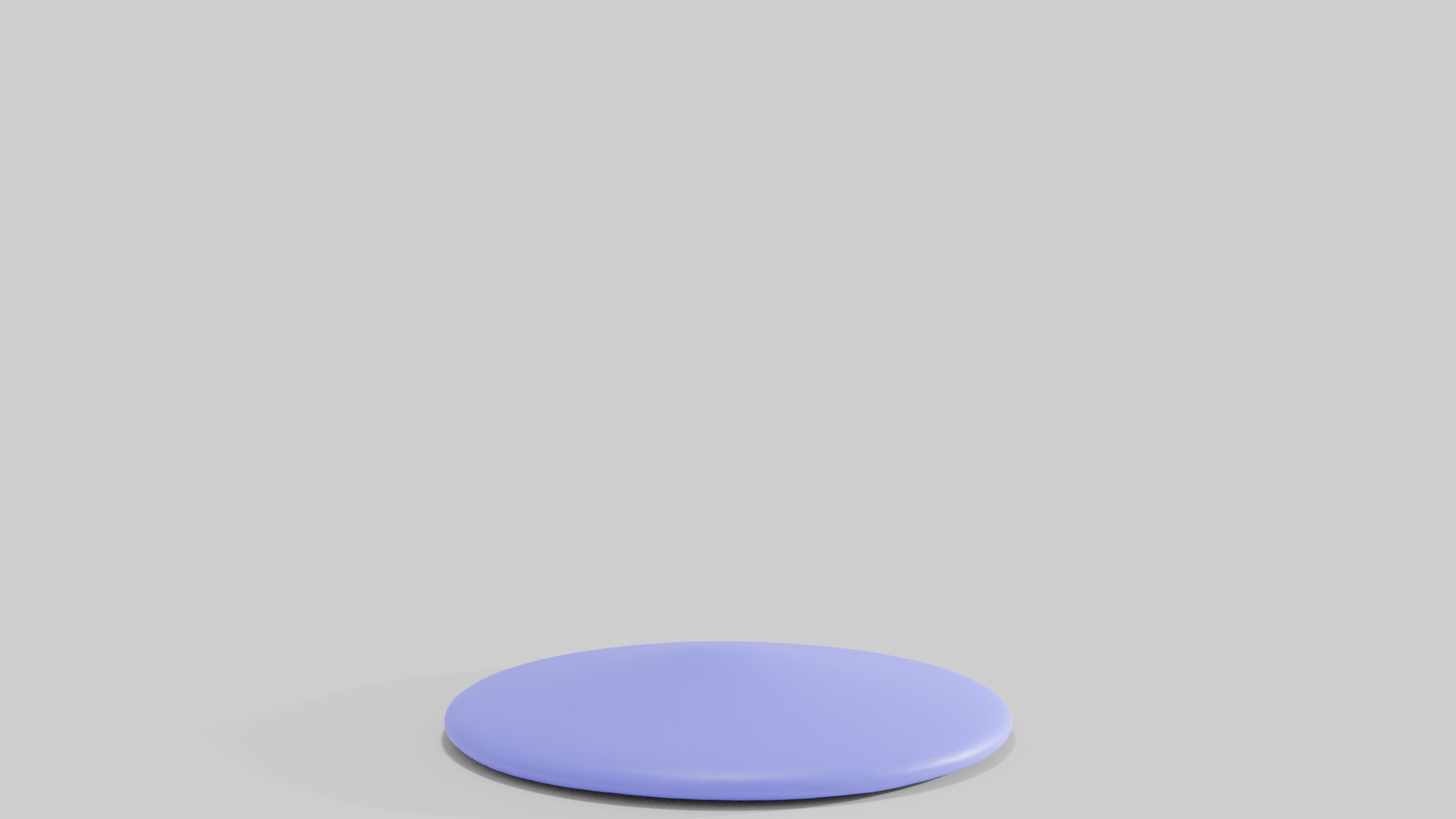}}
		%\caption*{(a3)}
		\label{sfig:ball-vc-3}
	\end{subfigure}%
	\begin{subfigure}{.14\linewidth}
		\centering
		\adjustbox{trim={.25\width} {.0\height} {.25\width} {.4\height},clip}%
		{\includegraphics[width=2.0\textwidth]{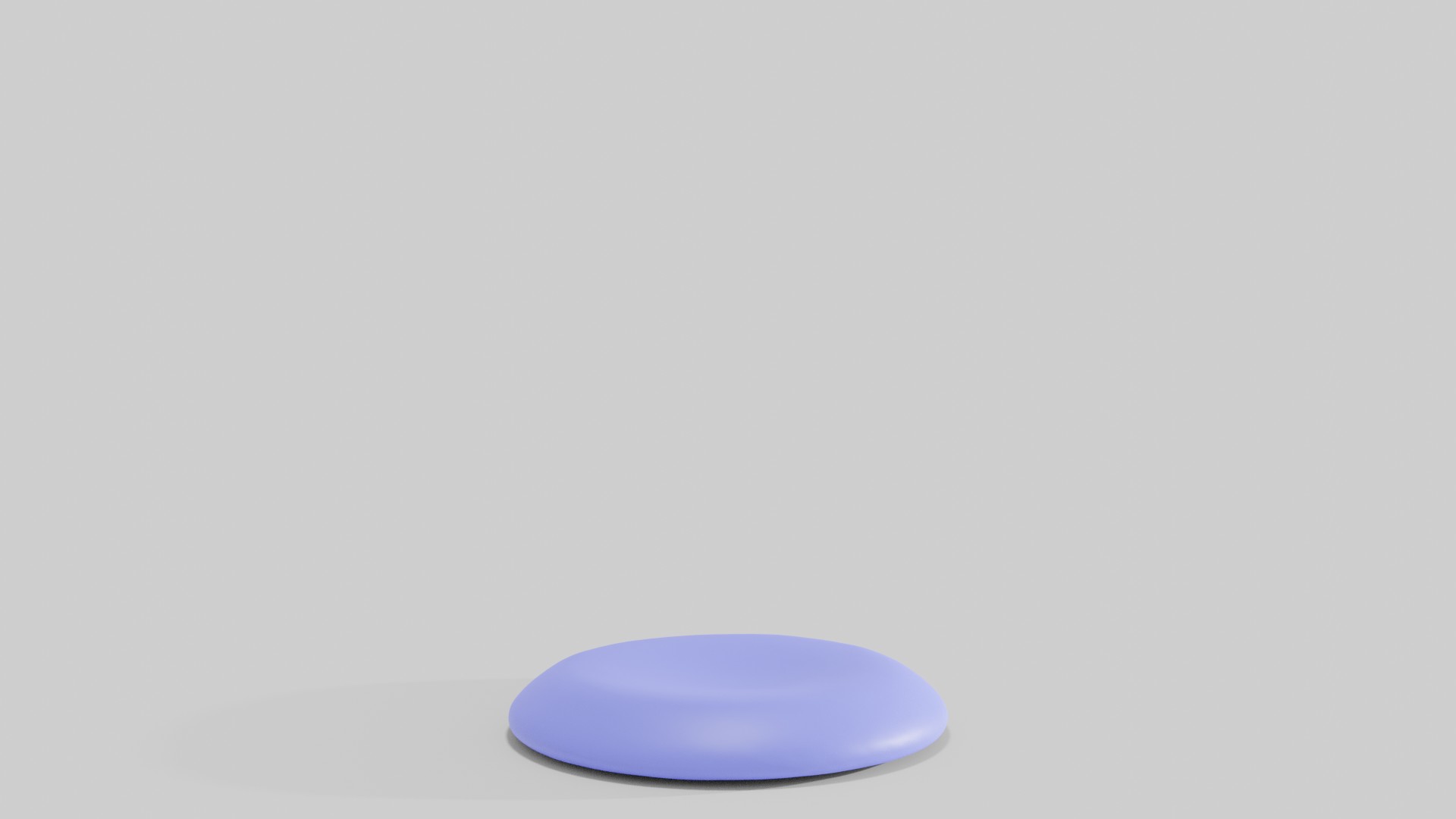}}
		%\caption*{(a4)}
		\label{sfig:ball-vc-4}
	\end{subfigure}%
	\begin{subfigure}{.14\linewidth}
		\centering
		\adjustbox{trim={.25\width} {.0\height} {.25\width} {.4\height},clip}%
		{\includegraphics[width=2.0\textwidth]{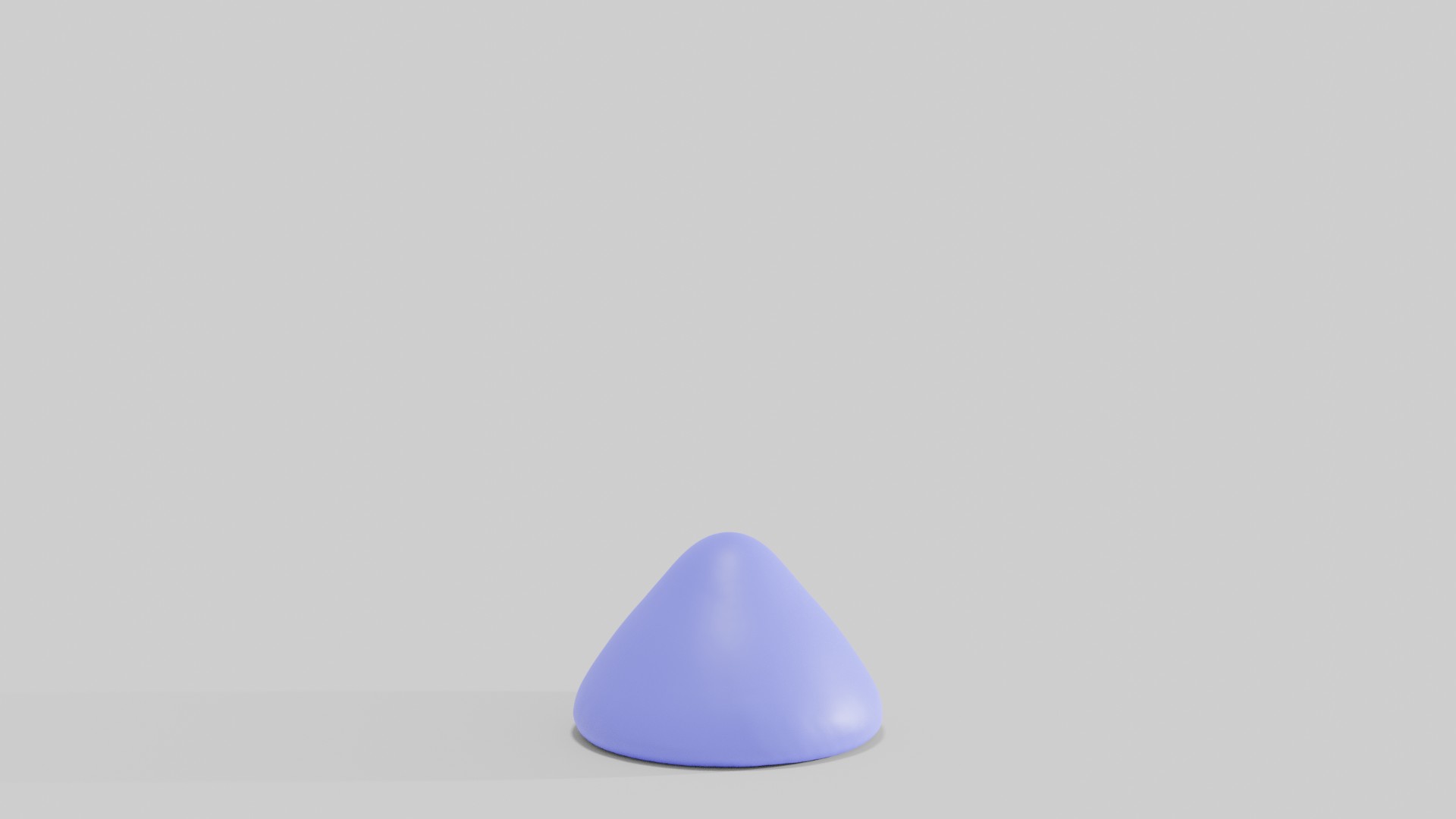}}
		%\caption*{(a5)}
		\label{sfig:ball-vc-5}
	\end{subfigure}%
	\begin{subfigure}{.14\linewidth}
		\centering
		\adjustbox{trim={.25\width} {.0\height} {.25\width} {.4\height},clip}%
		{\includegraphics[width=2.0\textwidth]{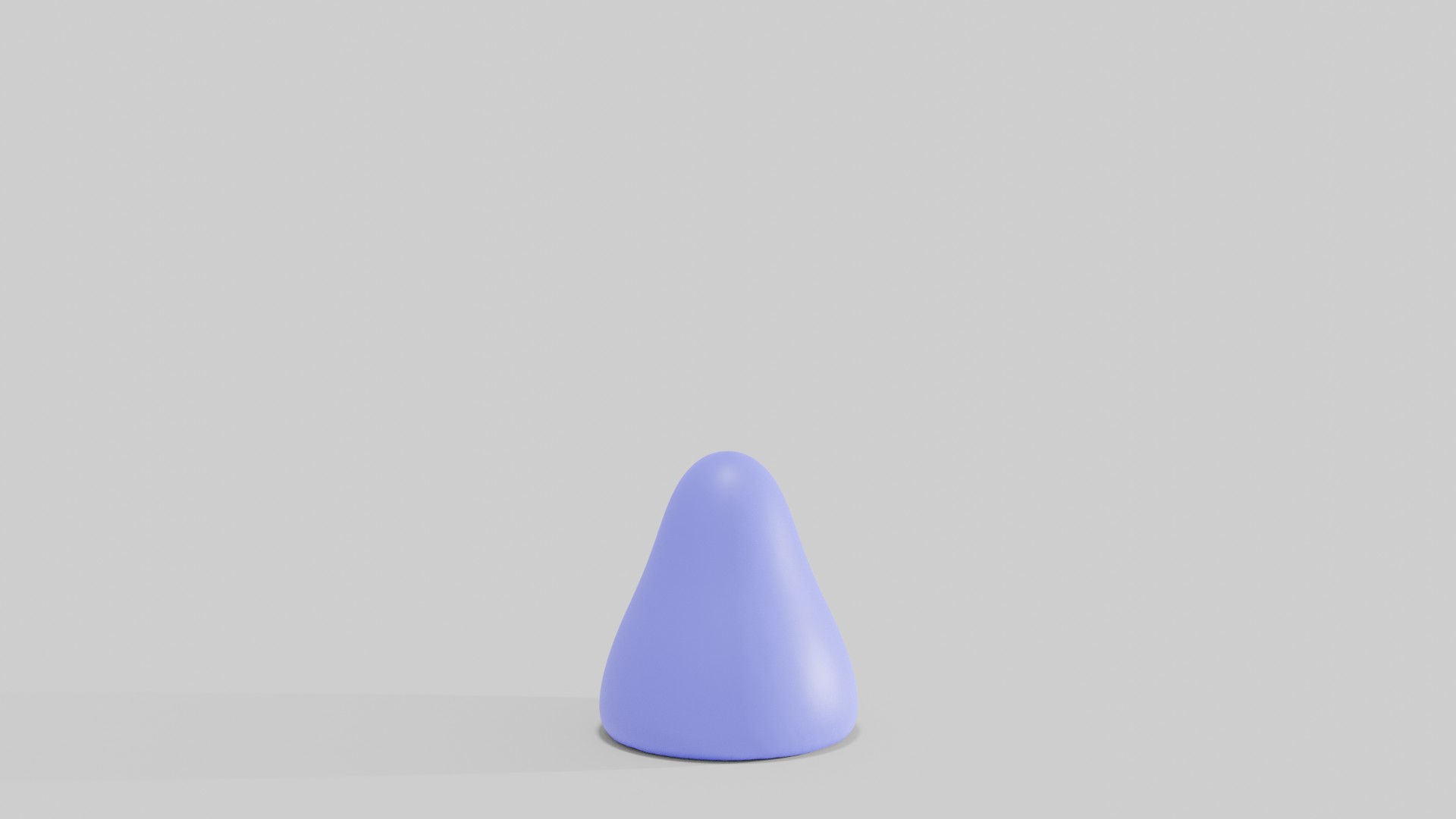}}
		%\caption*{(a6)}
		\label{sfig:ball-vc-6}
	\end{subfigure}\hfill
	\begin{subfigure}{.15\linewidth}
		\rotatebox[origin=c]{0}{\scriptsize{Frame}}
	\end{subfigure}%
	\begin{subfigure}{.14\linewidth}
		\centering
		200
	\end{subfigure}%
	\begin{subfigure}{.14\linewidth}
		\centering
		250
	\end{subfigure}%
	\begin{subfigure}{.14\linewidth}
		\centering
		300
	\end{subfigure}%
	\begin{subfigure}{.14\linewidth}
		\centering
		350
	\end{subfigure}%
	\begin{subfigure}{.14\linewidth}
		\centering
		400
	\end{subfigure}%
	\begin{subfigure}{.14\linewidth}
		\centering
		450
	\end{subfigure}%
	\caption{\textbf{Ball Drop}: An elastic sphere consisting of 64K tets is dropped to the ground. (a) shows the result with the standard UNH model with per-tet Poisson's ratio $\nu = 0.45$, where the ball loses more than half its volume in the second column. (b) is the UNH result with $\nu = 0.495$, where the volumetric locking makes the ball appear unnaturally stiff. Notice how the ball always retains its spherical shape and just gets flattened and stretched in the vertical direction. (c) is the result using our CNH model with global volume constraint and local compression penalty with $\lambda$ equivalent to $\nu = 0.45$. Note that the volume of the sphere is preserved, producing a nice ``squash-and-stretch'' effect, and the artificial stiffness is removed.} \label{fig:fine-ball}
\end{figure*}

We tested a simple dynamic result of a soft ball consisting of 64K tets dropped on the ground (Figure~\ref{fig:fine-ball}). We used a timestep of 1ms and $\mu = 16.0$ KPa. Using a per-tet Poisson's ratio $\nu = 0.495$ results in the sphere behaving much stiffer than what the material parameters would suggest, while still losing up to 12.7\% of its volume. When using a per-tet Poisson's ratio of $\nu = 0.45$, the ball retains its appearance of soft elastic deformation, but loses up to 51\% of its original volume. Using our method, we are able to simulate the soft elastic deformations while preserving the volume down to solver accuracy, while being 5.7\% faster than the high Poisson's ratio case and only 3.7\% slower than the $\nu=0.45$ case.

\begin{figure*}[h!]
	\centering
	\begin{subfigure}{.15\linewidth}
		\rotatebox[origin=c]{0}{\scriptsize{(a) UNH, $\nu=0.45$}}
	\end{subfigure}%
	\begin{subfigure}{.14\linewidth}
		\centering
		\adjustbox{trim={.25\width} {.0\height} {.25\width} {.4\height},clip}%
		{\includegraphics[width=2.0\textwidth]{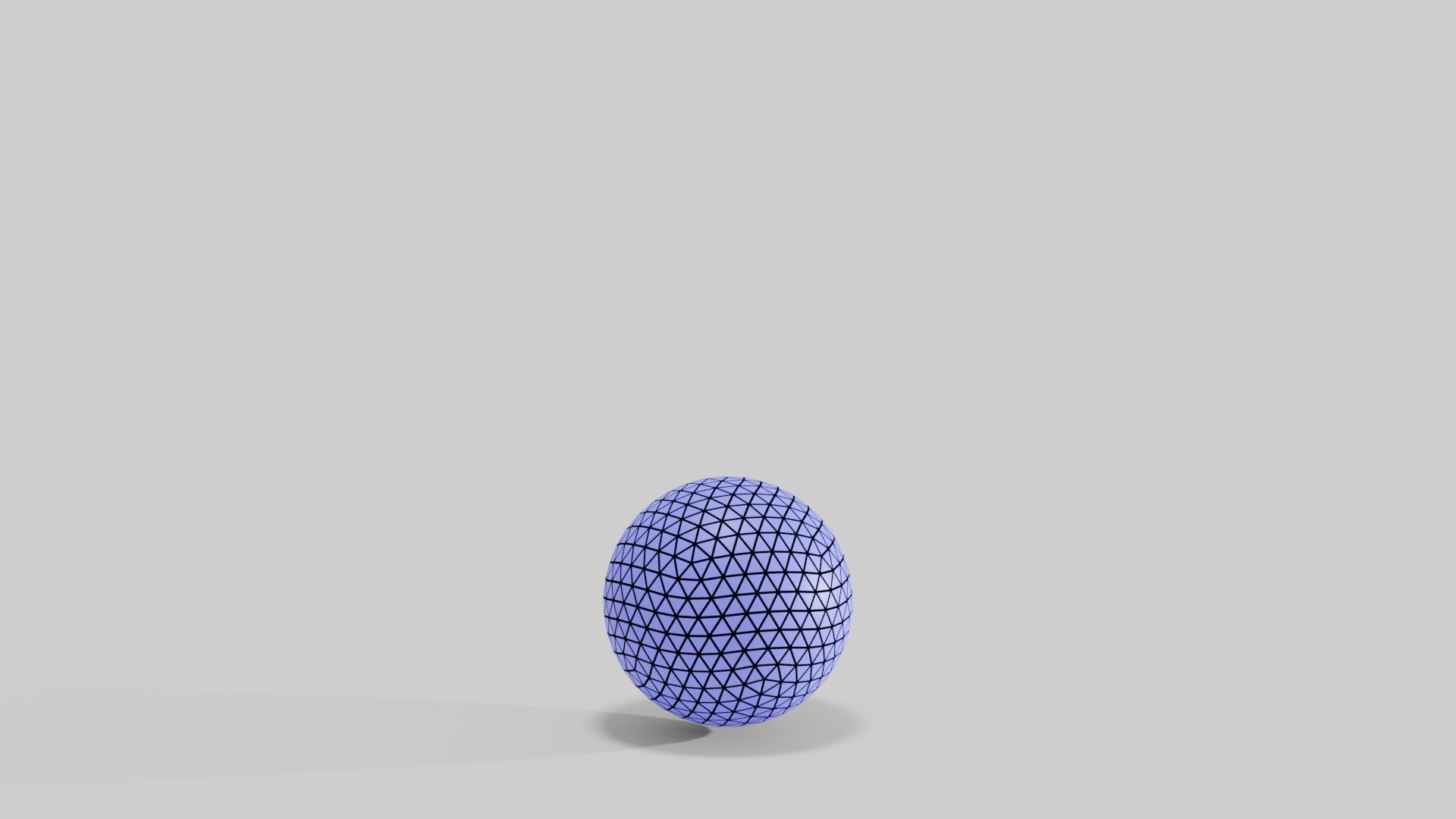}}
		%\caption*{(a1)}
		\label{sfig:ball-045-1}
	\end{subfigure}%
	\begin{subfigure}{.14\linewidth}
		\centering
		\adjustbox{trim={.25\width} {.0\height} {.25\width} {.4\height},clip}%
		{\includegraphics[width=2.0\textwidth]{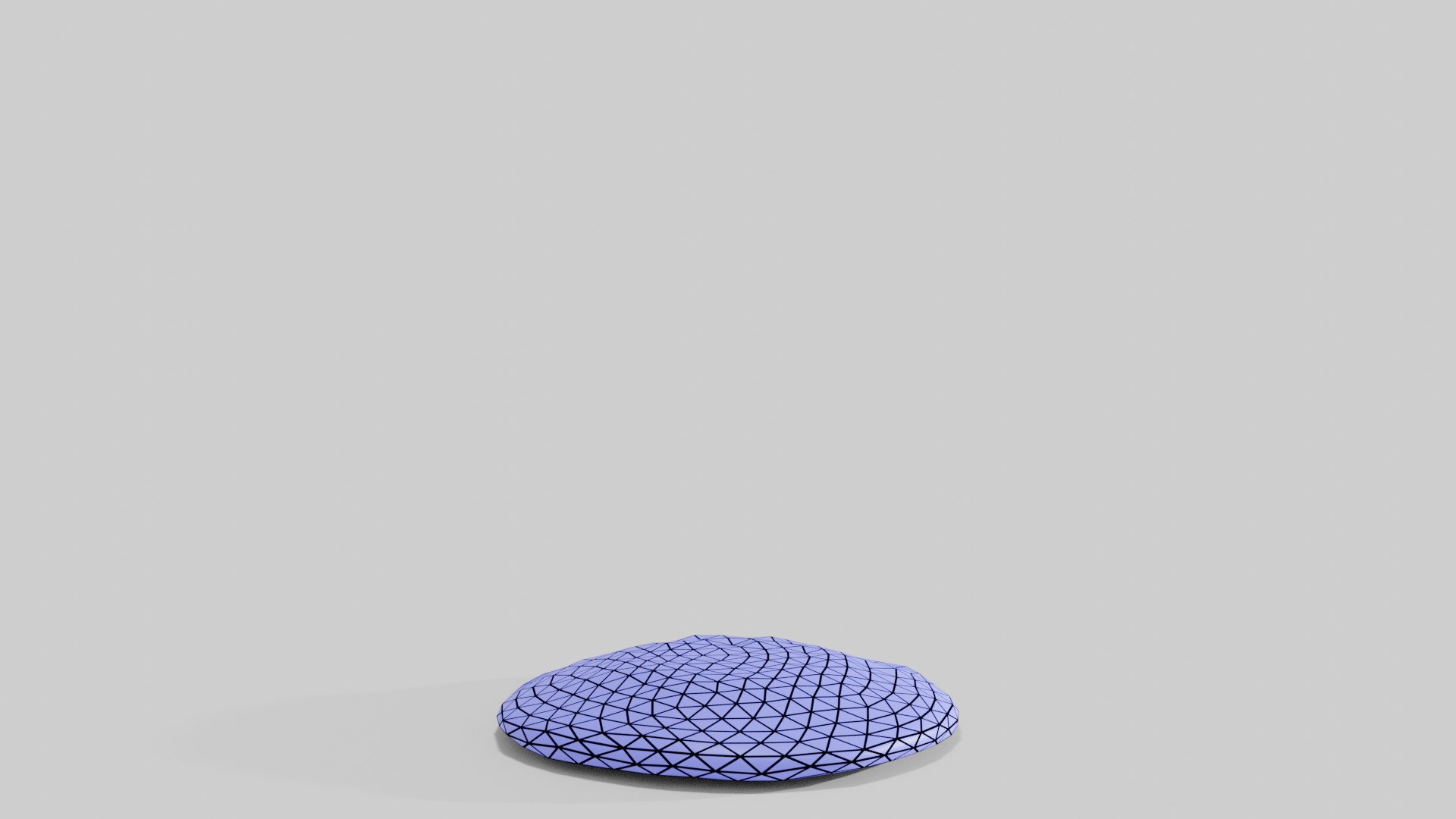}}
		%\caption*{(a2)}
		\label{sfig:ball-045-2}
	\end{subfigure}%
	\begin{subfigure}{.14\linewidth}
		\centering
		\adjustbox{trim={.25\width} {.0\height} {.25\width} {.4\height},clip}%
		{\includegraphics[width=2.0\textwidth]{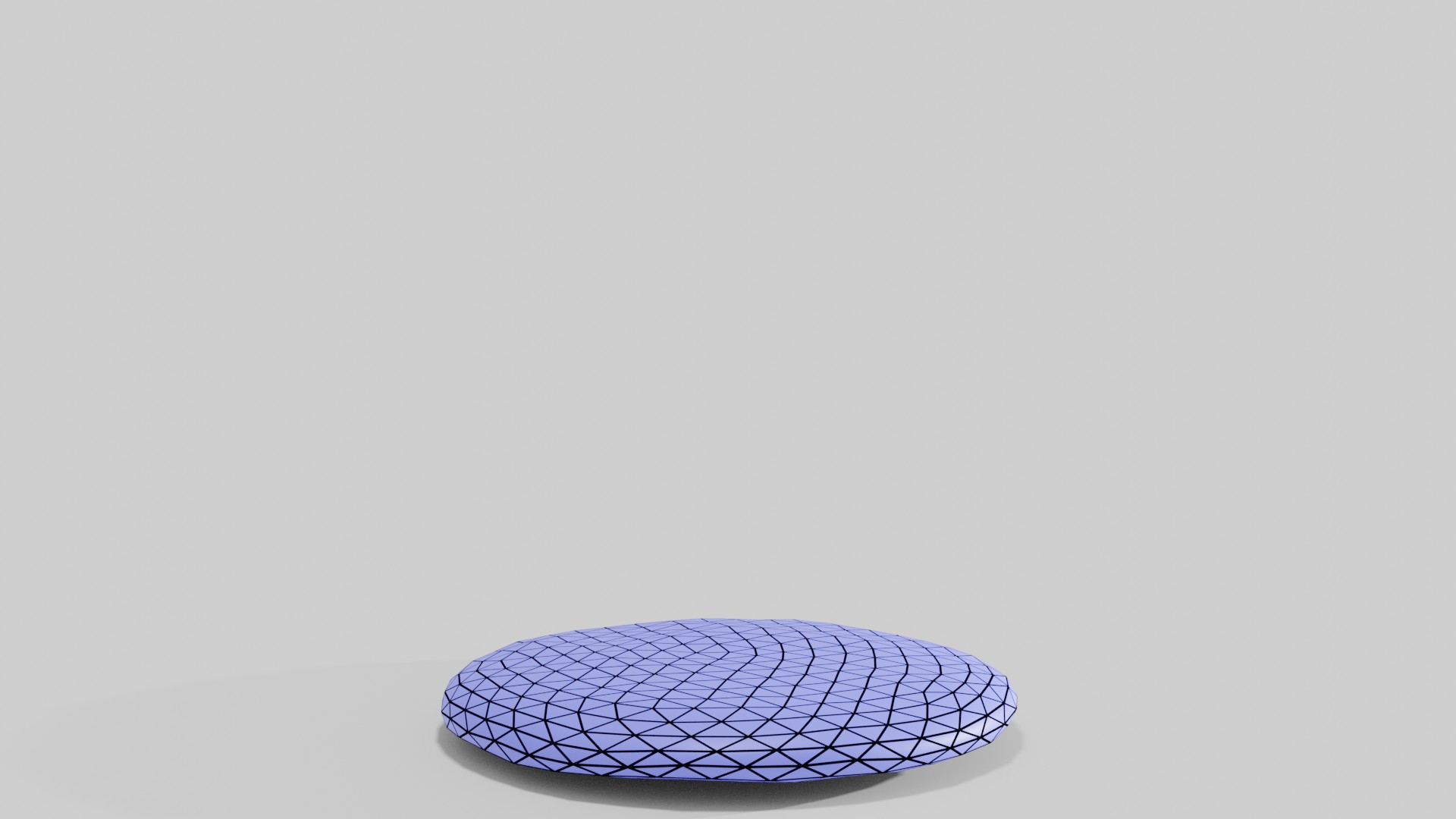}}
		%\caption*{(a3)}
		\label{sfig:ball-045-3}
	\end{subfigure}%
	\begin{subfigure}{.14\linewidth}
		\centering
		\adjustbox{trim={.25\width} {.0\height} {.25\width} {.4\height},clip}%
		{\includegraphics[width=2.0\textwidth]{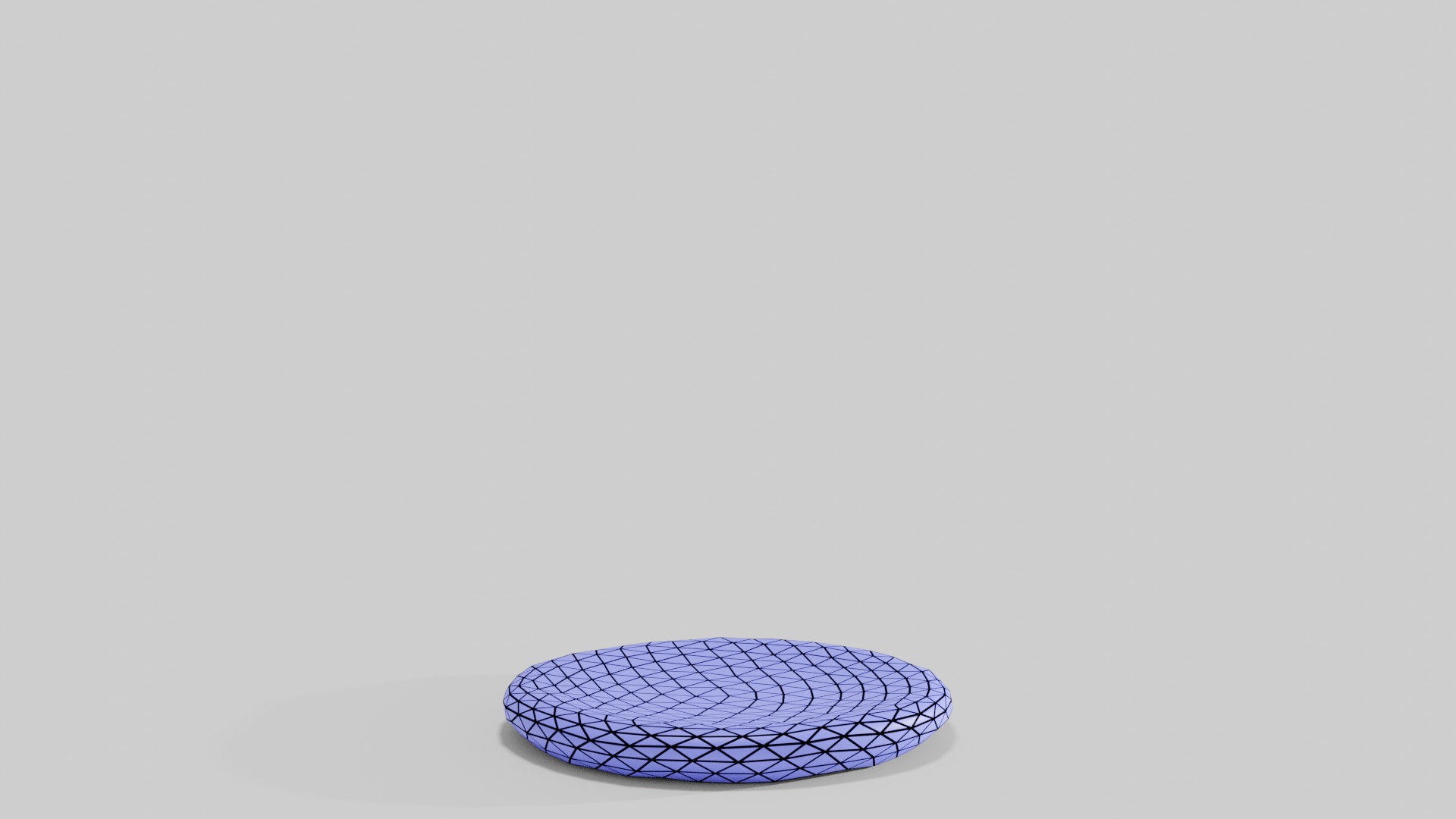}}
		%\caption*{(a4)}
		\label{sfig:ball-045-4}
	\end{subfigure}%
	\begin{subfigure}{.14\linewidth}
		\centering
		\adjustbox{trim={.25\width} {.0\height} {.25\width} {.4\height},clip}%
		{\includegraphics[width=2.0\textwidth]{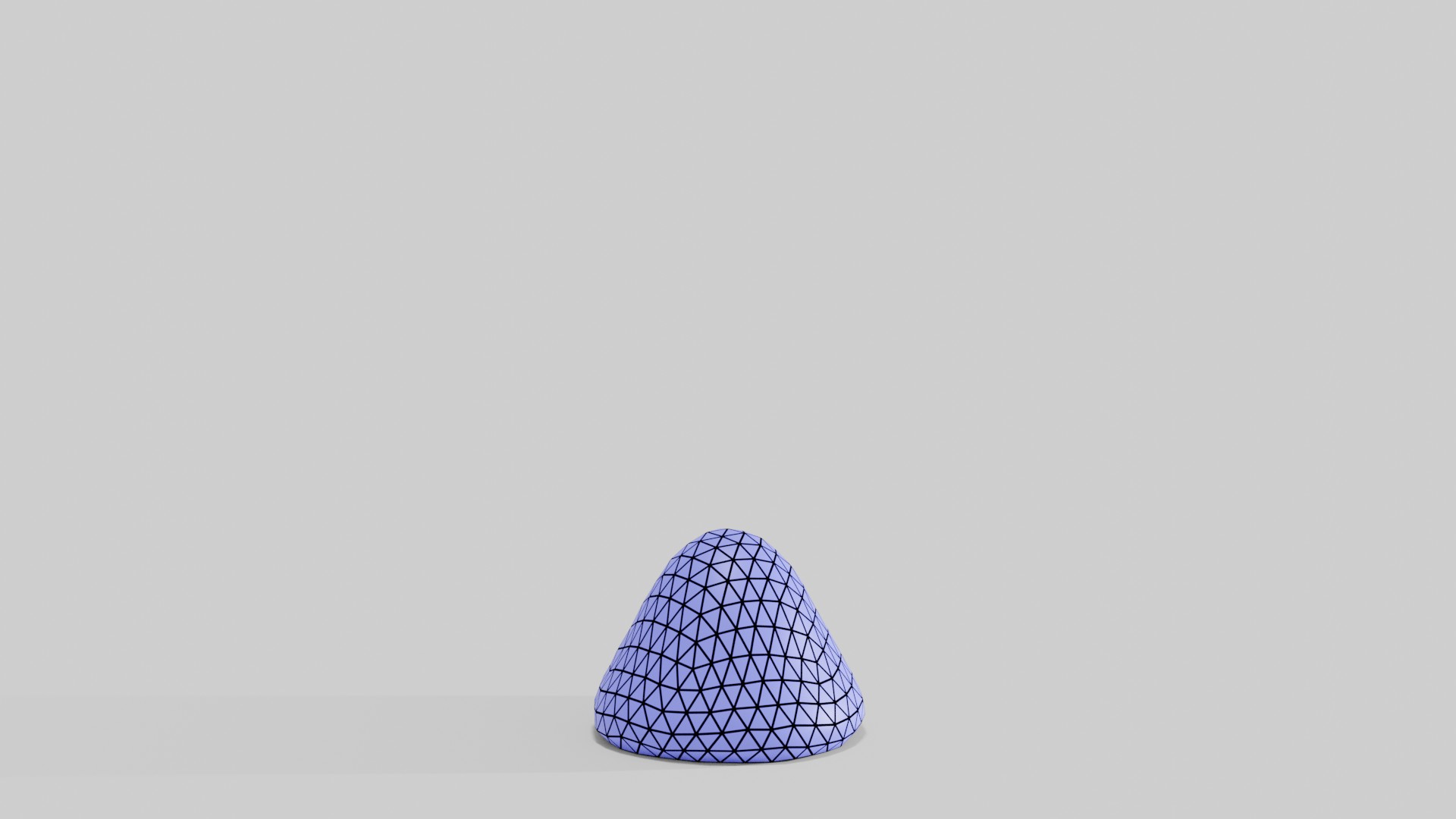}}
		%\caption*{(a5)}
		\label{sfig:ball-045-5}
	\end{subfigure}%
	\begin{subfigure}{.14\linewidth}
		\centering
		\adjustbox{trim={.25\width} {.0\height} {.25\width} {.4\height},clip}%
		{\includegraphics[width=2.0\textwidth]{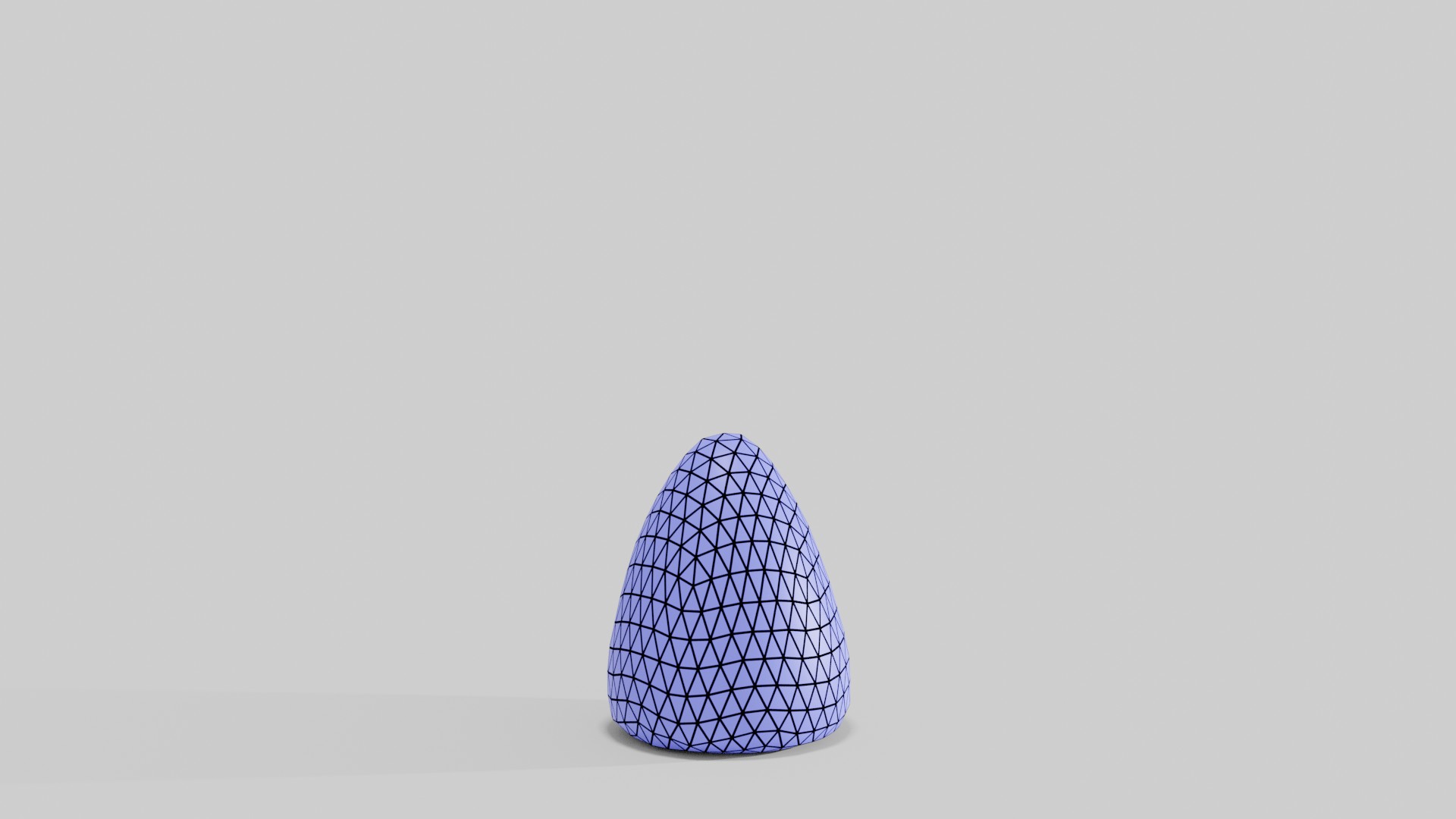}}
		%\caption*{(a6)}
		\label{sfig:ball-045-6}
	\end{subfigure}\hfill
	\begin{subfigure}{.15\linewidth}
		\rotatebox[origin=c]{0}{\scriptsize{(b) UNH, $\nu=0.495$}}
	\end{subfigure}%
	\begin{subfigure}{.14\linewidth}
		\centering
		\adjustbox{trim={.25\width} {.0\height} {.25\width} {.4\height},clip}%
		{\includegraphics[width=2.0\textwidth]{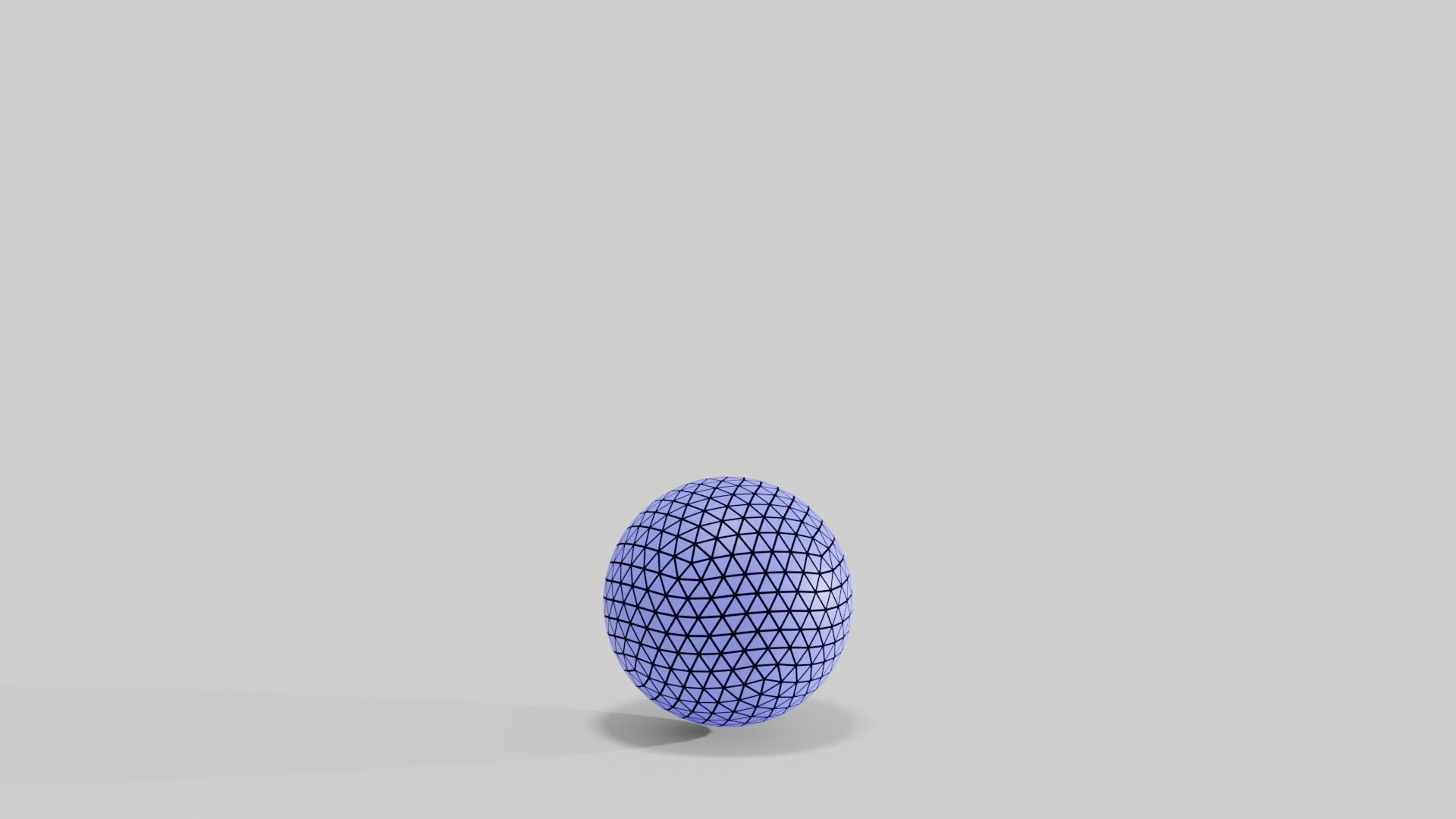}}
		%\caption*{(a1)}
		\label{sfig:ball-0495-1}
	\end{subfigure}%
	\begin{subfigure}{.14\linewidth}
		\centering
		\adjustbox{trim={.25\width} {.0\height} {.25\width} {.4\height},clip}%
		{\includegraphics[width=2.0\textwidth]{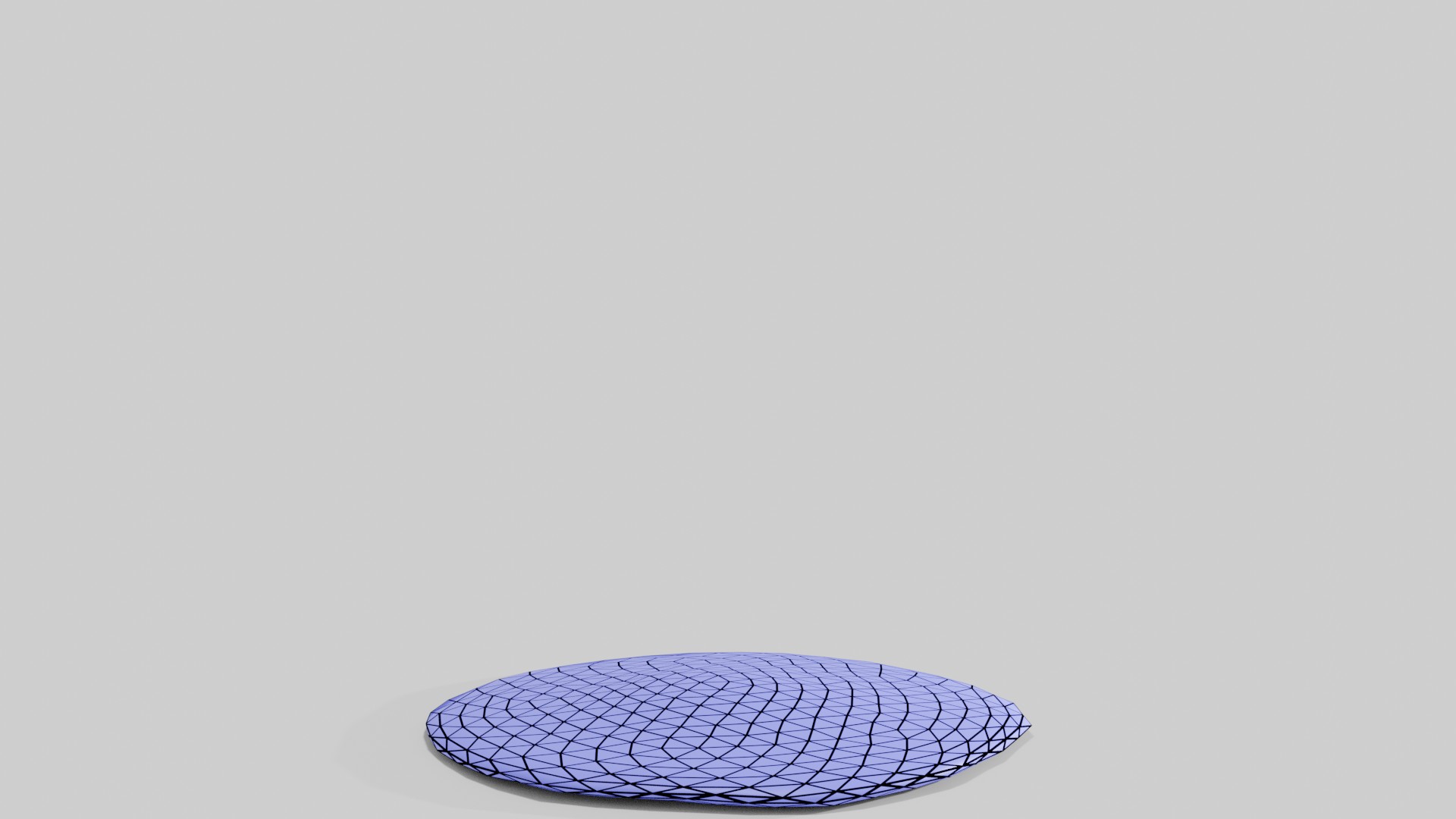}}
		%\caption*{(a2)}
		\label{sfig:ball-0495-2}
	\end{subfigure}%
	\begin{subfigure}{.14\linewidth} \begin {tikzpicture}
		\node [opacity=0.5,inner sep=0pt,anchor=south west,cross out,draw=red] at (0,0) {\adjincludegraphics[trim={.25\width} {.0\height} {.25\width} {.4\height},width=1.0\textwidth,clip]{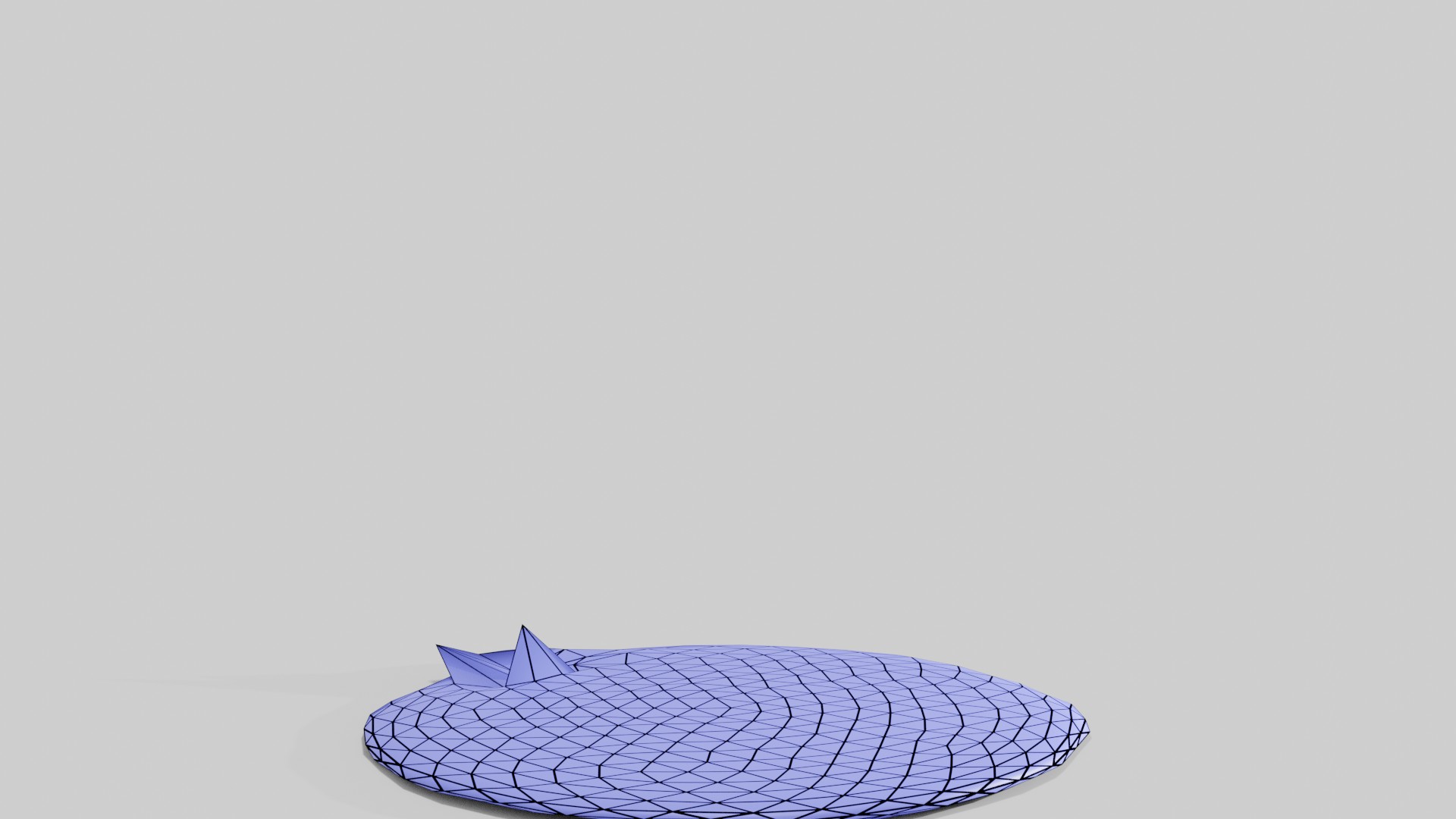}};
		\end {tikzpicture}
		\label{sfig:ball-0495-3}
	\end{subfigure}
	\begin{subfigure}{.42\linewidth}
		\centering
		\textcolor{red}{\footnotesize{\textbf{Simulation failed at the 275th frame.}}}
	\end{subfigure}\hfill
	\begin{subfigure}{.15\linewidth}
		\rotatebox[origin=c]{0}{\scriptsize{(c) Ours}}
	\end{subfigure}%
	\begin{subfigure}{.14\linewidth}
		\centering
		\adjustbox{trim={.25\width} {.0\height} {.25\width} {.4\height},clip}%
		{\includegraphics[width=2.0\textwidth]{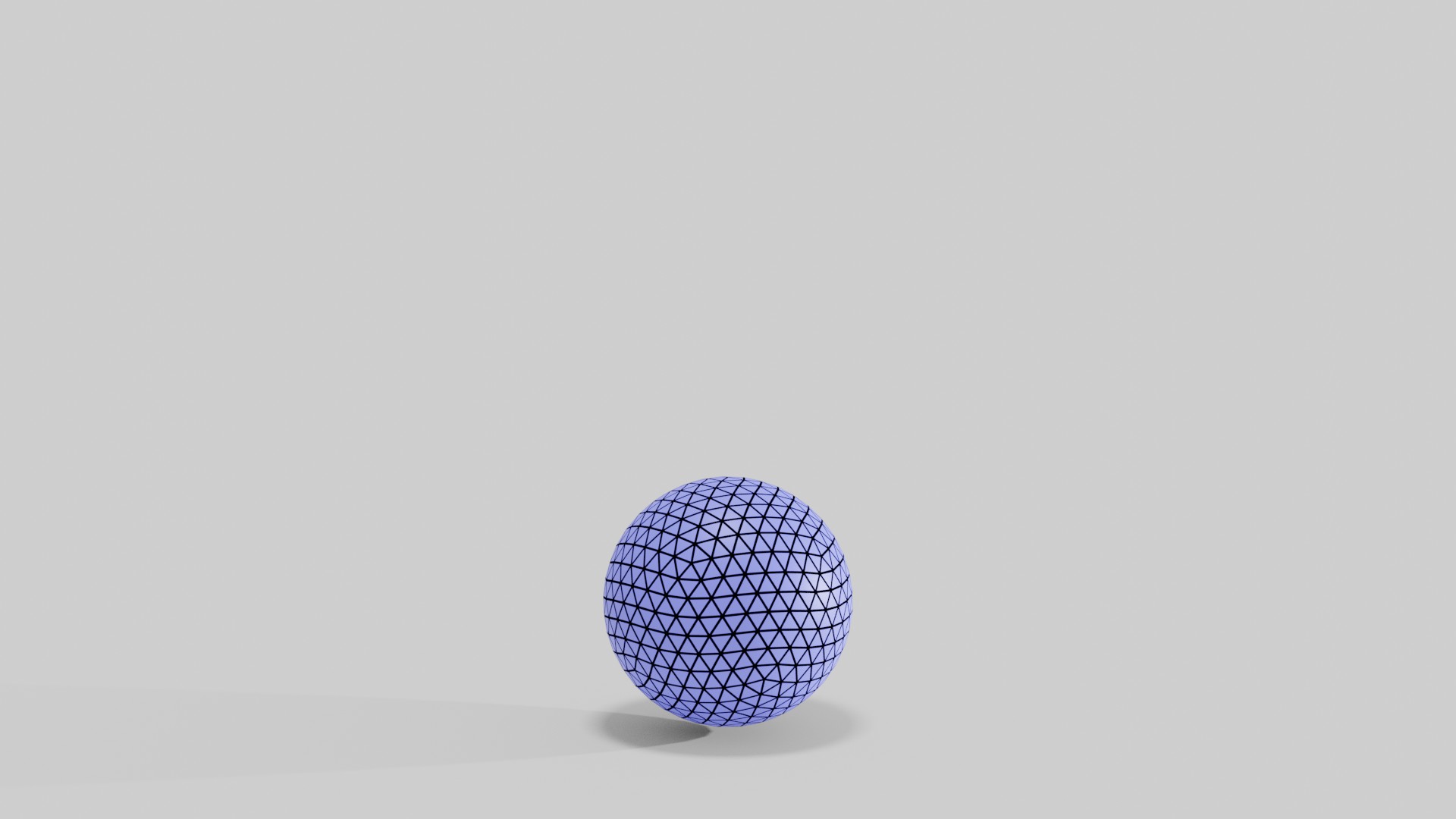}}
		%\caption*{(a1)}
		\label{sfig:ball-vc-1}
	\end{subfigure}%
	\begin{subfigure}{.14\linewidth}
		\centering
		\adjustbox{trim={.25\width} {.0\height} {.25\width} {.4\height},clip}%
		{\includegraphics[width=2.0\textwidth]{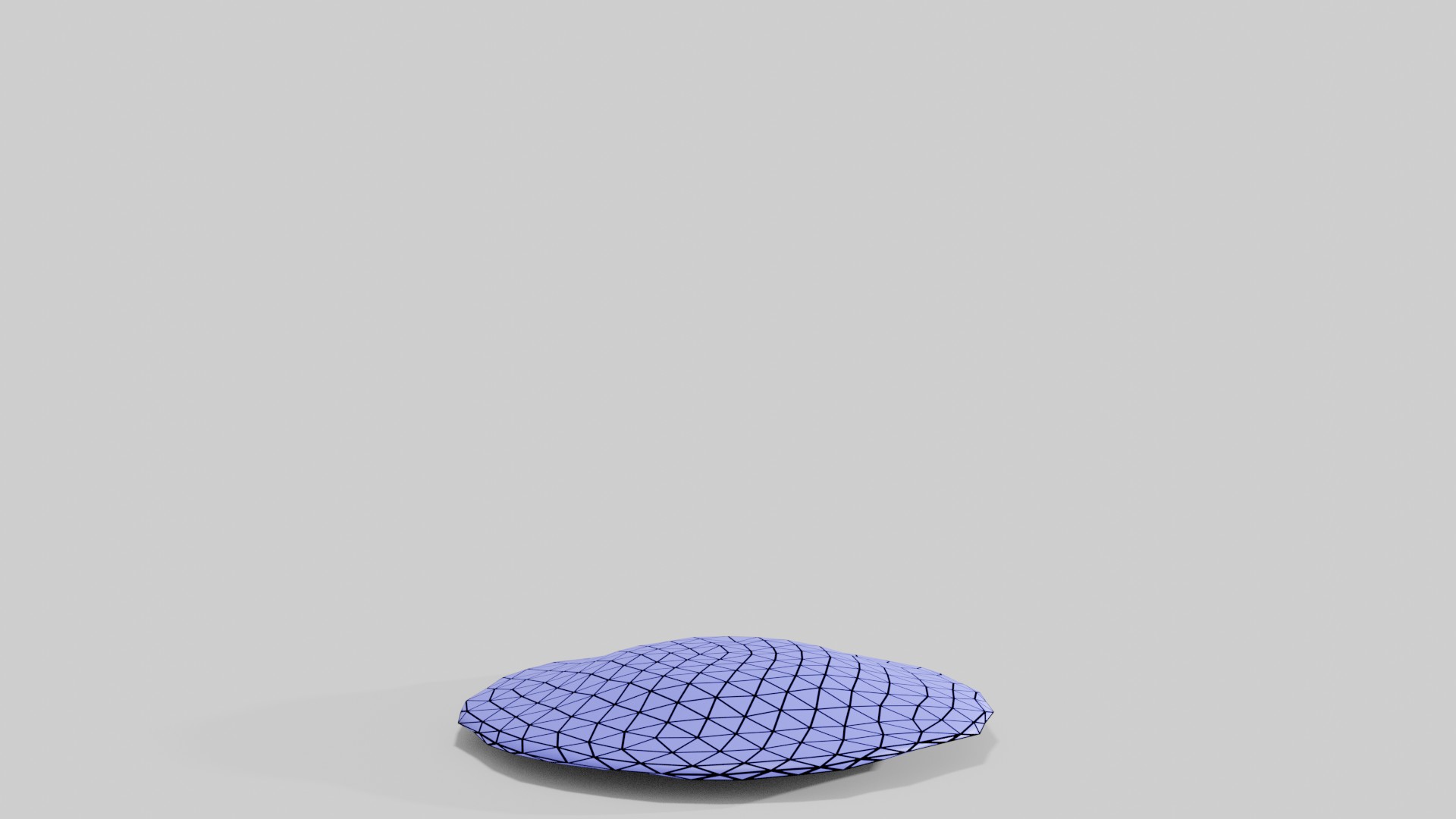}}
		%\caption*{(a2)}
		\label{sfig:ball-vc-2}
	\end{subfigure}%
	\begin{subfigure}{.14\linewidth}
		\centering
		\adjustbox{trim={.25\width} {.0\height} {.25\width} {.4\height},clip}%
		{\includegraphics[width=2.0\textwidth]{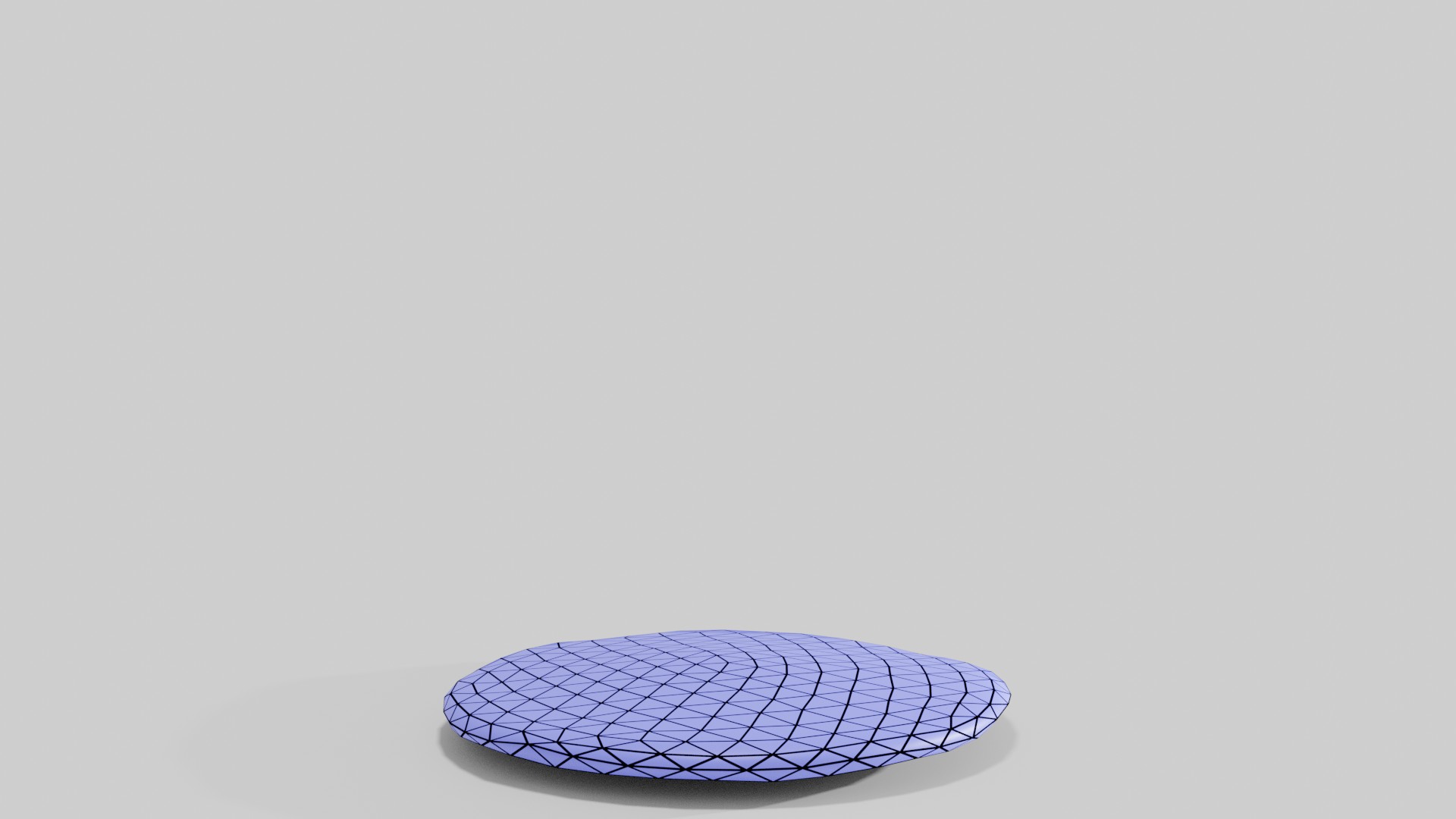}}
		%\caption*{(a3)}
		\label{sfig:ball-vc-3}
	\end{subfigure}%
	\begin{subfigure}{.14\linewidth}
		\centering
		\adjustbox{trim={.25\width} {.0\height} {.25\width} {.4\height},clip}%
		{\includegraphics[width=2.0\textwidth]{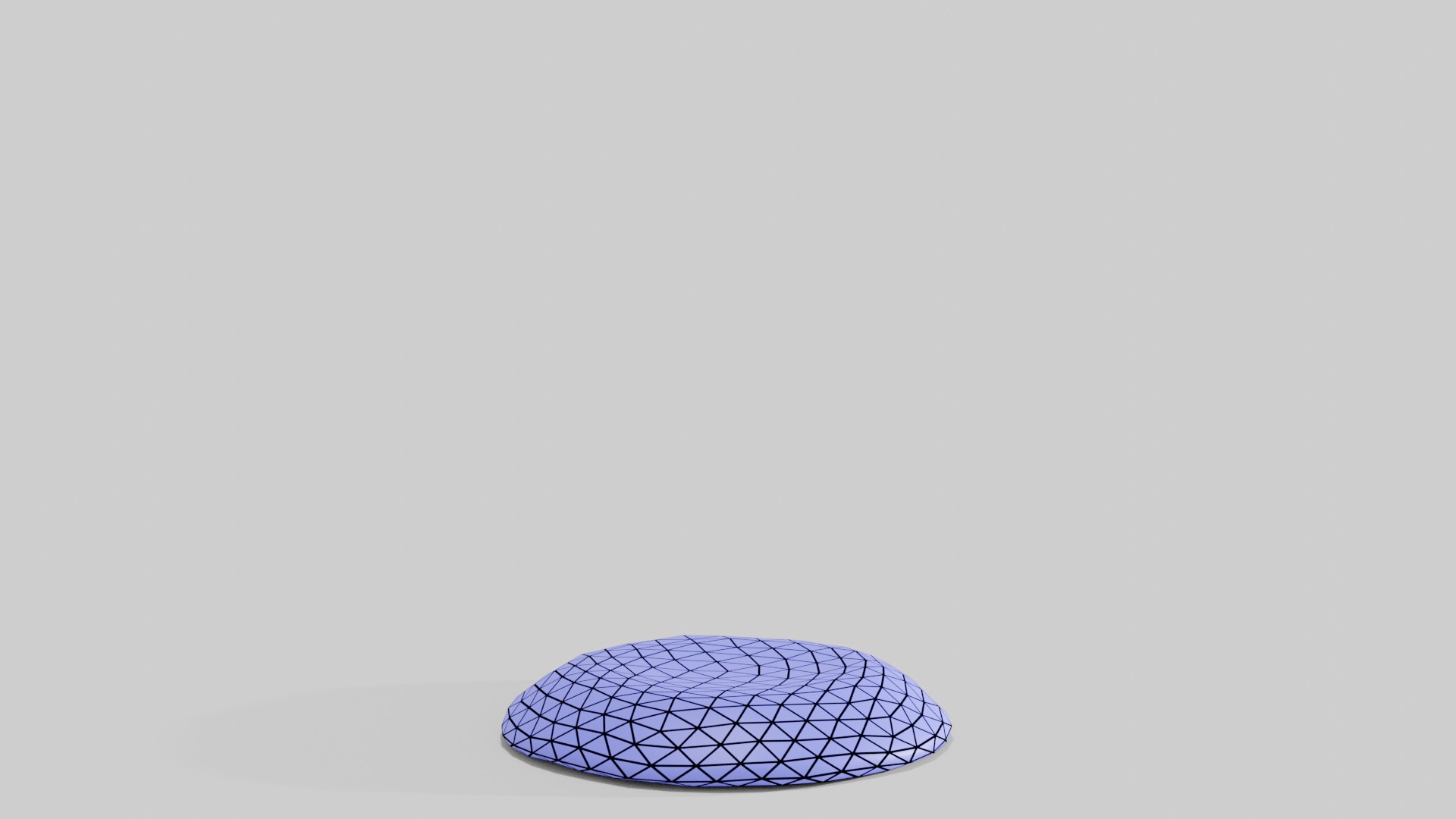}}
		%\caption*{(a4)}
		\label{sfig:ball-vc-4}
	\end{subfigure}%
	\begin{subfigure}{.14\linewidth}
		\centering
		\adjustbox{trim={.25\width} {.0\height} {.25\width} {.4\height},clip}%
		{\includegraphics[width=2.0\textwidth]{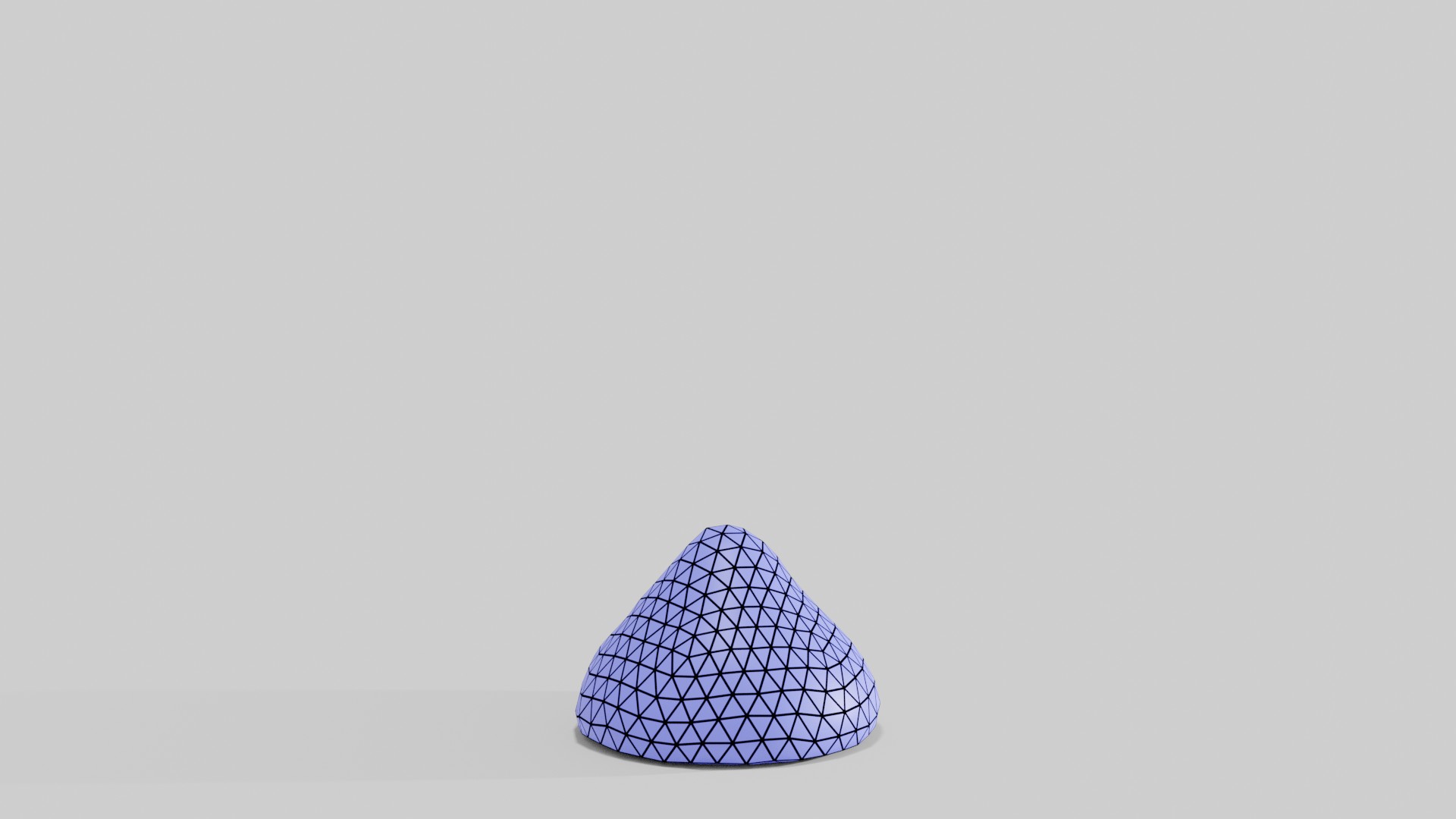}}
		%\caption*{(a5)}
		\label{sfig:ball-vc-5}
	\end{subfigure}%
	\begin{subfigure}{.14\linewidth}
		\centering
		\adjustbox{trim={.25\width} {.0\height} {.25\width} {.4\height},clip}%
		{\includegraphics[width=2.0\textwidth]{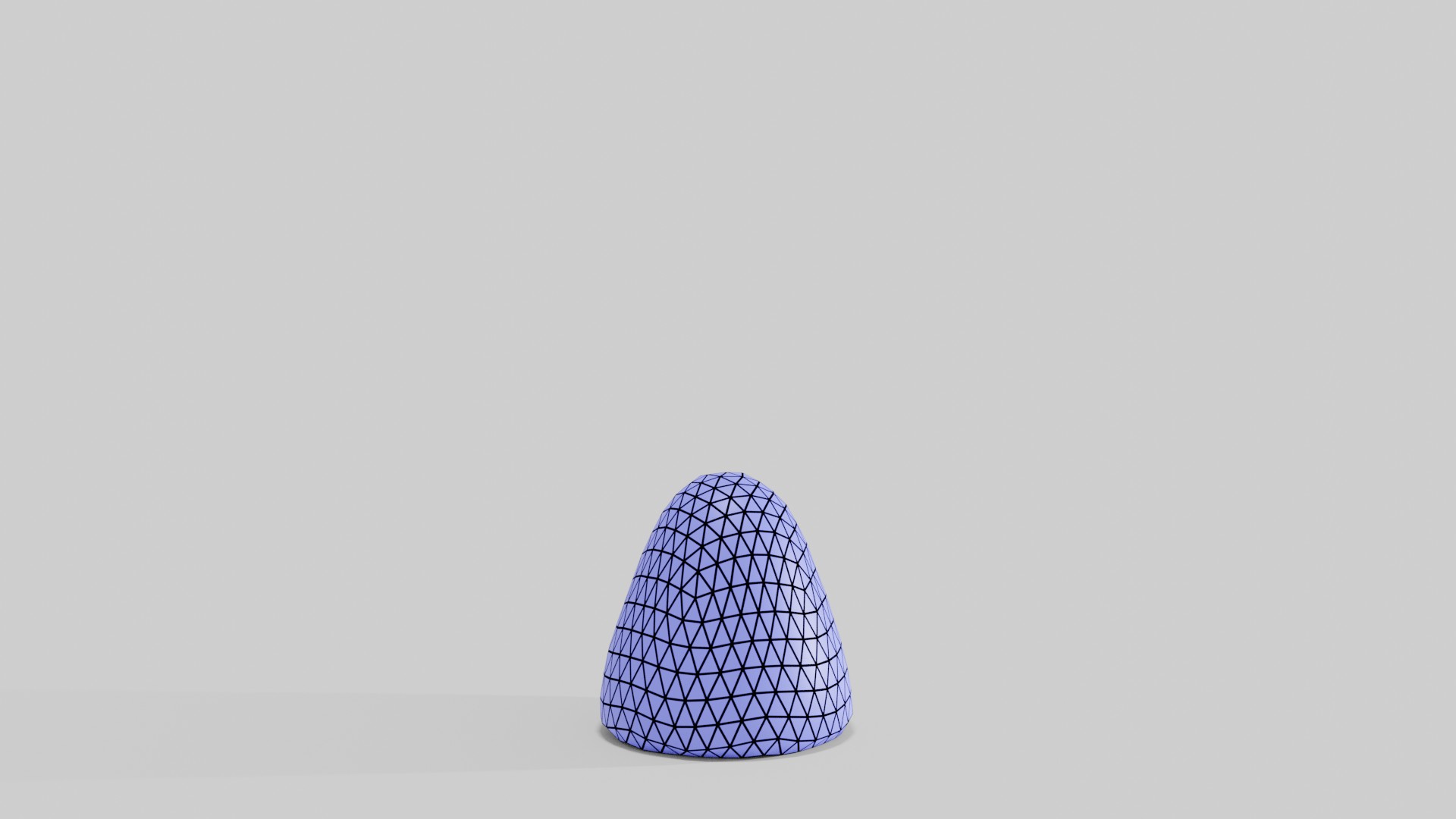}}
		%\caption*{(a6)}
		\label{sfig:ball-vc-6}
	\end{subfigure}\hfill
	\begin{subfigure}{.15\linewidth}
		\rotatebox[origin=c]{0}{\scriptsize{Frame}}
	\end{subfigure}%
	\begin{subfigure}{.14\linewidth}
		\centering
		200
	\end{subfigure}%
	\begin{subfigure}{.14\linewidth}
		\centering
		250
	\end{subfigure}%
	\begin{subfigure}{.14\linewidth}
		\centering
		300
	\end{subfigure}%
	\begin{subfigure}{.14\linewidth}
		\centering
		350
	\end{subfigure}%
	\begin{subfigure}{.14\linewidth}
		\centering
		400
	\end{subfigure}%
	\begin{subfigure}{.14\linewidth}
		\centering
		450
	\end{subfigure}%
	\caption{\textbf{Coarse ball}: Similar to Figure~\ref{fig:fine-ball}, a much coarser sphere with 4.7K tets is dropped to the ground. (a) shows the UNH result  with $\nu = 0.45$, where the ball has lost 52\% of its original volume at the 250th Frame, but gain 36\% volume at the 300th frame. (b) is the result of using UNH with $\nu = 0.495$, where due to using a coarser mesh the issue of locking is exacerbated, and the simulation fails to converge at the 275th frame. (c) is our result, where the simulation is stable and consistent with the result when using a much finer mesh, demonstrating our advantage of resolution consistency. } \label{fig:coarse-ball}
\end{figure*}

\subsection{Resolution Consistency}

An important advantage of enforcing volume preservation with zonal constraints is that it allows a way of simulating incompressible
objects using a much coarser mesh than by using a traditional 1-field method. C\'ea's Lemma already couples the quasi-best approximation error
with mesh resolution, and since a 1-field FE solutions also couple the bulk modulus to the upper bound of the approximation error, it makes it
even harder to use a coarser mesh when bulk modulus is high. However, when incompressibility is decoupled from the bulk modulus, and we can
use much smaller $\lambda$, we are able to achieve simulation results of a fine-mesh simulation that is consistent with a much coarser mesh.
%Figure~\ref{fig:fine-ball} shows a simple dynamic simulation with a very fine mesh, and Figure~\ref{fig:coarse-ball} shows the same simulation
%using a much coarser mesh. We see that both the low Poisson ratio and our method achieves consistent visual results between the fine and coarse
%case, but the high Poisson ratio case fails to converge very early in the simulation.

When a coarser mesh (4.7K tets) is used, the advantage of our method becomes even clearer (see Figure~\ref{fig:coarse-ball}). For the low Poisson's ratio example, the maximum volume loss is almost equal to when using a finer mesh (52.8\%). But after its impact with the ground, the ball actually gains volume due to the severe volumetric deformations resulting in extremely high volumetric elastic force, and the ball gains up to 36.1\% of its initial volume. The high Poisson's ratio case fails to converge after the 275th frame (corresponds to the 0.275th second). This failure to converge when using a coarse mesh demonstrates how locking is aggravated when the simulation mesh is coarser, leading to a extremely high approximation error as predicted by C\'ea's Lemma. However, using our method allows a simulation of a completely volume preserving soft elastic ball even with a very coarse mesh. The visual result is consistent with the result using finer resolution, demonstrating that our method allows a resolution-consistent simulation of volume preserving soft objects.

\subsection{Additional Results}
In addition to the examples above, we show how our method performs in large scale deformations and further demonstrate its robustness. 
We present three large scale simulations of soft tissue with skin, one is shown in Figure 1, and two examples show a deformation of an armadillo model when pressed between cylinders. 
To demonstrate robustness. we stretch and twist an elastic cube producing large deformations, which can pose problems for traditional simulators.

\scriptsize
\begin{table}[]
	\centering
	\begin{tabular}{@{}rcccccccc@{}}
		\toprule
		Example       & Model & VC  & $\lambda$ & $\beta$ & $\lambda_e$ & $t_r$  & $ i_{\textbf{avg}}$ \\
		\midrule
		Cloth-Body    & SNH   & no  & 400.0     &         & 0.0         & 12.10  & 21.32               \\
		Cloth-Body    & SNH   & no  & 40.0      &         & 0.0         & 12.93  & 23.09               \\
		Cloth-Body    & Ours  & yes & 120.0     & 6.0     & 0.0         & 12.82  & 17.98               \\
		Cloth-Body    & Ours  & yes & 120.0     & 6.0     & 100.0       & 14.63  & 19.18               \\
		\midrule
		Puck          & SNH   & no  & 40.0      &         & 0.0         & 1.55   & 3.00                \\
		Puck          & SNH   & no  & 400.0     &         & 0.0         & 1.58   & 3.14                \\
		Puck          & Ours  & yes & 100.0     & 0.0     & 0.0         & 2.25   & 4.00                \\
		Puck          & Ours  & yes & 100.0     & 9.0     & 0.0         & 1.93   & 3.38                \\
		Puck          & Ours  & yes & 100.0     & 9.0     & 100.0       & 2.25   & 3.51                \\
		\midrule
		Suspend       & SNH   & no  & 400.0     &         & 0.0         & 1.02   & 7.28                \\
		Suspend       & Ours  & yes & 40.0      & 0.0     &             & 1.02   & 6.52                \\
		\midrule
		Ball (Fine)   & NH    & no  & 120.0     &         & 0.0         & 17.27  & 14.64               \\
		Ball (Fine)   & NH    & no  & 400.0     &         & 0.0         & 19.00  & 15.90               \\
		Ball (Fine)   & Ours  & yes & 120.0     & 1.0     & 0.0         & 17.91  & 12.29               \\
		Ball (Coarse) & NH    & no  & 120.0     &         & 0.0         & 1.81   & 15.08               \\
		Ball (Coarse) & NH    & no  & 400.0     &         & 0.0         & 2.50   & 21.84               \\
		Ball (Coarse) & Ours  & yes & 120.0     & 1.0     & 0.0         & 2.71   & 22.10               \\
		\midrule
		Armadillo     & SNH   & no  & 400.0     &         & 40.0        & 6.26   & 11.29               \\
		Armadillo     & Ours  & yes & 60.0      & 12.0    & 40.0        & 6.97   & 10.84               \\
		\midrule
		Stretch       & SNH   & no  & 40.0      &         & 0.0         & 2.80   & 4.51                \\
		Stretch       & SNH   & no  & 400.0     &         & 0.0         & 9.91   & 14.33               \\
		Stretch       & Ours  & yes & 25.0      & 0.0     & 0.0         & 7.89   & 9.92                \\
		Stretch       & Ours  & yes & 25.0      & 1.0     & 0.0         & 7.28   & 9.24                \\
		Stretch       & Ours  & yes & 25.0      & 1.0     & 10.0        & 6.25   & 7.33                \\
		\midrule
		Twist         & SNH   & no  & 100.0     &         & 0.0         & 86.06  & 138.33              \\
		Twist         & Ours  & no  & 100.0     & 9.0     & 0.0         & 37.32  & 50.33               \\
		Twist         & Ours  & yes & 100.0     & 6.0     & 0.0         & 62.15  & 90.33               \\
		Twist         & Ours  & yes & 100.0     & 6.0     & 15.0        & 103.10 & 92                  \\
		\bottomrule
	\end{tabular}%
	\caption{\textbf{Performance Analysis}: Performance table representing average run times per frame, and average Ipopt iterations per frame for major examples. SNH is the Stable neo-Hookean
		% \eqref{e-snh}
		\cite{Smith:2018}. (VC: Volume Constraint, $t_r$: Run Times per Frame (sec.), $i_{\textbf{avg}}$: Avg. Ipopt Iterations.)}
	\label{tab:performance}
\end{table}

\normalsize

%%% Local Variables:
%%% mode: latex
%%% TeX-master: "sigconf"
%%% End:

\section{Limitations and Conclusion}

Although our method generates more realistic volume preservation and
reduces locking, for simulations without large deformations, the
standard Neo-Hookean model may be sufficient due to its
simplicity. This is especially true when no other constraints, such as due to external contact, are present in the simulation. Constrained optimization adds
some complexity to the simulation, though our results in Table
\ref{tab:performance} show that the increase is computation times is
not prohibitive in most cases. Overall, our findings demonstrate an inexpensive
extension to existing FEM systems can effectively solve the problem of volumetric
locking while simulating incompressible materials such as the human
body.

\paragraph{Performance} Our formulation uses exact non-linear volume constraints on a non-linear
optimization problem to preserve volume exactly in demanding applications like statics and dynamics
with large time steps.  This limits the choice of optimization solvers to ones that
support non-linear equality constraints (e.g., Interior Point or SQP solvers).  However, for dynamics problems
with smaller time steps, or in applications with tolerance for volume loss/gain, 
linearizing the volume constraint can drastically simplify the problem. This can reduce the
overhead of enforcing equality constraints, while still avoiding locking.

\paragraph{Choosing $\lambda$} By decoupling $\lambda$ from its interpretation as a material parameter, our formulation is
faced with an additional challenge, which is to determine how exactly $\lambda$ affects the outcome
of a simulation.  Fortunately, this is not a significant drawback since material parameters for
standard Neo-Hookean FEM simulations also deviate from their measured values due to numerical
stiffenning. This means that even the parameters of standard models require manual tuning to
reproduce real phenomena in simulation.  Luckily data-driven methods for determining simulation
parameters (which has seen significant attention in recent literature) are generally agnostic to the
true physical meaning of these parameters, and thus are equally as compatible with our method.

In conclusion, we presented a general method for realistic volumetric FEM simulations of human soft tissue.
Our method provides exact volume preservation without the artificial stiffness due to volumetric
locking using zonal volume constraints. This method gives modelers the ability to define volume
preserving zones that conform to anatomical compartments and automatically produces
``squash-and-stretch'' effects. In addition, we introduced an epidermis model for simulating
skin mechanics, as an additional surface area-preserving potential. We also proposed a
modification to the energy potential to provide control over local volume flow
that results in improved
recovery during extreme compression and inversion. 
We also demonstrated how our method allows stable simulations of volume preserving materials with
coarse meshes that are consistent with high-resolution simulations.
Our approach can be applied to a variety of energy
models. In particular, we have demonstrated the effectiveness of these simple modifications to the
invariant-based non-linear hyperelastic energies such as the Neo-Hookean and Stable Neo-Hookean energy
models.

\section*{Acknowledgements}
This research was supported in part NSERC, the Canada Research Chairs program, and an NSERC Idea-to-Innovation project co-sponsored by Vital Mechanics. The authors would like to thank Ye Fan for his early contributions to this research.

\clearpage

\bibliographystyle{ACM-Reference-Format}
\bibliography{bibliography}

\end{document}